\documentclass{article}
\usepackage{titling}
\usepackage{datetime}
\usepackage{float}
\usepackage{pifont}
\usepackage{minitoc}

\usepackage[dvipsnames]{xcolor}
\usepackage{booktabs}
\usepackage{colortbl}
\usepackage{tabularx}
\definecolor{myorange}{rgb}{1, 0.647, 0}
\definecolor{mckblue}{RGB}{48,57,171}
\usepackage{hyperref}
\hypersetup{
    colorlinks,
    linkcolor={red!50!black},
    citecolor={blue!50!black},
    urlcolor={blue!80!black}
}

\usepackage{enumitem}
\makeatletter
\def\namedlabel#1#2{\begingroup
    #2%
    \def\@currentlabel{#2}%
    \phantomsection\label{#1}\endgroup
}
\makeatother

\usepackage[top=2.5cm,bottom=2.5cm,textwidth=180mm]{geometry}

\usepackage{amsmath, amssymb, mathtools, amsthm}
\usepackage{mathrsfs}
\usepackage{empheq}
\usepackage[thinc]{esdiff}
\usepackage{accents}

\usepackage{ragged2e}

\newtheorem{exmp}{Example}[section]

\DeclareMathOperator*{\argmin}{arg\,min}

\DeclareMathOperator*{\Rplus}{\mathbb{R}_{+}}

\DeclareMathOperator*{\Natural}{\mathbb{N}}

\usepackage{tikz}
\usepackage{bbold}
\usepackage[font=small]{caption}
\DeclareCaptionLabelSeparator{pipe}{ \textbar{} }
\usepackage[font=small]{subcaption}

\usepackage[section]{placeins}
\usepackage{siunitx}
\usepackage{algpseudocode}
\usepackage{algorithm}

\allowdisplaybreaks

\usepackage{mdframed}
\newmdenv[
  topline=false,
  bottomline=false,
  rightline=false,
  skipabove=\topsep,
  skipbelow=\topsep,
  leftmargin=2pt,
  rightmargin=0pt,
  innertopmargin=0pt,
  innerbottommargin=0pt
]{sideline}

\usepackage[version=4]{mhchem}
\def\arrvline{\hfil\kern\arraycolsep\vline\kern-\arraycolsep\hfilneg}

\usepackage[sorting=none, style=nature]{biblatex}
\AtBeginEnvironment{proof}{\footnotesize \linespread{0.7}}

\title{Interpretable Neural Approximation of Stochastic Reaction Dynamics \\ with Guaranteed Reliability}

\author{Quentin Badolle$^{1,*}$, Arthur Theuer$^{1,*}$, Zhou Fang$^{2}$, Ankit Gupta$^1$, Mustafa Khammash$^{1,\dag}$}

\usepackage[para]{footmisc}
\renewcommand{\footnoterule}{%
  \kern 3mm  %
  \hrule width 0.4\textwidth
  \kern 2mm %
}

\date{}

\begin{document}

\maketitle

\begingroup
\renewcommand{\thefootnote}{} %
\renewcommand{\footnotelayout}{\noindent} %
\footnotetext{$^1$Department of Biosystems Science and Engineering, ETH Zurich, 4056 Basel, Switzerland. $^2$State Key Laboratory of Mathematical Sciences, Academy of Mathematics and Systems Science, Chinese Academy of Sciences, Beijing 100190, China. $^*$These authors contributed equally: Quentin Badolle, Arthur Theuer. $^\dag$To whom correspondence should be addressed: \href{mailto:mustafa.khammash@bsse.ethz.ch}{mustafa.khammash@bsse.ethz.ch}.}
\addtocounter{footnote}{-1} %
\endgroup

\vspace{-2cm}
\pdfbookmark[1]{Abstract}{abstract}
\begin{abstract}
Stochastic Reaction Networks (SRNs) are a fundamental modeling framework for systems ranging from chemical kinetics and epidemiology to ecological and synthetic biological processes. A central computational challenge is the estimation of expected outputs across initial conditions and times, a task that is rarely solvable analytically and becomes computationally prohibitive with current methods such as Finite State Projection or the Stochastic Simulation Algorithm. Existing deep learning approaches offer empirical scalability, but provide neither interpretability nor reliability guarantees, limiting their use in scientific analysis and in applications where model outputs inform real-world decisions. Here we introduce DeepSKA, a neural framework that jointly achieves interpretability, guaranteed reliability, and substantial computational gains. DeepSKA yields mathematically transparent representations that generalise across states, times, and output functions, and it integrates this structure with a small number of stochastic simulations to produce unbiased, provably convergent, and dramatically lower-variance estimates than classical Monte Carlo. We demonstrate these capabilities across nine SRNs, including nonlinear and non-mass-action models with up to ten species, where DeepSKA delivers accurate predictions and orders-of-magnitude efficiency improvements. This interpretable and reliable neural framework offers a principled foundation for developing analogous methods for other Markovian systems, including stochastic differential equations.
\end{abstract}

\vspace{0.5cm}
\begin{refsection}
\pdfbookmark[1]{Introduction}{introduction}
Reaction networks provide a convenient language to mechanistically describe the evolution of systems across the life sciences~\cite{allen2008introduction,irurzun2020beyond,erdi1989mathematical}. Such systems are composed of \emph{species}, which may represent types of individuals in an epidemic, animals in an ecosystem, cells in an organism, ion channels in a neuron, or molecules in a cell. The \emph{state} of the system is described at any given time by the abundance of these species, which evolves through events referred to as \emph{reactions}. Each reaction event results in the consumption of specific reactants and the formation of corresponding products.\newline

When such reaction networks describe intracellular biochemical processes, stochasticity and discreteness become defining features. Cell-to-cell variability within populations of genetically identical cells is omnipresent throughout molecular biology~\cite{elowitz2002stochastic,swain2002intrinsic}, and it can have profound functional and therapeutic implications~\cite{briat2023noise,filo2023biomolecular}. This phenomenon has received renewed attention in the growing body of single-cell studies, facilitated by advanced measurement techniques such as flow cytometry, time-lapse microscopy, and single-cell RNA sequencing~\cite{munsky2009listening,golding2005real,la2018rna}. Part of this variability can be attributed to the random occurrence of chemical reactions due to the low copy number of some molecular species~\cite{schwanhausser2011global}. Moreover, the abundance of molecular species fundamentally takes discrete values. Stochastic Reaction Networks~(SRNs) are reaction network models that capture both the stochastic occurrence of chemical reactions and the discreteness of molecular abundances~\cite{mazza2014stochastic,wilkinson2018stochastic,anderson2015stochastic}. Mathematically, SRNs are Continuous-Time Markov Chains (CTMCs), a modeling framework widely used both within and beyond the life sciences, where they are also known as discrete-event systems~\cite{plyasunov2007efficient,asmussen2007stochastic}.\newline

A central aim in the study of SRNs is to predict the temporal evolution of the probability distribution of the system given an initial distribution. Letting $X(t)$ denote the state of the SRN at time $t$, the problem can be formally stated as determining the following:
\begin{equation}
p(x,t) \coloneqq \mathbb{P}(X(t)=x), \quad \forall x, \; \forall t.
\end{equation}

The distribution $p$ evolves according to the Kolmogorov forward equation, a system of coupled Ordinary Differential Equations~(ODEs) referred to in the chemical literature as the Chemical Master Equation~(CME)~\cite{gardiner1985handbook,van1992stochastic,del2015biomolecular}. In all but a few cases, analytical solutions to the CME are unknown~\cite{schnoerr2017approximation,jahnke2007solving}. Consequently, the time evolution of $p$ is typically investigated using numerical methods. Each reachable state $x$ of the SRN gives rise to one ODE. Because the state space of $X(t)$ is generally countably infinite, the CME comprises infinitely many ODEs. To make the problem tractable, some computational methods truncate the state space of $X(t)$, resulting in a finite number of ODEs which can be integrated numerically~\cite{munsky2006finite,burrage2006krylov,kazeev2014direct}. These methods are collectively referred to as Finite State Projection~(FSP) methods, and they come with error guarantees in terms of the size of the truncated state space considered. However, they become computationally inefficient when the dimensionality of the truncation is increased, typically limiting their application to systems with only a few species.\newline

Alternatively, the CME can be investigated over its full, generally unbounded domain using stochastic simulations. The dynamics of the SRN are given by a stochastic evolution equation that can be simulated exactly with a Stochastic Simulation Algorithm~(SSA) and incorporated into Monte Carlo schemes~\cite{gillespie1976general,gillespie1977exact,anderson2007modified}. In practice, Monte Carlo methods are typically used to estimate expected network outputs, and the task reduces to estimating:
\begin{equation}
\mathbb{E}[f(X(t))] \coloneqq \sum\nolimits_{x} f(x) p(x,t), \quad \forall t,
\end{equation}
where $p(\cdot,t)$ is the distribution at time $t$ (determined by the initial condition), and $f$ is a function that reflects some outputs of interest, such as the average abundance of some species (see panel~\textbf{a} in Fig.~\ref{main_figure:deepska}). Although sophisticated Monte Carlo schemes have been devised, they can still suffer from slow convergence rates~\cite{anderson2012multilevel,anderson2022conditional}.\newline

Recent advances and successes in deep learning and scientific machine learning have motivated the use of neural networks for studying SRNs. The goal is to compute estimates more rapidly than with conventional methods~(time efficiency) and to obtain a compact representation of the reaction dynamics~(memory efficiency). Most existing work focuses on approximating the distribution $p$ of the system. Ref.~\cite{tang2023neural} uses a sequence of Variational Autoregressive Networks~(VANs) to represent the solution of the CME over a finite-time horizon. These networks are trained in the Neural-Network Chemical Master Equation~(NNCME) framework by constraining them to satisfy the ODEs in the CME, in the spirit of the Physics-Informed Neural Network~(PINN) paradigm~\cite{raissi2019physics,lu2021deepxde}. Ref.~\cite{liu2024distilling} leverages the VAN representations obtained from the NNCME framework to train a Master Equation Transformer~(MET). Alternatively, refs.~\cite{bortolussi2018deep,repin2021automated,sukys2022approximating} use approximations of the distribution derived from FSP or Monte Carlo estimators to train Mixture Density Networks~(MDNs).\newline

The DeepCME methodology adopts a different perspective: it employs trainable variables to directly approximate multiple expected outputs at a specific time~$t$ for a given initial state~$x$~\cite{gupta2021deepcme}. Conceptually, this is similar to the Deep Backward Stochastic Differential Equation~(DeepBSDE) framework, where trainable quantities are required to satisfy an almost sure relationship as part of a Reinforcement Learning~(RL) formulation~\cite{han2018solving}. In this setting, each simulated trajectory contributes an additional constraint to the approximation, in contrast to standard Monte Carlo methods, which discard the detailed information contained in a trajectory when only an expectation at a chosen time $t$ is sought. Unlike other approaches, DeepCME does not require any state space truncation. Moreover, the associated identification problem is well-posed: the only quantity that satisfies the almost sure relationship is the exact expected output.\newline

Although existing deep learning methods for SRNs have demonstrated good numerical accuracy across a range of benchmark examples, they currently lack the reliability guarantees available for methods such as Monte Carlo simulation. This limitation reflects a broader challenge in the field of deep learning, where guarantees beyond universal approximation theorems remain scarce and where methods for quantifying prediction uncertainty, such as conformal prediction, are the focus of active research~\cite{leshno1993multilayer,pinkus1999approximation,vovk2005algorithmic,balasubramanian2014conformal,fontana2023conformal}. The absence of reliability guarantees hinders the deployment of neural approaches to SRNs in scientific analysis and in decision-making settings, particularly in safety-critical domains such as therapeutic interventions, where stochastic models and SRNs play an increasingly instrumental role~\cite{briat2023noise,filo2023biomolecular}.\newline

Even when accurate, existing neural approximations of stochastic reaction dynamics operate as black-box models and therefore lack biological or mathematical interpretability. This limitation stems from the broader challenge of developing deep learning models whose components are interpretable and whose computations are transparent~\cite{doshi2017towards,murdoch2019definitions,du2019techniques,roscher2020explainable}. The opacity hampers integration with established numerical approaches, complicates the diagnosis of model failures, and impedes the extraction of biological or mathematical insights into the underlying systems.\newline

In this paper, we introduce an interpretable neural architecture for representing expected outputs, which we term the Spectral Decomposition-based network~(SDnet). Thanks to its explicit connection to a spectral decomposition, SDnet provides an interpretable and transparent representation of expected outputs, in contrast to conventional black-box neural networks. SDnet shares its grounding in spectral decomposition, together with its associated theoretical guarantees, with the Stochastic Koopman Approximation (SKA). This non-deep-learning method computes the spectral decomposition of SRNs from a frequency-domain perspective by exploiting the properties of the resolvent operator of the Markov transition semigroup~(also called the stochastic Koopman operator)~\cite{gupta2025sparse}. Building on SDnet, we develop two hybrid estimators that combine Monte Carlo simulations with the trained neural architecture to provide reliable estimates of expected outputs: SSA with Deep Importance Sampling~(SSA with DeepIS) and SSA with Deep Control Variates~(SSA with DeepCV). Together, these form the Deep Learning/Monte Carlo~(DLMC) estimators. They combine the reliability missing from purely neural network-based approaches with the capacity to deliver precise estimates from substantially fewer samples than conventional Monte Carlo schemes. This improved sample efficiency arises from the ability of SDnet to steer the stochastic simulations and reduce estimator variance. In SSA with DeepIS, SDnet informs a change of measure, whereas in SSA with DeepCV, it provides deep control variates along stochastic trajectories. Owing to its connection with SKA, we refer to the resulting framework as DeepSKA~(see panel~\textbf{b} in Fig.~\ref{main_figure:deepska}). We demonstrate the efficiency of SDnet and the DLMC estimators through numerical experiments on systems with up to ten species (state space up to $\mathbb{N}^{10}$), including cases with nonlinear and non-mass-action kinetics.

\begin{figure}[H]
\centering
\captionsetup{labelfont=bf}
\includegraphics[width=\textwidth]{MAIN-main01.pdf}
\caption{\textbf{The DeepSKA framework.} \textbf{a,} Stochastic Reaction Networks~(SRNs) are mechanistic models describing the time evolution of systems across the life sciences. In this context, functions $f$ of the random state are studied to capture some properties of the system of interest. A central goal is the estimation of expected outputs.
\textbf{b,} The DeepSKA framework is composed of two complementary components: the Spectral Decomposition-based network~(SDnet), which is an interpretable architecture to represent expected outputs; and two Deep Learning/Monte Carlo~(DLMC) estimators, SSA with DeepCV and SSA with DeepIS, to provide provably reliable estimates of expected outputs.}
\label{main_figure:deepska}
\end{figure}

\pdfbookmark[1]{Results}{results}
\section*{Results}
\label{main_section:results}

\pdfbookmark[2]{Stochastic reaction networks}{stochastic-reaction-networks}
\subsection*{Stochastic reaction networks}

Let us consider a system with $N$ types of components, called species, and denoted by $\mathbf{S_1}, \dots, \mathbf{S_N}$. The state of the system at any given time is described by a discrete vector $x = (x_1,\dots,x_N)$ whose $i$-th component corresponds to the abundance of species $\mathbf{S_i}$. The state changes due to $M$ types of events, called reactions. When a reaction occurs, certain species interact and are consumed while others are produced. This can be expressed as:
\begin{equation}
\label{eq:reaction_graph}
\nu_{1k} \mathbf{S_1} + \ldots + \nu_{Nk} \mathbf{S_N} \xlongrightarrow[]{} \nu'_{1k} \mathbf{S_1} + \ldots + \nu'_{Nk} \mathbf{S_N},\quad k\in [\![1, M]\!].
\end{equation}

This reaction graph indicates that the $k$-th reaction consumes $\nu_{ik}$ copies of the species $\mathbf{S_i}$ and produces $\nu'_{ik}$ of them, leading to a net change of $\nu'_{ik} -\nu_{ik}$ copies in the abundance of $\mathbf{S_i}$.\newline

The dynamics of the system are described by a Continuous-Time Markov Chain~(CTMC) $(X(t))$. This stochastic process is fully characterised by its generator $\mathbb{A}$ that is specified by the stoichiometry vectors $\zeta_k \coloneqq \nu'_{\cdot k} - \nu_{\cdot k}$ and the state-dependent propensities $\lambda_k$. The dynamics of the CTMC are given by the stochastic evolution equation~\cite{anderson2015stochastic}:
\begin{equation}
X(t) = x + \sum_{k=1}^{M} \zeta_{k} R_{k}(t),
\end{equation}

where $x$ is the initial state, $(R_{k}(t))$ is a random time-changed unit Poisson process that counts the number of times reaction $k$ has happened~(see section~\ref{supp_section:srns} of the supplementary material). Given an output function $f$, our aim is to compute expected outputs $U$ defined as:
\begin{equation}
U(x, f, t) \coloneqq \mathbb{E}_{x}[f(X(t))] =  \mathbb{E}[f(X(t)) | X(0)=x],\quad \forall x, \; \forall t.
\end{equation}

\pdfbookmark[2]{The DeepSKA framework}{the-xxx-framework}
\subsection*{The DeepSKA framework}

The DeepSKA framework comprises two complementary components: (i) an interpretable architecture to represent expected outputs, which we call the {\em Spectral Decomposition-based network}~(SDnet); and (ii) {\em Deep Learning/Monte Carlo}~(DLMC) estimators, which combine Monte Carlo simulations with the trained SDnet to provide reliable estimates of expected outputs.\newline

\begin{figure}[H]
\centering
\captionsetup{labelfont=bf}
\includegraphics[width=\textwidth]{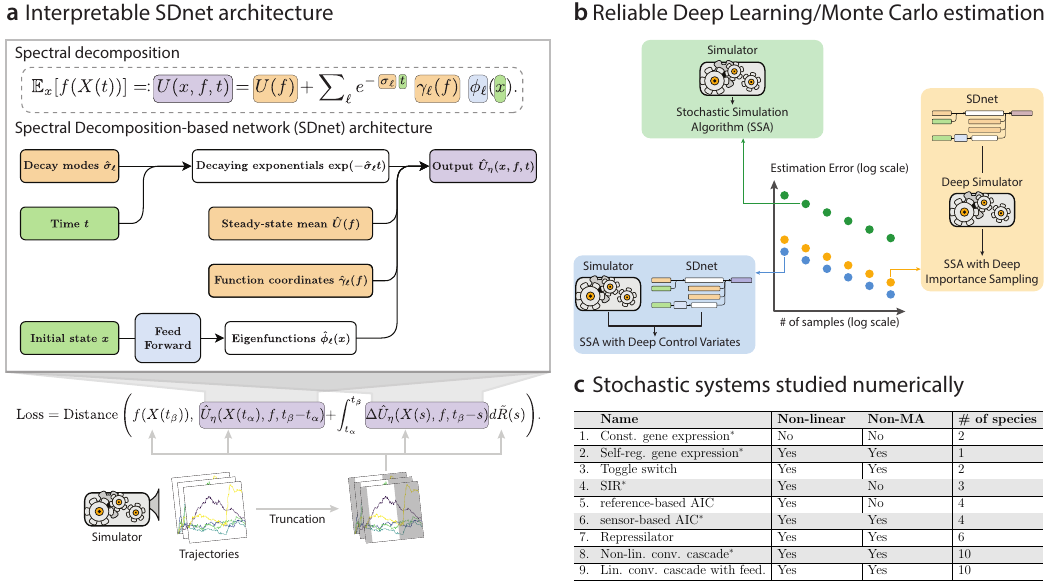}
\caption{\textbf{Components of the DeepSKA framework.} 
\textbf{a,} The spectral decomposition of $U(x,f,t)$ leads to a representation which is used to define the Spectral Decomposition-based network~(SDnet). The neural network is interpretable and its computations are transparent. In the schematic, inputs to SDnet are shown in green, outputs in purple, trainable variables in orange, and the feedforward neural network in blue. The architecture is trained with an extension of the Reinforcement Learning~(RL) procedure of ref.~\cite{gupta2021deepcme}, whose loss is motivated by an almost sure relationship and is computed using simulated trajectories.
\textbf{b,} The SSA with Deep Importance Sampling~(SSA with DeepIS) and the SSA with Deep Control Variates~(SSA with DeepCV) estimators combine Monte Carlo simulations with the trained SDnet. These Deep Learning/Monte Carlo~(DLMC) estimators are reliable, with an estimation error which is expected to be lower than that of the standard SSA.
\textbf{c,} We study nine Stochastic Reaction Networks (SRNs) in the DeepSKA framework. They are formally defined in section~\ref{supp_section:srns} of the supplementary material. The presence of nonlinear and non-mass-action~(MA) kinetics, as well as the species number, are used as a measure of the complexity of the networks. $^{*}$: results presented in sections~\ref{supp_section:nonlinear} and~\ref{supp_section:linear} of the supplementary material.}
\label{main_figure:components_deepska}
\end{figure}

\textbf{The Spectral Decomposition-based network (SDnet).} Previous neural approaches to representing expected outputs have suffered from a lack of interpretability~(see section~\ref{supp_section:sdnet} of the supplementary material). In contrast, SDnet is grounded in the spectral decomposition of $U$, and inherits a mathematically interpretable structure~(see panel~\textbf{a} in Fig.~\ref{main_figure:components_deepska}).\newline

Let us denote $(\tau_{\ell})_{\ell}$ the sequence of eigenvalues of $\mathbb{A}$, and introduce $\sigma_\ell$ as the negative of $\tau_{\ell}$, \emph{i.e.} $\sigma_{\ell} \coloneqq -\tau_{\ell}$. We call $\sigma_{\ell}$ a decay mode, and write $\phi_{\ell}$ the corresponding eigenfunction. Expanding the output function $f$ in the eigenbasis $(\phi_{\ell})_{\ell}$ gives:
\begin{equation}
f (x) =  \gamma_0(f) + \sum_{\ell=1}^{\infty}\gamma_{\ell}(f)\phi_{\ell}(x),
\end{equation}

where $\gamma_{\ell}(f)$ is the coordinate of $f$ associated to the function $\phi_{\ell}$. Using this expansion, $U(x,f,t)$ can be expressed as:
\begin{equation}
\label{eq:complex_decomposition_raw_main}
U(x,f,t) = U(f) + \sum_{\ell=1}^{\infty} e^{-\sigma_{\ell}t}\gamma_{\ell}(f)\phi_{\ell}(x),
\end{equation}
where $U(f)$ denotes the steady-state mean of $f$~(see section~\ref{supp_section:sdnet} of the supplementary material and ref.~\cite{gupta2025sparse}).\newline

As shown in ref.~\cite{gupta2025sparse}~(see appendix A1 and the numerical results), the relation remains accurate under finite truncation of the right-hand-side expansion, forming the basis of SDnet. Let $\widehat{U}_{\eta}$ denote the corresponding approximation of $U$, with $\eta$ collecting all trainable parameters. SDnet is then defined as:
\begin{equation}
\label{eq:sdnet_main}
\widehat{U}_{\eta}(x,f,t) = \hat{U}(f) + \sum_{\ell=1}^{r} e^{-\hat{\sigma}_{\ell}t}\hat{\gamma}_{\ell}(f)\hat{\phi}_{\ell}(x),
\end{equation}

where $r$ is a truncation order adapted to the complexity of the CTMC. The term $\hat{U}(f)$ is estimated via an ergodic mean and kept fixed during training. The parameters $\hat{\sigma}_{\ell}$ and $\hat{\gamma}_{\ell}(f)$ are trainable variables, and $(\hat{\phi}_{\ell})_{\ell}$ is a feedforward neural network~(see section~\ref{supp_section:sdnet} of the supplementary material).\newline

Thanks to its explicit connection to the spectral decomposition, SDnet is interpretable and its computations are transparent, in contrast to conventional black-box neural networks. Notably, this interpretability goes beyond the structural correspondence  between equations~\eqref{eq:complex_decomposition_raw_main} and~\eqref{eq:sdnet_main}: when SDnet matches the exact expectation $U(x,f,t)$ for all times~$t$ and initial states $x$, the quantities $\hat{\sigma}_{\ell}$ and $\hat{\phi}_{\ell}$ provably coincide with the decay modes and eigenfunctions of the generator $\mathbb{A}$, respectively~(see theorem 3.1 and equation~(23) in ref.~\cite{gupta2025sparse}). The interpretability of SDnet enables scalable learning of $U$ across multiple outputs: each function receives distinct coordinates $\gamma_{\ell}(f)$ while sharing the remaining spectral components, allowing additional functions to be studied with minimal additional parameters, avoiding network size growth and maintaining computational efficiency.\newline

SDnet is trained using an extension of the Reinforcement Learning~(RL) procedure introduced in ref.~\cite{gupta2021deepcme}~(see panel~\textbf{a} in Fig.~\ref{main_figure:components_deepska}). As in the original approach, the training loss is based on an equality that holds trajectory by trajectory:
\begin{equation}
\label{eq:almost_sure_relationship_main}
f(X(t_\beta)) = U(X(t_{\alpha}), f, t_{\beta} - t_{\alpha}) + \int_{t_{\alpha}}^{t_{\beta}}\sum_{k=1}^{M} \Delta_{\zeta_k} U(X(s), f, t_{\beta}-s)d\tilde{R}_k(s),
\end{equation}

where $\Delta_{\zeta_k} U(x, f, t)= U(x+\zeta_k, f, t) -  U(x, f, t)$,  the time points $t_{\alpha}$ and $t_{\beta}$ satisfy $t_{\alpha} \leq t_{\beta}$, and $(\tilde{R}_k(t))$ is a martingale derived from $(R_k(t))$~(see section~\ref{supp_section:training} of the supplementary material). Motivated by equation~\eqref{eq:almost_sure_relationship_main}, the approximation $\widehat{U}_{\eta}$ is trained to minimise the distance between $f(X(t_\beta))$ and:
\begin{equation}
\label{eq:almost_sure_relationship_approx_main}
\widehat{U}_{\eta}(X(t_{\alpha}), f, t_{\beta} - t_{\alpha}) + \int_{t_{\alpha}}^{t_{\beta}}\sum_{k=1}^{M} \Delta_{\zeta_k} \widehat{U}_{\eta}(X(s), f, t_{\beta}-s)d\tilde{R}_k(s).
\end{equation}

When the distance is zero for each trajectory, the true solution $U$ is guaranteed to be recovered (see theorem~3.3 in ref.~\cite{gupta2021deepcme}). The training loss is evaluated using trajectories simulated up to a terminal time $T$ with a Stochastic Simulation Algorithm~(SSA) known as the modified next reaction method \cite{anderson2007modified}. Unlike in the original method, the trajectories are considered on a different time interval $[t_{\alpha}, t_{\beta}]$ at each optimisation step, thereby enforcing a greater number of almost sure relationships without incurring the cost of generating new trajectories~(see section~\ref{supp_section:training} of the supplementary material). In contrast to previous deep learning approaches for SRNs, this extension directly yields an approximation of $U(x,f,t)$ at any time~$t$ for any initial state $x$, enabling a complete characterisation of the dependence of reaction network outputs on these variables.\newline

\textbf{Deep Learning/Monte Carlo estimators (DLMC).} Methods relying solely on neural networks generally lack reliability guarantees in non-ideal settings. To overcome this limitation, we augment SDnet with stochastic simulations, giving rise to two hybrid methods that possess such guarantees: SSA with Deep Importance Sampling~(SSA with DeepIS) and SSA with Deep Control Variates~(SSA with DeepCV), which we collectively refer to as Deep Learning/Monte Carlo~(DLMC) methods~(see panel~\textbf{b} in Fig.~\ref{main_figure:components_deepska}). The hybrid methods exhibit the unbiasedness and convergence guarantees of classical Monte Carlo approaches: as the number of samples increases, estimates converge to the exact value.\newline

At the same time, incorporating SDnet into the DLMC estimators is expected to markedly accelerate convergence relative to crude Monte Carlo by reducing variance. In this way, the hybrid methods possess the reliability guarantees missing from purely neural network-based approaches while yielding precise estimates with substantially fewer samples than conventional Monte Carlo schemes.\newline

In the DeepIS method, the propensities of the original process $(X(t))$ are adjusted according to the output of SDnet, yielding an auxiliary process $(X^{\text{IS}}_{\eta}(t))$. A likelihood process $(Z^{\text{IS}}_{\eta}(t))$ is introduced to compensate for the resulting change of measure~(see equations~\eqref{eq:x_is_methods} and~\eqref{eq:z_is_methods} in the \nameref{main_section:methods} for the expression of these two variables). Together, $(X^{\text{IS}}_{\eta}(t))$ and $(Z^{\text{IS}}_{\eta}(t))$ are used to construct the random variable $E_{\eta}^{\text{IS}}$ as:
\begin{equation}
E_{\eta}^{\text{IS}} \coloneqq f(X^{\text{IS}}_{\eta}(T))\,Z_{\eta}^{\text{IS}}(T).
\end{equation}

The corresponding Monte Carlo estimator remains unbiased, since~(see section~\ref{supp_section:dlmc} of the supplementary material):
\begin{equation}
\mathbb{E}_{x}\!\left[E_{\eta}^{\text{IS}}\right] \;=\; \mathbb{E}_x[f(X(T))].
\end{equation}

Hence, DeepIS inherits the reliability guarantees of classical Monte Carlo methods. Moreover, when SDnet provides an exact representation of $U$, a single simulation of $E_{\eta}^{\text{IS}}$ yields the exact expectation~(see section~\ref{supp_section:dlmc} of the supplementary material):
\begin{equation}
E_{\eta}^{\text{IS}} \;=\; \mathbb{E}_x[f(X(T))], \quad \text{almost surely}.
\end{equation}

In practice, even when SDnet is imperfect, the random variable $E_{\eta}^{\text{IS}}$ remains unbiased and is expected to deliver precise estimates with markedly fewer samples than estimators based on the original process $(X(t))$ alone~(see the numerical results).\newline

In the DeepCV method, the original process $(X(t))$ is simulated without alteration, while SDnet is used to construct an auxiliary control variate $Z_{\eta}^{\text{CV}}(t)$ that is correlated with the target functional $f(X(t))$ and has mean zero~(see equation~\eqref{eq:z_cv_methods} in the \nameref{main_section:methods} for a definition of this variable). We then define the random variable $E_{\eta}^{\text{CV}} $ as:
\begin{equation}
E_{\eta}^{\text{CV}} \coloneqq f(X(T)) - Z_{\eta}^{\text{CV}}(T).
\end{equation}

By construction, the corresponding estimator is also unbiased~(see section~\ref{supp_section:dlmc} of the supplementary material):
\begin{equation}
\mathbb{E}_x\left[E_{\eta}^{\text{CV}}\right] = \mathbb{E}_x[f(X(T))].
\end{equation}

When SDnet exactly represents $U$, the random variables $f(X(T))$ and $Z_{\eta}^{\text{CV}}(T)$ are perfectly correlated (Pearson correlation coefficient of $1$), and the control variate $Z_{\eta}^{\text{CV}}(T)$ exactly cancels the randomness in $f(X(t))$. As a result, a single simulation of $E_{\eta}^{\text{CV}}$ yields the exact expectation~(see section~\ref{supp_section:dlmc} of the supplementary material):
\begin{equation}
E_{\eta}^{\text{CV}} = \mathbb{E}_x[f(X(T))], \quad \text{almost surely}.
\end{equation}

Again, even when SDnet is imperfect, the estimator remains unbiased and typically achieves precise results with far fewer samples than those based solely on the original process $(X(t))$~(see the numerical results). The relative advantages of SSA with DeepIS and SSA with DeepCV are compared in section~\ref{supp_section:dlmc} of the supplementary material.

\pdfbookmark[2]{Examples}{examples}
\subsection*{Examples}

We validate the DeepSKA framework on nine SRNs covering a broad range of complexities~(see panel~\textbf{c} in Fig.~\ref{main_figure:components_deepska}). The examples include the constitutive gene expression network, the self-regulatory gene expression network, the genetic toggle switch network, the susceptible–infected–recovered network, the reference-based antithetic integral control of gene expression network, the sensor-based antithetic integral control of gene expression network, the repressilator network, the nonlinear conversion cascade network, and the linear conversion cascade with feedback network. 
Four representative examples are presented in the main text, with the remaining systems detailed in the supplementary material. Together, these results illustrate that the proposed framework applies to both linear and nonlinear, mass-action and non-mass-action systems, and remains effective across dimensions from one to ten species.\newline

\begin{figure}[H]
\centering
\captionsetup{labelfont=bf}
\includegraphics[width=\textwidth]{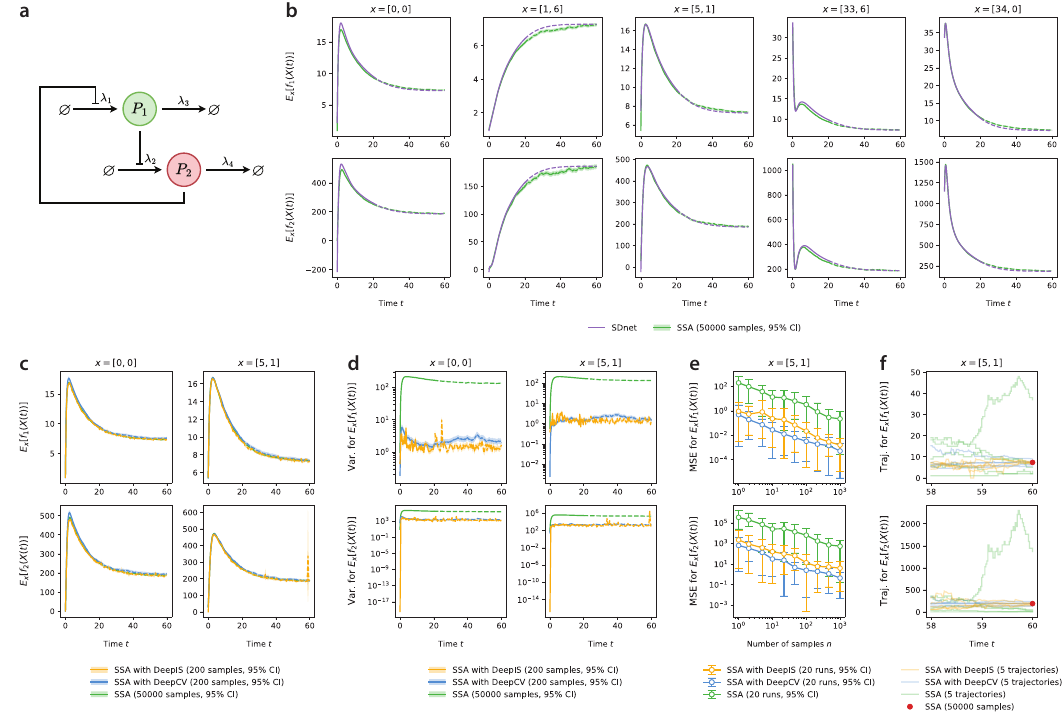}
\caption{\textbf{Results for the toggle switch.} \textbf{a}, Reaction graph of the system. The toggle switch consists of two mutually repressing proteins, $\mathbf{P_1}$ and $\mathbf{P_2}$. Each protein inhibits the production of the other through a Hill-type repression, and both species undergo degradation. \textbf{b}, Mean dynamics for different initial states and output functions obtained with the Spectral Decomposition-based network~(SDnet) and the Stochastic Simulation Algorithm~(SSA). Each plot shows the temporal evolution of the mean from time~0 to~60 for a given initial state and output function. Purple lines indicate SDnet predictions, and green bands show the mean and 95\% confidence interval~(CI) computed from 50 000 SSA samples. As in all other plots, solid lines correspond to the training interval $[0,20]$, and dashed lines indicate times beyond the training interval. Initial states vary horizontally from left to right, output functions vary vertically from top to bottom. The first output function corresponds to the first moment of protein $\mathbf{P_1}$~($f_1(x) = x_1$), and the second to the second moment~($f_2(x) = (x_1)^2$).
\textbf{c}, Mean dynamics under different initial states and output functions obtained with SSA with Deep Importance Sampling~(SSA with DeepIS), SSA with Deep Control Variates~(SSA with DeepCV), and SSA. The layout follows panel~b. Curves for the Deep Learning/Monte Carlo (DLMC) estimators are computed from 200 samples and displayed with 95\% CIs.
\textbf{d}, Temporal evolution of the variance of the estimators for different initial states and output functions. The panel layout mirrors panels~b and~c.
\textbf{e}, Mean-squared error~(MSE) of the estimators at time $t=5$ for the initial state $[5,1]$ as a function of sample size. Reference values are computed from SSA with DeepCV using 100 000 samples. The squared error is averaged over 20 independent runs.
\textbf{f}, Five sample paths obtained with SSA with DeepIS, SSA with DeepCV, and SSA. The red dot indicates the reference mean abundance at $t=60$, computed from 50 000 SSA samples.}
\label{main_figure:tsw}
\end{figure}

We systematically compared the predictions of SDnet and the performance of DLMC estimators against the SSA baseline. Results for each system are presented in Figs.~\ref{main_figure:tsw}–\ref{main_figure:lcf} of the main text and in Figs.~\ref{supp_figure:slf}–\ref{supp_figure:cge} of the supplementary material (see sections~\ref{supp_section:nonlinear} and~\ref{supp_section:linear}). In each figure, panel~\textbf{a} depicts the reaction graph corresponding to the studied SRN.\newline

The mathematically interpretable SDnet achieves high predictive accuracy across different initial conditions and output functions~(in panel \textbf{b}). Notably, SDnet exhibits strong temporal generalisation, maintaining accuracy well beyond the training time horizon~(see the dashed lines).\newline

The DLMC estimators achieve the accuracy and reliability guaranteed by their theoretical construction~(in panel~\textbf{c}). Comparison with the performance of SDnet~(in panel~\textbf{b}) shows that incorporating stochastic simulations effectively corrects the residual approximation error of SDnet. At the same time, the hybrid estimators substantially outperform classical Monte Carlo methods, achieving several orders of magnitude lower variance than standard SSA across all examples~(in panel~\textbf{d}), producing markedly tighter confidence intervals for a given number of samples.\newline

As the number of samples increases, the estimation error decreases, as expected by the inverse square root law~(in panel~\textbf{e}). The DLMC estimators inherit the reliability guarantees of standard Monte Carlo approaches, with estimates converging to the exact value. Moreover, for a fixed sample size, the error of the hybrid estimators is orders of magnitude smaller than that of the SSA alone across all examples (reduction up to 10 000-fold). This improvement stems directly from the reduced variance shown in panel~\textbf{d}. Incorporating SDnet therefore accelerates convergence substantially, and precise estimations are obtained from far fewer simulations. In the examples considered, SSA with DeepCV tends to achieve slightly higher efficiency than SSA with DeepIS.\newline

Finally, panel~\textbf{f} provides an intuitive visualisation of the source of variance and error reduction. The DLMC estimators guide simulated paths toward the expected outcome~(see the red dot). If SDnet provided an exact representation of~$U$, each trajectory would terminate precisely at the true expectation. Here, SDnet yields a close approximation, and the trajectories cluster tightly around the exact value, producing lower variance~(in panel~\textbf{d}) and consequently lower estimation error~(in panel~\textbf{e}).

\pdfbookmark[1]{Discussion}{discussion}
\section*{Discussion}

We have introduced the DeepSKA framework, a method for interpretable and reliable computation of expected outputs in Stochastic Reaction Networks~(SRNs). This framework integrates a mathematically interpretable neural network, the Spectral Decomposition-based network~(SDnet), with reliable Deep Learning/Monte Carlo~(DLMC) estimators. This combination enables a trustworthy and efficient characterisation of the temporal dynamics of SRNs.\newline

SDnet is grounded in the spectral decomposition of expected outputs, conferring a mathematically interpretable structure that yields several benefits. By establishing a direct correspondence between the network components and known spectral quantities, the architecture connects to an existing body of well-established analytical and numerical methods, enabling their direct application. Leveraging the underlying mathematical structure of expected outputs also allows SDnet to be substantially smaller than generic black-box models. Moreover, SDnet enables scalable learning across multiple output functions through the reuse of shared spectral elements. Once trained, the components of SDnet can also be transferred to new output functions, eliminating the need to retrain the full model, or used to compute related quantities. Furthermore, SDnet enforces physically consistent steady-state behaviour by ensuring that its output converges, in the large-time limit, to the steady-state mean or its accurate ergodic approximation. The benefits of interpretability of SDnet are discussed in detail in section~\ref{supp_section:sdnet} of the supplementary material. Across diverse systems, SDnet accurately captured complex temporal dynamics, including those with nonlinear interactions and non-mass-action kinetics. SDnet consistently maintained predictive accuracy under varying initial conditions and output functions and generalised well beyond the training horizon across the nine SRNs tested.

The DLMC estimators augment SDnet with stochastic simulations, thereby preserving the unbiasedness and convergence guarantees of classical Monte Carlo approaches, which are absent from purely neural methods such as SDnet. Across the tested systems, these hybrid estimators demonstrated the theoretically guaranteed reliability and achieved substantial variance reduction, enabling faster convergence and precise estimation with significantly fewer simulations than conventional Monte Carlo approaches. Although the DLMC estimators could, in principle, be employed with other approximation methods of expected outputs, such as moment closure techniques~\cite{sukys2022momentclosure}, neural networks provide the advantage of rapid evaluation for any initial state and time, highlighting the complementarity between SDnet and the DLMC estimators in the DeepSKA framework.

\begin{figure}[H]
\centering
\captionsetup{labelfont=bf}
\includegraphics[width=\textwidth]{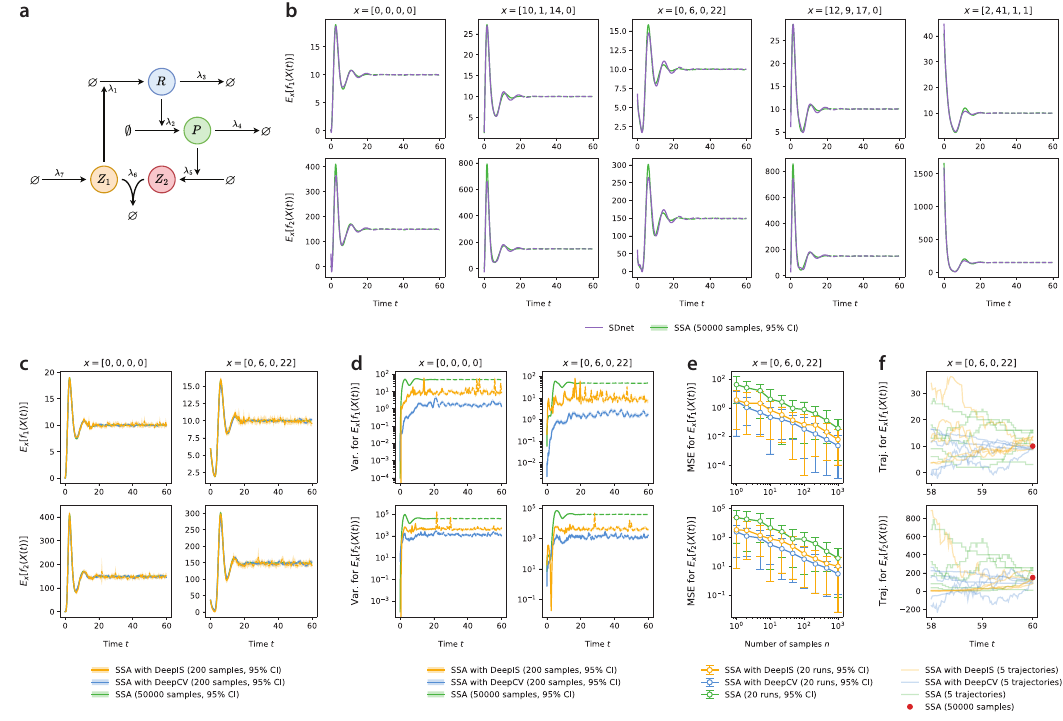}
\caption{\textbf{Results for the reference-based antithetic integral control of gene expression.} The conventions are the same as in Fig.~\ref{main_figure:tsw}. In this system, the protein $\mathbf{P}$ is regulated by two controller species, $\mathbf{Z_1}$ and $\mathbf{Z_2}$. The mRNA $\mathbf{R}$ is transcribed in proportion to $\mathbf{Z_1}$~(actuation), while $\mathbf{Z_2}$ is produced in proportion to $\mathbf{P}$~(sensing). Translation of $\mathbf{R}$ yields $\mathbf{P}$, both $\mathbf{R}$ and $\mathbf{P}$ undergo degradation, and the controller species $\mathbf{Z_1}$ and $\mathbf{Z_2}$ are jointly removed through an annihilation reaction. The reaction graph of this system is displayed in panel~\textbf{a}, the mean dynamics for different initial states and output functions obtained with the Spectral Decomposition-based network~(SDnet) and the Stochastic Simulation Algorithm~(SSA) in panel~\textbf{b}, 
the mean dynamics under different initial states and output functions obtained with SSA with Deep Importance Sampling~(SSA with DeepIS), SSA with Deep Control Variates~(SSA with DeepCV), and SSA in panel~\textbf{c}, 
the temporal evolution of the variance of the estimators for different initial states and output functions in panel~\textbf{d}, the mean-squared error~(MSE) of the estimators at time $t=5$ for the initial state $[0, 6, 0, 22]$ as a function of sample size in panel~\textbf{e}, 
and five sample paths obtained with SSA with DeepIS, SSA with DeepCV, and SSA in panel~\textbf{f}. The first output function corresponds to the first moment of protein $\mathbf{P}$~($f_1(x) = x_2$), and the second to the second moment~($f_2(x) = (x_2)^2$).}
\end{figure}

In the SKA method, the decay modes are recovered through a convex optimisation procedure~\cite{gupta2025sparse}. We propose leveraging the decay-mode estimates from SKA as an informed initialisation for the DeepSKA framework, potentially accelerating its convergence. In SKA, the function coordinate-eigenfunction products in the spectral decomposition need to be estimated afresh for each initial state. In contrast, in DeepSKA, the function coordinates and eigenfunctions are learned separately, with eigenfunctions directly represented as a mapping. Moreover, the SKA estimation of expected outputs lacks the guaranteed unbiasedness and convergence properties that are intrinsic to DLMC estimators.

\begin{figure}[H]
\centering
\captionsetup{labelfont=bf}
\includegraphics[width=\textwidth]{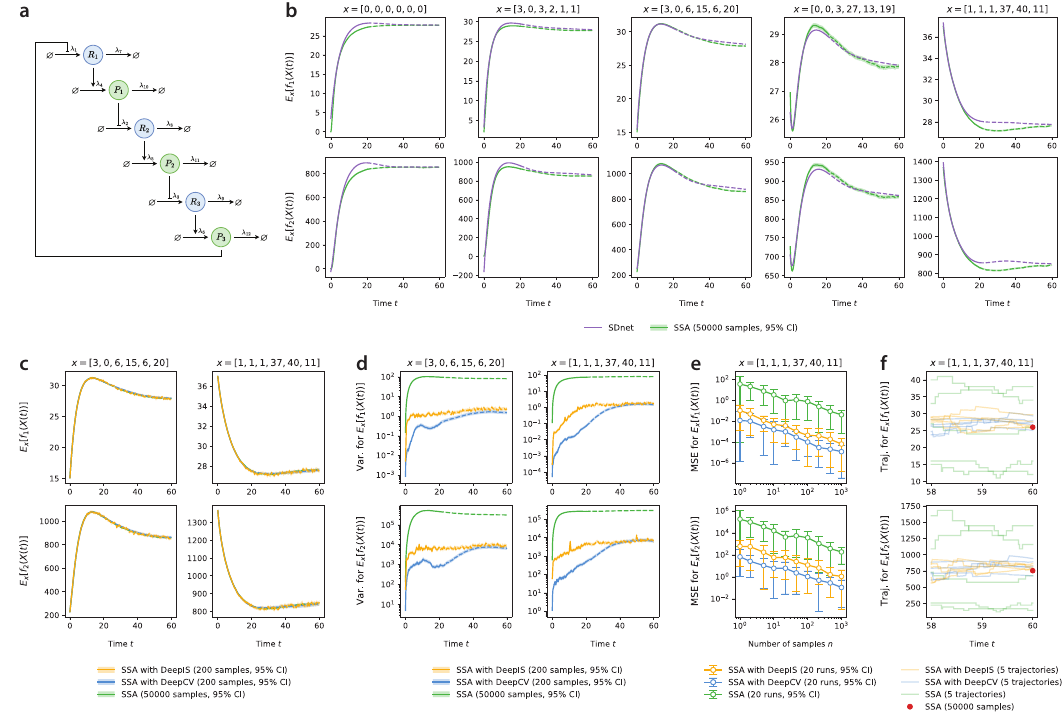}
\caption{\textbf{Results for the repressilator.} The conventions are the same as in Fig.~\ref{main_figure:tsw}. The repressilator consists of three genes forming a cyclic inhibitory network. The proteins $\mathbf{P_1}$, $\mathbf{P_2}$, and $\mathbf{P_3}$ are translated from the corresponding mRNA $\mathbf{R_1}$, $\mathbf{R_2}$, and $\mathbf{R_3}$. Each protein represses the transcription of the next gene in the cycle: $\mathbf{P_3}$ inhibits $\mathbf{R_1}$, $\mathbf{P_1}$ inhibits $\mathbf{R_2}$, and $\mathbf{P_2}$ inhibits $\mathbf{R_3}$. All mRNA and protein species undergo degradation. The reaction graph of the system is displayed in panel~\textbf{a},
the mean dynamics for different initial states and output functions obtained with the Spectral Decomposition-based network~(SDnet) and the Stochastic Simulation Algorithm~(SSA) in panel~\textbf{b}, 
the mean dynamics under different initial states and output functions obtained with SSA with Deep Importance Sampling~(SSA with DeepIS), SSA with Deep Control Variates~(SSA with DeepCV), and SSA in panel~\textbf{c}, 
the temporal evolution of the variance of the estimators for different initial states and output functions in panel~\textbf{d}, the mean-squared error~(MSE) of the estimators at time $t=5$ for the initial state $[1, 1, 1, 37, 40, 11]$ as a function of sample size in panel~\textbf{e}, 
and five sample paths obtained with SSA with DeepIS, SSA with DeepCV, and SSA in panel~\textbf{f}.  The first output function corresponds to the first moment of protein $\mathbf{P_1}$~($f_1(x) = x_4$), and the second to the second moment~($f_2(x) = (x_4)^2$).}
\end{figure}

To enable a comprehensive characterisation of SRNs, the DeepSKA framework has been extended to compute steady-state means and variances. Detailed descriptions of these extensions, together with illustrative numerical examples, are provided in the \nameref{main_section:methods} and the supplementary material. Briefly, the Ergodic Mean with Deep Control Variates~(EM with DeepCV) is a DLMC estimator of the steady-state mean that retains the reliability guarantees of classical ergodic averaging while achieving substantial variance reduction, allowing precise estimations from shorter time averaging. EM with DeepCV leverages a solution to the Poisson equation, represented via the Poisson Spectral Decomposition-based network~(P-SDnet). The Deep Integral Path Algorithm (DeepIPA) is a DLMC estimator of the variance, which is generally biased in contrast to the previous DLMC estimators. The numerical experiments demonstrate that it remains accurate and achieves high precision even with a limited number of samples. Importantly, it can be computed using the same small set of trajectories employed for SSA with DeepCV.\newline

The DeepSKA framework enables systematic investigation of extrinsic noise arising from variability in initial states. In particular, it can be used for variance decomposition via the law of total variance or for global sensitivity analysis through the computation of Sobol indices~\cite{swain2002intrinsic,le2015variance,navarro2016global}. The framework can also be extended to explore another source of extrinsic variability, namely parameter variability. By learning the dependency of expected outputs on parameters, the framework could facilitate studies of system robustness to parametric perturbations, an avenue we reserve for immediate future work.

\begin{figure}[H]
\centering
\captionsetup{labelfont=bf}
\includegraphics[width=\textwidth]{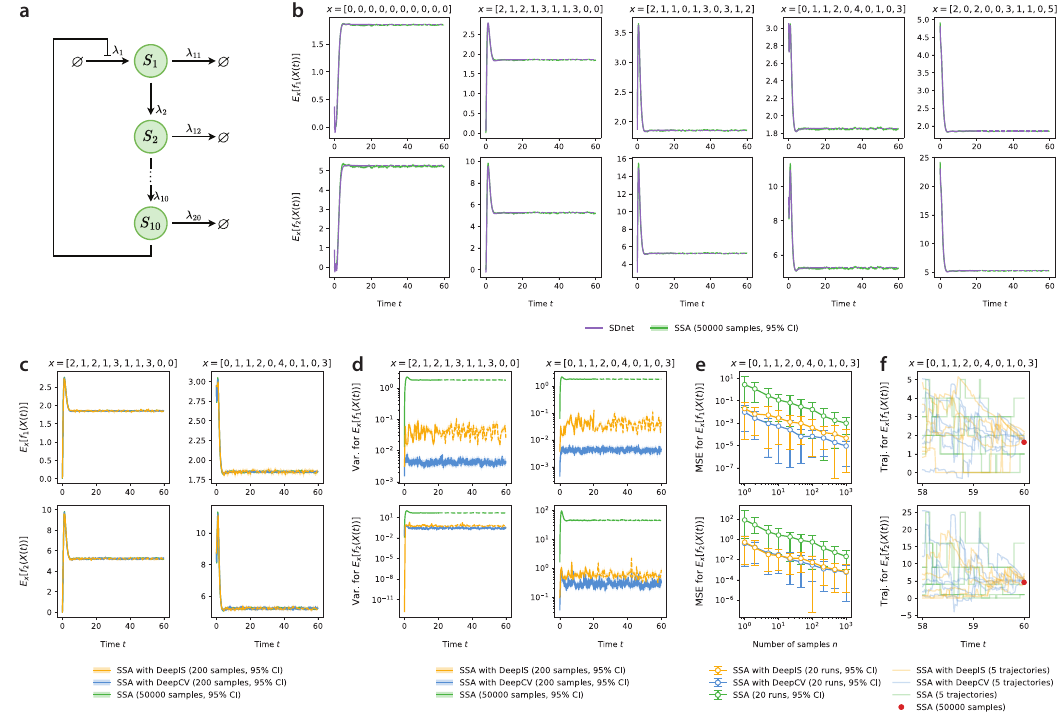}
\caption{\textbf{Results for the linear conversion cascade with feedback.} The conventions are the same as in Fig.~\ref{main_figure:tsw}. The first species $\mathbf{S_1}$ is produced under inhibition by the last species $\mathbf{S_{10}}$ through a Hill-type repression. Each species $\mathbf{S_1}, \dots, \mathbf{S_9}$ is sequentially converted into $\mathbf{S_2}, \dots, \mathbf{S_{10}}$, and all species are subject to degradation. The reaction graph of the system is displayed in panel~\textbf{a},
the mean dynamics for different initial states and output functions obtained with the Spectral Decomposition-based network~(SDnet) and the Stochastic Simulation Algorithm~(SSA) in panel~\textbf{b}, 
the mean dynamics under different initial states and output functions obtained with SSA with Deep Importance Sampling~(SSA with DeepIS), SSA with Deep Control Variates~(SSA with DeepCV), and SSA in panel~\textbf{c}, 
the temporal evolution of the variance of the estimators for different initial states and output functions in panel~\textbf{d}, the mean-squared error~(MSE) of the estimators at time $t=5$ for the initial state $[0, 1, 1, 2, 0, 4, 0, 1, 0, 3]$ as a function of sample size in panel~\textbf{e}, 
and five sample paths obtained with SSA with DeepIS, SSA with DeepCV, and SSA in panel~\textbf{f}. The first output function corresponds to the first moment of species $\mathbf{S_{10}}$~($f_1(x) = x_{10}$), and the second to the second moment~($f_2(x) = (x_{10})^2$).}
\label{main_figure:lcf}
\end{figure}

In practice, current observation techniques often capture only a subset of species in biological systems, which may not include the species of primary interest. Estimating the conditional mean of unobserved species based on partial observations corresponds to the stochastic filtering problem, which is generally computationally challenging~\cite{alt2023entropic,rathinam2021state,fang2024advanced}. We defer the extensive development required for this extension to future work.\newline

Looking forward, the DeepSKA framework directly extends to other Markovian models, including Stochastic Differential Equations~(SDEs) driven by Brownian motion. A relationship analogous to equation~\eqref{eq:almost_sure_relationship_main} has already been employed to estimate expected outputs in SDEs~\cite{han2018solving}. The SDnet architecture and DLMC estimators developed for SRNs could be readily adapted to diffusion processes, offering interpretable and reliable computation of expectations in these continuous-state stochastic systems.

\pdfbookmark[1]{Methods}{methods}
\section*{Methods}
\label{main_section:methods}

Derivations that are not provided in context are collected in section~\ref{app:proof} of the supplementary material.\newline

\textbf{Details on the Deep Learning/Monte Carlo estimators.} We now give precise definitions of the auxiliary process $(X_{\eta}^{\text{IS}}(t))$ and the likelihood process $(Z_{\eta}^{\text{IS}}(t))$ used in SSA with Deep Importance Sampling (SSA with DeepIS), as well as the control variate process~$(Z_{\eta}^{\text{CV}}(t))$ used in SSA with Deep Control Variates (SSA with DeepCV).\newline

The auxiliary process $(X_{\eta}^{\text{IS}}(t))$ shares the same stoichiometry vectors $\zeta_k$ as the original process $(X(t))$ but employs modified, time-dependent propensities $\lambda_{k}^{\text{IS}}$ derived from the Spectral Decomposition-based network~(SDnet). Its dynamics are given by the following stochastic evolution equation:
\begin{equation}
\label{eq:x_is_methods}
X_{\eta}^{\text{IS}}(t) \coloneqq x + \sum_{k=1}^{M} \zeta_k R^{\text{IS}}_{k}(t),
\end{equation}

where $(R^{\text{IS}}_{k}(t))$ counts the number of times reaction $k$ has happened under the adjusted propensities~(see section~\ref{supp_section:dlmc} of the supplementary material). The likelihood process $(Z_{\eta}^{\text{IS}}(t))$, introduced to compensate for the change of measure, is the stochastic exponential given by:
\begin{equation}
\label{eq:z_is_methods}
\begin{aligned}
Z_{\eta}^{\text{IS}}(t) \coloneqq \exp\Bigg(
    \sum_{k=1}^{M}
    \int_{0}^{t} &\log \Bigg(\frac{\lambda_k(X_{\eta}^{\text{IS}}(s))}{\lambda_k^{\text{IS}}(X_{\eta}^{\text{IS}}(s), t-s)} \Bigg)dR_{k}^{\text{IS}}(s)\\ &- \sum_{k=1}^{M} \int_{0}^{t}\big(\lambda_k(X_{\eta}^{\text{IS}}(s))
   - \lambda_k^{\text{IS}}(X_{\eta}^{\text{IS}}(s), t-s)\big) ds \Bigg).
\end{aligned}
\end{equation}

For SSA with DeepCV, the control variate process $(Z_{\eta}^{\text{CV}}(t))$, constructed using SDnet, is defined as:
\begin{equation}
\label{eq:z_cv_methods}
Z^{\text{CV}}_{\eta}(t) \coloneqq \sum_{k=1}^{M}\int_{0}^{t} \Delta_{\zeta_k} \widehat{U}_{\eta}(X(s), f, T-s)d\tilde{R}_k(s).
\end{equation}

As mentioned in the~\nameref{main_section:results}, a single sample of $E_{\eta}^{\text{CV}}$ is enough to recover $\mathbb{E}_{x}[f(X(t))]$ when SDnet provides an exact representation of $U$. To analyse the case where SDnet is imperfect, we introduce the error function $\varepsilon$ defined as $\varepsilon(x, f, t) \coloneqq U(x, f, t) - \widehat{U}_{\eta}(x, f, t)$. The variance of SSA with DeepCV is given by:
\begin{equation}
\text{Var}_{x}\Big(E_{\eta}^{\text{CV}}\Big) = \mathbb{E}_x\Bigg[\sum_{k=1}^{M}\int_{0}^{t} \Big(\Delta_{\zeta_k}\varepsilon(X(s), f, t-s)\Big)^2\lambda_{k}(X(s))ds\Bigg],
\end{equation}

showing that it depends ``smoothly'' on the approximation error $\varepsilon$~(see section~\ref{supp_section:dlmc} of the supplementary material).\newline

\textbf{Extension to the steady-state mean.} The supplementary material presents extensions of the DeepSKA framework to the computation of the stationary mean and the variance, which we summarise here. We first introduce a DLMC estimator of the steady-state mean, the Ergodic Mean with Deep Control Variates~(EM with DeepCV). This method retains the guarantees of classical ergodic averaging while offering the potential to improve convergence speed.\newline

Let $\mathcal{E}(f,t)$ denote the finite-time average given by:
\begin{equation}
\mathcal{E}(f,t) \coloneqq \frac{1}{t}\int_{0}^{t}f(X(s))ds.
\end{equation}

While $\mathcal{E}(f,t)$ converges to the steady-state mean $U(f)$ as $t \to \infty$, its variance can decay slowly. To address this, EM with DeepCV introduces an auxiliary time average that is correlated with $\mathcal{E}(f,t)$ and asymptotically centred at zero. This auxiliary term is constructed using a neural approximation of a solution $F$ to the Poisson equation given by:
\begin{equation}
\label{eq:poisson_equation_main}
\mathbb{A}F = U(f) - f,
\end{equation}

where $\mathbb{A}$ is the generator of the process $(X(t))$~(see section~\ref{supp_section:sdnet} of the supplementary material). The function $F$ admits the representation:
\begin{equation}
F(x) = \sum_{\ell=1}^{\infty}\frac{\gamma_{\ell}(f)}{\sigma_{\ell}}\phi_{\ell}(x),
\end{equation}

where $\gamma_{\ell}(f)$, $\sigma_{\ell}$ and $\phi_{\ell}$ are the coordinates, decay modes, and eigenfunctions, respectively, from the spectral decomposition of $U$ in equation~\eqref{eq:complex_decomposition_raw_main}~(see section~\ref{supp_section:sdnet} of the supplementary material). To approximate $F$, we define the Poisson Spectral Decomposition-based network~(P-SDnet) architecture, defined as a truncated version of this expansion:
\begin{equation}
\widehat{F}_{\eta}(x) \coloneqq \sum_{\ell=1}^{r}\frac{\hat{\gamma}_{\ell}(f)}{\hat{\sigma}_{\ell}}\hat{\phi}_{\ell}(x),
\end{equation}

where $\hat{\gamma}_{\ell}(f)$, $\hat{\sigma}_{\ell}$ and $\hat{\phi}_{\ell}(x)$ are obtained directly from a trained SDnet~(see section~\ref{supp_section:sdnet} of the supplementary material). Using P-SDnet, the auxiliary time average is constructed as:
\begin{equation}
\frac{1}{t}\int_{0}^{t}[\mathbb{A}\widehat{F}_{\eta}](X(s))ds,
\end{equation}

which leads to the finite-time average $\mathcal{E}^{\text{CV}}_{\eta}(f,t)$ defined as:
\begin{equation}
\mathcal{E}^{\text{CV}}_{\eta}(f,t) \coloneqq \frac{1}{t}\int_{0}^{t}[f+\mathbb{A}\widehat{F}_{\eta}](X(s))ds.
\end{equation}

By design, the DLMC estimator $\mathcal{E}^{\text{CV}}_{\eta}(f,t)$ converges to the steady-state mean $U(f)$ in the large-time limit~(see section~\ref{supp_section:dlmc} of the supplementary material):
\begin{equation}
\lim_{t\to\infty} \mathcal{E}^{\text{CV}}_{\eta}(f,t)  = U(f).
\end{equation}

Hence, EM with DeepCV inherits the reliability of standard ergodic averages. In the case where P-SDnet exactly matches $F$, $\mathcal{E}^{\text{CV}}_{\eta}(f,t)$ coincides with $U(f)$ at all finite times $t$~(see section~\ref{supp_section:dlmc} of the supplementary material). More generally, even with an imperfect approximation, $\mathcal{E}^{\text{CV}}_{\eta}(f,t)$ is expected to reduce variance relative to the standard ergodic mean, allowing for precise estimates from shorter time horizons $t$. We demonstrate the performance of EM with DeepCV through numerical experiments, reported in section~\ref{supp_section:ergodic_mean} of the supplementary material.\newline

\textbf{Extension to the variance.} We next extend the framework to the variance $\mathcal{W}$, defined as:
\begin{equation}
\mathcal{W}(x,f, t) \coloneqq \text{Var}_{x}\Big(f(X(t))\Big) \coloneqq \mathbb{E}_{x}\Bigg[\Big(f(X(t)) - U(x,f,t)\Big)^2\Bigg].
\end{equation}

In the same way that expected outputs $U$ can be learnt from equation~\eqref{eq:almost_sure_relationship_main}, the variance $\mathcal{W}$ can be learnt from the following almost sure relationship~(see section~\ref{supp_section:training} of the supplementary material):
\begin{equation}
\begin{split}
 \sum_{k=1}^{M} \int_{t_{\alpha}}^{t_{\beta}} \lambda_{k}(X(s))\Big(\Delta_{\zeta_k}U(X(s), f, t_{\beta}-s)\Big)^2ds &= \mathcal{W}(X(t_{\alpha}), f, t_{\beta} - t_{\alpha}) \\
 &+ \sum_{k=1}^{M} \int_{t_{\alpha}}^{t_{\beta}}  \Delta_{\zeta_k} \mathcal{W}(X(s), f, t_{\beta}-s)d\tilde{R}_k(s).
 \end{split}
\end{equation}

This identity further yields the following integral representation for the variance $\mathcal{W}$~(see section~\ref{supp_section:training} of the supplementary material):
\begin{equation}
\label{eq:variance_int_rep_main}
\mathcal{W}(x,f, t) = \mathbb{E}_x\Bigg[\sum_{k=1}^{M}\int_{0}^{t} \lambda_{k}(X(s))\Big(\Delta_{\zeta_k}U(X(s), f, t-s)\Big)^2ds\Bigg].
\end{equation}

Based on this formula, we introduce a DLMC estimator of the variance, the Deep Integral Path Algorithm (DeepIPA). Here, the unknown $U$ is replaced by SDnet when evaluating the integral over paths in equation~\eqref{eq:variance_int_rep_main}~(see section~\ref{supp_section:dlmc} of the supplementary material). Unlike the previous DLMC estimators, the DeepIPA is generally biased, since errors in SDnet propagate into the variance estimate. Nonetheless, it may lead to precise estimates even when only a limited number of samples are available. Numerical experiments illustrating the performance of DeepIPA are presented in section~\ref{supp_section:variance} of the supplementary material.\newline

Beyond finite-time variances $\mathcal{W}$, one is often interested in the variance of ergodic averages. This asymptotic variance is governed by the Time Average Variance Constant~(TAVC), denoted $\sigma_{\text{TAVC}}^2$. The TAVC admits an integral representation analogous to equation~\eqref{eq:variance_int_rep_main}, involving the solution $F$ to the Poisson equation~\eqref{eq:poisson_equation_main}. This representation motivates a DLMC estimator of the TAVC, described in section~\ref{supp_section:dlmc} of the supplementary material.\newline

\textbf{Details on the numerical results.} The systems studied numerically are formally defined in section~\ref{supp_section:srns} of the supplementary material.\newline

For SDnet, the steady-state mean approximation $\hat{U}(f)$ was obtained as the average of five ergodic means computed between time $100$ and $100\,000$ to obtain precise estimates. The variable $\hat{U}(f)$ remained fixed throughout the training. The real parts of the decay modes $\hat{\sigma}_{\ell}$ were obtained as the softplus of learnable variables initialised uniformly in $[0, 1]$, ensuring nonnegativity~(see section~\ref{supp_section:sdnet} of the supplementary material). Their imaginary parts were also initialised uniformly in $[0,1]$, while both the real and imaginary components of the coordinates $\hat{\gamma}_{\ell}(f)$ were initialised to $1$. The feedforward neural network approximating the eigenfunctions $\phi_{\ell}$ in SDnet had 2 hidden layers with 24 neurons each, and it was initialised using the Glorot uniform method~\cite{glorot2010understanding}.\newline

Training employed $20\,000$ trajectories over the time interval $[0,20]$, except for the susceptible–infected–recovered network, where the upper bound was $30$. Initial states were sampled uniformly from the hypercube $[0,3]^{N}$, where $N$ is the number of species, to encourage state space coverage, except for three systems: the self-regulatory network used $[0,30]$, the antithetic integral control of gene expression networks used $[0,10]^4$, and the susceptible–infected–recovered network sampled from $[40, 60] \times [5, 15] \times [0, 2]$. Each trajectory was discretised on a uniform time grid to enable efficient integration during training, using a step size of $0.02$~(see section~\ref{supp_section:training} of the supplementary material). The loss was normalised across all systems using the sample average, except for the toggle switch and the repressilator, for which the ergodic mean was employed~(see section~\ref{supp_section:training} of the supplementary material). The same step size as used during training was applied to the piecewise-constant propensities of SSA with DeepIS and the time integration of SSA with DeepCV~(see section~\ref{supp_section:dlmc} of the supplementary material). Exact pathwise integration was employed for SSA with DeepCV in panels~\textbf{e}. The total solving time, along with its decomposition into training and simulation components, is reported in section~\ref{supp_section:solving_times} of the supplementary material.\newline

All numerical experiments were carried out on a High-Performance Computing~(HPC) cluster using 40 Central Processing Unit~(CPU) cores from an AMD EPYC 9655 processor.

\pdfbookmark[1]{Data availability}{data-availability}
\section*{Data availability}

The data used in the study were generated directly from the code.

\pdfbookmark[1]{Code availability}{code-availability}
\section*{Code availability}

The code used in the study is publicly available on GitHub (\href{https://github.com/arthur-theuer/deep-ska}{{\tt https://github.com/arthur-theuer/deep-ska}}).

\pdfbookmark[1]{References}{references}
\printbibliography[title={References}]
\end{refsection}

\pdfbookmark[1]{Acknowledgements}{Acknowledgements}
\addtocontents{toc}{\protect\setcounter{tocdepth}{0}}
\section*{Acknowledgements}
\addtocontents{toc}{\protect\setcounter{tocdepth}{2}}

This research was funded in whole or in part by the Swiss National Science Foundation (SNSF) grant number 216505.

\clearpage
\appendix
\label{appendix}
\setcounter{page}{1}
\setcounter{figure}{0}
\setcounter{table}{0}
\setcounter{equation}{0}
\renewcommand{\theequation}{S\arabic{equation}}

\title{Interpretable neural approximation of stochastic reaction dynamics \\ with guaranteed reliability}
\author{Quentin Badolle, Arthur Theuer, Zhou Fang, Ankit Gupta, Mustafa Khammash}
\date{}
\maketitle
\captionsetup[figure]{
  name=Supplementary Fig.,
  labelsep=pipe
}
\captionsetup[table]{
  name=Supplementary Table,
  labelsep=pipe
}
\tableofcontents

\begin{refsection}
\clearpage
\section{List of the main abbreviations and notations}

\begin{table}[h]
\centering
\captionsetup{labelfont=bf}
\begin{tabularx}{\textwidth}{l|X|X}
\hline
\rowcolor{gray!20} \textbf{Abbreviation} & \textbf{Meaning} & \textbf{First appearance in the SM$^*$} \\
\hline
SRN & Stochastic Reaction Network & page~\pageref{first:srn}\\
\hline
\rowcolor{gray!20} CTMC &  Continuous-Time Markov Chain & page~\pageref{first:ctmc} \\
\hline
SDnet & Spectral Decomposition-based network & page~\pageref{first:sd} \\
\hline
\rowcolor{gray!20} SL & Supervised Learning  & page~\pageref{first:sl} \\
\hline
RL & Reinforcement Learning & page~\pageref{first:rl} \\
\hline
\rowcolor{gray!20} EM & Ergodic Mean & page~\pageref{first:em} \\
\hline
DeepIS & Deep Importance Sampling & page~\pageref{first:is} \\
\hline
\rowcolor{gray!20}DLMC & Deep Learning/Monte Carlo & page~\pageref{first:dlmc} \\
\hline
DeepCV & Deep Control Variates & page~\pageref{first:cv} \\
\hline
\end{tabularx}
\caption{\textbf{List of the main abbreviations.} $^*$SM: Supplementary Material.}
\end{table}

\begin{table}[h]
\centering
\captionsetup{labelfont=bf}
\begin{tabularx}{\textwidth}{X|X}
\hline
\rowcolor{gray!20} \textbf{Notation} & \textbf{First appearance in the SM$^*$} \\
\hline
$N$ & page~\pageref{first:n} \\
\hline
\rowcolor{gray!20} $M$ & page~\pageref{first:m} \\
\hline
$\zeta_k$ & page~\pageref{first:zeta} \\
\hline
\rowcolor{gray!20} $\lambda_k(x)$ & page~\pageref{first:lambda} \\
\hline
$(X(t))$ & page~\pageref{first:x} \\
\hline
\rowcolor{gray!20} $\mathbb{A}$ & page~\pageref{first:a} \\
\hline
$(R_k(t))$ & page~\pageref{first:r} \\
\hline
\rowcolor{gray!20} $U(x,f,t)$ & page~\pageref{first:u} \\
\hline
$\pi$ & page~\pageref{first:pi} \\
\hline
\rowcolor{gray!20} $\phi_{\ell}$ & page~\pageref{first:phi} \\
\hline
$\gamma_{\ell}(f)$ & page~\pageref{first:gamma} \\
\hline
\rowcolor{gray!20} $\sigma_{\ell}$ & page~\pageref{first:sigma} \\
\hline
$U(f)$ & page~\pageref{first:vstat} \\
\hline
\rowcolor{gray!20} $a_{\ell}$, $b_{\ell}$ & page~\pageref{first:ab} \\
\hline
 $c_{\ell}$, $d_{\ell}$ & page~\pageref{first:cd} \\
\hline
\rowcolor{gray!20} $g_{\ell}(f)$, $h_{\ell}(f)$ & page~\pageref{first:gh} \\
\hline
$(\tilde{R}_k(t))$ & page~\pageref{first:tilder} \\
\hline
\rowcolor{gray!20} $E_{\eta}^{\text{IS}}$ & page~\pageref{first:eis} \\
\hline
$E_{\eta}^{\text{CV}}$ & page~\pageref{first:ecv} \\
\hline
\end{tabularx}
\caption{\textbf{List of the main notations.} $^*$SM: Supplementary Material.}
\end{table}

\section{Stochastic reaction networks}
\label{supp_section:srns}

\subsection{Mathematical definition of stochastic reaction networks}
\label{supp_subsection:srns}

Stochastic Reaction Networks~(SRNs)\label{first:srn} are extensively studied in refs.~\cite{mazza2014stochastic,wilkinson2018stochastic,anderson2015stochastic}. In what follows, we introduce the key concepts and tools relevant to this work. Consider a network consisting of a finite number $N\in\Natural^{*}$\label{first:n} of \emph{species}. The \emph{state} of the system at any given time is represented by a discrete vector in $\Natural^N$. The species interact through a finite number $M\in\Natural^{*}$\label{first:m} of \emph{reactions}, also referred to as reaction channels. Every time the $k$-th reaction occurs, the state of the system is displaced by the stoichiometric vector $\zeta_{k} \in \mathbb{Z}^{N}$\label{first:zeta} defined as:
\begin{equation}
\zeta_{k} \coloneqq \nu'_{\cdot k} - \nu_{\cdot k},
\end{equation}

where the reactant vector $\nu_{\cdot k}$ and the product vector $\nu'_{\cdot k}$ were introduced in the reaction graph~\eqref{eq:reaction_graph}. Species that do not participate as reactants~(resp.\ as products) in reaction $k$ are typically omitted from the left-hand side~(resp.\ right-hand side) of the reaction graph. If no species acts as a reactant~(resp.\ as a product) in reaction $k$, the left-hand side~(resp.\ the right-hand side) is replaced by the empty set $\varnothing$ in the reaction graph.\newline

We define a propensity function $\lambda = (\lambda_{k})_{k\in [\![1, M]\!]}$\label{first:lambda} that depends on the state of the system $x \in \Natural^N$. In other fields, $\lambda$ is also referred to as the intensity or rate function. The propensity $\lambda_k$ of reaction $k$~(or the associated parameter) is typically shown above the arrow in the reaction graph~\eqref{eq:reaction_graph}:
\begin{equation*}
\nu_{1k} \mathbf{S_1} + \ldots + \nu_{Nk} \mathbf{S_N} \xrightarrow[]{\lambda_k} \nu'_{1k} \mathbf{S_1} + \ldots + \nu'_{Nk} \mathbf{S_N}.
\end{equation*}

The dynamics of the system are described by a Continuous-Time Markov Chain~(CTMC)\label{first:ctmc} $(X(t))_{t\in\Rplus}$\label{first:x}, fully characterised by the stoichiometry vectors $\zeta_k$ and the propensities $\lambda_k$. These can be represented in several equivalent ways, which we introduce below and refer to throughout.\newline

\textbf{Transition rate matrix $\mathbb{Q}$.} Since the state space of the CTMC is countable, we can construct a bijection $\mathcal{B}$ between the state space and $\mathbb{N}$. Any two states $x$ and $\tilde{x}$ in the state space can then be written as $x = \mathcal{B}^{-1}(i)$ and $\tilde{x} = \mathcal{B}^{-1}(j)$, where $i$ and $j$ are two natural numbers. We define the transition rate matrix $\mathbb{Q}$\label{first:q} of the CTMC as the~(possibly bi-infinite) matrix defined by~\cite{anderson2015stochastic}:
\begin{equation}
[\mathbb{Q}]_{ij} \coloneqq \begin{cases} 
\begin{aligned}
\sum_{k\in\mathcal{T}_{x \rightarrow \tilde{x}}} \lambda_{k}(x)\quad &\text{if } x \neq \tilde{x}, \\
- \sum_{k=1}^{M} \lambda_{k}(x)\quad &\text{if } x=\tilde{x},
\end{aligned}
\end{cases}
\label{eq:infinit_def_generator}   
\end{equation}

where $\mathcal{T}_{x \rightarrow \tilde{x}} \coloneqq \{k\in [\![1, M]\!]\: |\: \tilde{x} = x + \zeta_k\}$ denotes the set of reactions that can move the system from state~$x$ to state~$\tilde{x}$. The transition rate matrix $\mathbb{Q}$ can also be interpreted as an operator acting on probability measures. It is used to formulate the Chemical Master Equation~(CME)\label{first:cme}, a linear Partial Differential Equation~(PDE) satisfied by the distribution $p$ of the process $(X(t))$, given by:
\begin{equation}
\label{eq:cme}
\frac{\partial p}{\partial t}(x,t) = \mathbb{Q}p(x,t), \quad \text{with } x\in\mathbb{N}^N \text{and } t \geq 0.
\end{equation}

In the broader stochastic process literature, $\mathbb{Q}$ is also referred to as an intensity~(or jump rate) matrix, and the CME as a Fokker–Planck equation.\newline

\textbf{Generator $\mathbb{A}$.} We now introduce the generator $\mathbb{A}$\label{first:a} of the CTMC as the operator defined by~\cite{anderson2015stochastic}:
\begin{equation}
\label{eq:def_generator}
 \mathbb{A}f(x) \coloneqq \sum_{k=1}^{M} \lambda_k(x)\Delta_{\zeta_k}f(x) = \sum_{k=1}^{M} \lambda_k(x) (f(x+\zeta_k)-f(x)),
\end{equation} 

for any bounded, real-valued function $f$ on $\mathbb{N}^{N}$. The operator $\mathbb{A}$ can also be viewed as a~(possibly bi-infinite) matrix, and it is the adjoint of $\mathbb{Q}$, \emph{i.e.} $\mathbb{A}^{*} = \mathbb{Q}$. It is used to express Itô's formula for SRNs and the Kolmogorov backward equation satisfied by $V(x, f, t, T) \coloneqq \mathbb{E}[f(X(T)) | X(t)=x]$, given by~\cite{anderson2015stochastic}:
\begin{equation}
\label{eq:kb_intro}
0 = \mathbb{A}V(x, f, t, T) + \frac{\partial V}{\partial t}(x, f, t, T), \quad \text{with } x\in\mathbb{N}^N \text{and } 0 \leq t \leq T.
\end{equation}

\textbf{Stochastic evolution equation.} Given a collection of independent, unit-rate Poisson processes $\{(Y_k(t))_{t \in \mathbb{R}_{+}}\}_{k \in [\![1, M]\!]}$, we associate to each reaction $k$ a counting process $(R_{k}(t))_{t \in \mathbb{R}_{+}}$\label{first:r} defined by:
\begin{equation}
\label{eq:reaction_count_process}
R_{k}(t) \coloneqq Y_{k}\bigg(\int_0^t \lambda_k(X(s)) ds\bigg).
\end{equation}

Its state increases by 1 every time the $k$-th reaction occurs. For an initial state $x \in \Natural^N$, the dynamics of the CTMC are given by the following stochastic evolution equation~\cite{anderson2015stochastic}:
\begin{equation}
\label{eq:rtc}    
X(t) = x + \sum_{k=1}^{M} \zeta_{k} R_{k}(t).
\end{equation}

Equations~\eqref{eq:reaction_count_process} and~\eqref{eq:rtc} define the so-called Random Time Change~(RTC) representation of $(X(t))$~\cite{kurtz1980representations}. The compensated version of $(R_k(t))$ is used to state Itô's formula for SRNs~(see equation~\eqref{eq:compensated_poisson} for the definition of the compensated process).\newline

\textbf{Quantity of interest.} Given an output function $f$, we are interested in the expected network output $U$ defined as:\label{first:u}
\begin{equation}
U(x, f, t) \coloneqq \mathbb{E}_{x}[f(X(t))] = \mathbb{E}[f(X(t)) | X(0)=x].
\end{equation}

By the time homogeneity of the process $(X(t))$, we have that: $V(x, f, t, T) = U(x, f, T-t)$.

\subsection{Stochastic reaction networks studied}

Below, we formally define all SRNs used throughout this work.

\subsubsection{Analytical running example}

As an analytical running example, we use the pure degradation network, a simple reaction network with a single species and reaction. This, in particular, allows us to omit the indexing of the species and reaction. The reaction graph of the network is given by:
\begin{equation}
\mathbf{S} \xrightarrow[]{\lambda} \varnothing,
\end{equation}

where $\lambda(x) = \theta x$. The stochastic dynamics of the network are described by:
\begin{equation}  
X(t) = x - R(t).
\end{equation}

Each time the reaction fires, the state of the network decreases by 1. For the output function $f$ defined as $f(x) = x$, the network is analytically tractable, and we have:
\begin{equation}
\label{eq:pure_death_first_moment}
U(x, f, t) =xe^{-\theta t}.
\end{equation}

Choosing the output function $f$ as $f(x)=x^2$ instead, we have:
\begin{equation}
\label{eq:pure_death_second_moment}
U(x, f, t) = x e^{-\theta t}+x(x-1)e^{-2\theta t}.
\end{equation}

\subsubsection{Stochastic reaction networks studied numerically}

The networks used in the numerical examples are introduced concisely by diagrams outside of this subsection. Here, they are fully specified. In that context, the generic label $\mathbf{S_i}$ for the $i$-th species is replaced by example-specific denominations.

\begin{exmp}[Constitutive gene expression network~\cite{thattai2001intrinsic}]
\begin{equation}
\varnothing \xrightarrow[]{\kappa_r} \mathbf{R},\quad \mathbf{R} \xrightarrow[]{\kappa_p} \mathbf{R} + \mathbf{P},\quad \mathbf{R} \xrightarrow[]{\gamma_r} \varnothing,\quad \mathbf{P} \xrightarrow[]{\gamma_p} \varnothing.
\end{equation}
\end{exmp}

To illustrate the notations introduced in subsection~\ref{supp_subsection:srns}, note that the network has two species~($N=2$), four reactions~($M=4$), and that:
\begin{equation}
\zeta_1 = \left[\begin{array}{c} 
1 \\
0
\end{array} \right] =  \left[\begin{array}{c} 
1 \\
0
\end{array} \right] - \left[\begin{array}{c} 
0 \\
0
\end{array} \right],\quad \text{and}\quad \zeta_2 = \left[\begin{array}{c} 
0 \\
1 
\end{array} \right] = \left[\begin{array}{c} 
1 \\
1 
\end{array} \right] - \left[\begin{array}{c} 
1 \\
0 
\end{array} \right] \quad \text{for example.}
\end{equation}

Species $\mathbf{S_1}$ corresponds to $\mathbf{R}$, and $\mathbf{S_2}$ to $\mathbf{P}$. The network has mass-action kinetics:
\begin{equation}
\lambda_1(x) = \kappa_r,\quad \lambda_2(x) = \kappa_p x_1,\quad \lambda_3(x) = \gamma_r x_1,\quad \lambda_4(x) = \gamma_p x_2.
\end{equation}

In this network, an mRNA $\mathbf{R}$ is constitutively transcribed at rate $\kappa_r$ and translated into a protein $\mathbf{P}$ at rate $\kappa_p$. Both molecules can be degraded, at rate $\gamma_r$ and $\gamma_p$, respectively. The propensities are linear in the abundance $x_i$ of each species, so we say that the network is \emph{linear}. We take: $\kappa_r=10$, $\kappa_p=2$, $\gamma_r=1$ and $\gamma_p=0.5$.

\begin{exmp}[Self-regulatory gene expression network~\cite{gupta2022frequency}]
\begin{equation}
\varnothing \xrightarrow[]{\lambda_1} \mathbf{P},\quad \mathbf{P} \xrightarrow[]{\gamma} \varnothing,
\end{equation}
\end{exmp}

where the propensities are given by:
\begin{equation}
\lambda_1(x) = \frac{\alpha}{K + x^n}, \quad\lambda_2(x) = \gamma x.
\end{equation}

In this network, a protein $\mathbf{P}$ inhibits its own production through a repressing Hill function and is degraded following mass-action kinetics. This network is nonlinear and involves a single species~($N=1$). We take: $\alpha=10,$ $K=10$, $n=0.5$ and $\gamma=0.05$.

\begin{exmp}[Genetic toggle switch network~\cite{gardner2000construction}]
\begin{equation}
\varnothing \xrightarrow[]{\lambda_1} \mathbf{P_1},\quad \varnothing \xrightarrow[]{\lambda_2} \mathbf{P_2},\quad \mathbf{P_1} \xrightarrow[]{\gamma_1} \varnothing,\quad \mathbf{P_2} \xrightarrow[]{\gamma_2} \varnothing.
\end{equation}
\end{exmp}

Species $\mathbf{S_1}$ corresponds to $\mathbf{P_1}$, and $\mathbf{S_2}$ to $\mathbf{P_2}$. The network has propensities given by:
\begin{equation}
\begin{aligned}
&\lambda_1(x) = b_1 + \frac{\alpha_1}{K_1 + (x_2)^{n_1}},&& \lambda_2(x) = b_2 + \frac{\alpha_2}{K_2 + (x_1)^{n_2}},\\
&\lambda_3(x) = \gamma_1 x_1,&& \lambda_4(x) = \gamma_2 x_2.
\end{aligned}
\end{equation}

The production of protein $\mathbf{P_1}$  is inhibited by protein $\mathbf{P_2}$ through a repressing Hill function, while the production of $\mathbf{P_2}$ is similarly inhibited by $\mathbf{P_1}$. Both molecules are degraded at rate $\gamma_1$ and $\gamma_2$, respectively. This network is nonlinear and involves two species~($N=2$). We take: $b_1=b_2=1$, $\alpha_1=50$, $\alpha_2=16$, $K_1=K_2=1$, $n_1= 2.5$, $n_2=1$ and $\gamma_1=\gamma_2=1$.

\begin{exmp}[Susceptible–infected–recovered network~\cite{thanh2015simulation}]
\begin{equation}
\begin{aligned}
&\mathbf{S} + \mathbf{I} \xrightarrow[]{\beta}  2\mathbf{I},& 
&\mathbf{I} \xrightarrow[]{\gamma} \mathbf{R},&
&\mathbf{R} \xrightarrow[]{\mu} \mathbf{S},&\\
&\mathbf{S} \xrightarrow[]{\delta_S} \varnothing,&
&\mathbf{I} \xrightarrow[]{\delta_I} \varnothing,&
&\mathbf{R} \xrightarrow[]{\delta_R} \varnothing.
\end{aligned}
\end{equation}
\end{exmp}

Species $\mathbf{S_1}$ corresponds to $\mathbf{S}$, $\mathbf{S_2}$ to $\mathbf{I}$, and $\mathbf{S_3}$ to $\mathbf{R}$. The network has mass-action kinetics:
\begin{equation}
\begin{aligned}
\lambda_1(x) &= \beta x_1 x_2, & \lambda_2(x) &= \gamma x_2,\\
\lambda_3(x) &= \mu x_3, & \lambda_4(x) &= \delta_S x_1,\\
\lambda_5(x) &= \delta_I x_2, & \lambda_6(x) &= \delta_R x_3.
\end{aligned}   
\end{equation}

An infected individual $\mathbf{I}$ can infect a susceptible person $\mathbf{S}$ at rate $\beta$. An infected patient can recover to become an individual $\mathbf{R}$ at rate $\gamma$, and recovered individuals can become susceptible again. The three types of individuals die at rates $\delta_S$, $\delta_I$, and $\delta_R$ respectively. This network is nonlinear and involves three species~($N=3$). We use: $\beta=0.0036$, $\gamma=0.02$, $\mu=0.007$, $\delta_S=0.002$, $\delta_I=0.05$ and $\delta_R=0.002$.

\begin{exmp}[Reference-based antithetic integral control of gene expression network~\cite{briat2016antithetic}]
\begin{equation}
\begin{aligned}
&\mathbf{Z_1} \xrightarrow[]{\kappa_r} \mathbf{Z_1} + \mathbf{R},
&& \mathbf{R} \xrightarrow[]{\kappa_p} \mathbf{R} + \mathbf{P},
&&\mathbf{R} \xrightarrow[]{\gamma_r} \varnothing,
&&\mathbf{P} \xrightarrow[]{\gamma_p} \varnothing,\\
&\mathbf{P} \xrightarrow[]{\theta} \mathbf{P} + \mathbf{Z_2},
&&\mathbf{Z_1} + \mathbf{Z_2} \xrightarrow[]{\eta} \varnothing,
&&\varnothing \xrightarrow[]{\mu} \mathbf{Z_1}.&&
\end{aligned}
\end{equation}
\end{exmp}

Species $\mathbf{S_1}$ corresponds to $\mathbf{R}$, $\mathbf{S_2}$ to $\mathbf{P}$, $\mathbf{S_3}$ to $\mathbf{Z_1}$, and $\mathbf{S_4}$ to $\mathbf{Z_2}$.  The network has mass-action kinetics:
\begin{equation}
\begin{aligned}
&\lambda_1(x)= \kappa_r x_3,&& \lambda_2(x) = \kappa_p x_1,&&
\lambda_3(x) = \gamma_r x_1,&& \lambda_4(x) = \gamma_p x_2,\\
&\lambda_5(x) = \theta x_2,&& \lambda_6(x) = \eta x_3 x_4,&&
\lambda_7(x) = \mu.
\end{aligned}
\end{equation}

The abundance of the protein $\mathbf{P}$ is regulated by the antithetic integral controller involving the regulatory species $\mathbf{Z_1}$ and $\mathbf{Z_2}$. The transcription rate of the mRNA $\mathbf{R}$ is proportional to the abundance of $\mathbf{Z_1}$~(actuation reaction), while the production rate of $\mathbf{Z_2}$ is proportional to the abundance of the protein $\mathbf{P}$~(sensing reaction). The propensity $\lambda_2$ corresponds to the translation of mRNA $\mathbf{R}$ into protein $\mathbf{P}$, and the propensities $\lambda_3$ and $\lambda_4$ correspond to their degradation. The species $\mathbf{Z_1}$ and $\mathbf{Z_2}$ are jointly degraded via an annihilation reaction with propensity $\lambda_6$. Here, $\theta$ is considered only one element of the parameter vector, consistent with ref.~\cite{briat2016antithetic}. This network is nonlinear and comprises four species~($N=4$). We take: $\kappa_r=5$, $\kappa_p=2$, $\gamma_r=5$, $\gamma_p=0.5$, $\theta=1$, $\eta=100$ and
$\mu=10$.

\begin{exmp}[Sensor-based antithetic integral control of gene expression network~\cite{filo2023hidden}]
\begin{equation}
\begin{aligned}
&\varnothing \xrightarrow[]{\lambda_1} \mathbf{R},
&& \mathbf{R} \xrightarrow[]{\kappa_p} \mathbf{R} + \mathbf{P},
&&\mathbf{R} \xrightarrow[]{\gamma_r} \varnothing,
&&\mathbf{P} \xrightarrow[]{\gamma_p} \varnothing,\\
&\mathbf{P} \xrightarrow[]{\theta} \mathbf{P} + \mathbf{Z_2},
&&\mathbf{Z_1} + \mathbf{Z_2} \xrightarrow[]{\eta} \varnothing,
&&\varnothing \xrightarrow[]{\mu} \mathbf{Z_1}.&&
\end{aligned}
\end{equation}
\end{exmp}

Species $\mathbf{S_1}$ corresponds to $\mathbf{R}$, $\mathbf{S_2}$ to $\mathbf{P}$, $\mathbf{S_3}$ to $\mathbf{Z_1}$, and $\mathbf{S_4}$ to $\mathbf{Z_2}$. The network has propensities given by:
\begin{equation}
\begin{aligned}
&\lambda_1(x) = b+\frac{\alpha}{K + (x_4)^{n}},&& \lambda_2(x) = \kappa_p x_1,&&
\lambda_3(x) = \gamma_r x_1,&& \lambda_4(x) = \gamma_p x_2,\\
&\lambda_5(x) = \theta x_2,&& \lambda_6(x) = \eta x_3 x_4,&&
\lambda_7(x) = \mu.
\end{aligned}
\end{equation}

As in the previous network, the abundance of protein $\mathbf{P}$ is regulated by the species $\mathbf{Z_1}$ and $\mathbf{Z_2}$. The only difference is in the first reaction, where $\mathbf{Z_2}$ inhibits the production of the mRNA $\mathbf{R}$. This network is nonlinear and comprises four species~($N=4$). We take: $b=10$, $\alpha=5$, $K=0.5$, $n=1$, $\kappa_p=2$, $\gamma_r=5 $, $\gamma_p=0.5 $, $\theta=1$, $\eta=100$ and $\mu=10$.
    
\begin{exmp}[Repressilator network~\cite{elowitz2000synthetic}]
\begin{equation}
\begin{aligned}
&\varnothing \xrightarrow[]{\lambda_1} \mathbf{R_1},& 
&\varnothing \xrightarrow[]{\lambda_2} \mathbf{R_2},&
&\varnothing \xrightarrow[]{\lambda_3} \mathbf{R_3},&\\
&\mathbf{R_1} \xrightarrow[]{\kappa_{p,1}} \mathbf{R_1} + \mathbf{P_1},& 
&\mathbf{R_2} \xrightarrow[]{\kappa_{p,2}} \mathbf{R_2} + \mathbf{P_2},&
&\mathbf{R_3} \xrightarrow[]{\kappa_{p,3}} \mathbf{R_3} + \mathbf{P_3},&\\
&\mathbf{R_1} \xrightarrow[]{\gamma_{r,1}} \varnothing,&
&\mathbf{R_2} \xrightarrow[]{\gamma_{r,2}} \varnothing,&
&\mathbf{R_3} \xrightarrow[]{\gamma_{r,3}} \varnothing,\\
&\mathbf{P_1} \xrightarrow[]{\gamma_{p,1}} \varnothing,&
&\mathbf{P_2} \xrightarrow[]{\gamma_{p,2}} \varnothing,&
&\mathbf{P_3} \xrightarrow[]{\gamma_{p,3}} \varnothing.
\end{aligned}
\end{equation}
\end{exmp}

Species $\mathbf{S_1}$ corresponds to $\mathbf{R_1}$, $\mathbf{S_2}$ to $\mathbf{R_2}$, $\mathbf{S_3}$ to $\mathbf{R_3}$, $\mathbf{S_4}$ to $\mathbf{P_1}$, $\mathbf{S_5}$ to $\mathbf{P_2}$ and $\mathbf{S_6}$ to $\mathbf{P_3}$.  The network has propensities given by:
\begin{equation}
\begin{aligned}
&\lambda_1(x) =  \beta_1 + \frac{\alpha_1}{K_1 + (x_6)^{n_1}}, 
&&\lambda_2(x) = \beta_2 + \frac{\alpha_2}{K_2 + (x_4)^{n_2}},
&&\lambda_3(x) = \beta_3 + \frac{\alpha_3}{K_3 + (x_5)^{n_3}},\\
&\lambda_4(x) = \kappa_{p,1}x_1,
&&\lambda_5(x) = \kappa_{p,2}x_2,
&&\lambda_6(x) = \kappa_{p,3}x_3,\\
&\lambda_7(x) = \gamma_{r,1}x_1,
&&\lambda_8(x) = \gamma_{r,2}x_2,
&&\lambda_9(x) = \gamma_{r,3}x_3,\\
&\lambda_{10}(x) = \gamma_{p,1}x_4,
&&\lambda_{11}(x) = \gamma_{p,2}x_5,
&&\lambda_{12}(x) = \gamma_{p,3}x_6.
\end{aligned}
\end{equation}

Some mRNA molecules $\mathbf{R_1}$, $\mathbf{R_2}$, and $\mathbf{R_3}$ are produced constitutively at rates $\beta_1$, $\beta_2$, and $\beta_3$. The protein $\mathbf{P_3}$ translated from $\mathbf{R_3}$ inhibits the production of $\mathbf{R_1}$ through a repressing Hill function, while $\mathbf{P_1}$ inhibits the production of $\mathbf{R_2}$, and $\mathbf{P_2}$ that of $\mathbf{R_3}$. The mRNA molecules are then translated at rates $\kappa_{p,1}$, $\kappa_{p,2}$, and $\kappa_{p,3}$, respectively. The six types of molecules are degraded at rates $\gamma_{r,1}$, $\gamma_{r,2}$, $\gamma_{r,3}$, $\gamma_{p,1}$, $\gamma_{p,2}$, and $\gamma_{p,3}$. This network is nonlinear and has 6 species~($N=6$). We take: $\beta_1=\beta_2=\beta_3=1$, $\alpha_1=\alpha_2=\alpha_3=10$, $K_1=K_2=K_3=1$, $n_1=n_2=n_3=1$, $\kappa_{p,1}=\kappa_{p,2}=\kappa_{p,3}=2$, $\gamma_{r,1}=\gamma_{r,2}=\gamma_{r,3}=1$ and $\gamma_{p,1}=\gamma_{p,2}=\gamma_{p,3}=0.1$.

\begin{exmp}[Nonlinear conversion cascade network~\cite{tang2023neural}]
\begin{equation}
\varnothing \xrightarrow[]{\kappa} \mathbf{S_1},\quad \mathbf{S_i} \xrightarrow[]{\lambda_{i+1}}  \mathbf{S_{i+1}} \text{ \emph{for} } i \in [\![1,N-1]\!],\quad \mathbf{S_i} \xrightarrow[]{\gamma} \varnothing  \text{ \emph{for} } i \in [\![1,N]\!],
\end{equation}
\end{exmp}

where the propensities are given by:
\begin{equation}
\begin{aligned}
&\lambda_1(x) = \kappa,&& \lambda_{i+1}(x) = \begin{cases} 
b + \dfrac{\alpha (x_i)^n}{K + (x_i)^{n}} & \text{if } x_{i} > 0, \\
0 & \text{otherwise,}
\end{cases}
\quad \text{ for }i \in [\![1,N-1]\!],\\
&\lambda_{N+i}(x) = \gamma x_i \text{ for } i \in [\![1,N]\!].&&
\end{aligned}
\end{equation}

The first species $\mathbf{S_1}$ is produced constitutively at rate $\kappa$. Each species $\mathbf{S_i}$ for $i \in [\![1,N-1]\!]$ gets converted to the next species $\mathbf{S_{i+1}}$ through a Hill activation function. All species are degraded at rate $\gamma$. This network is nonlinear and we choose $N=10$. We take: $\kappa=10$, $b=1$, $\alpha=100$, $K=10$, $n=1$ and $\gamma=1$.

\begin{exmp}[Linear conversion cascade with feedback network~\cite{tang2023neural}]
\begin{equation}
\varnothing \xrightarrow[]{\lambda_1} \mathbf{S_1},\quad \mathbf{S_i} \xrightarrow[]{\kappa}  \mathbf{S_{i+1}} \text{ \emph{for} } i \in [\![1,N-1]\!],\quad \mathbf{S_i} \xrightarrow[]{\gamma} \varnothing  \text{ \emph{for} } i \in [\![1,N]\!],
\end{equation}
\end{exmp}

where the propensities are given by:
\begin{equation}
\begin{aligned}
&\lambda_1(x) = b + \frac{\alpha}{K + (x_N)^{n}},&& \lambda_{i+1}(x) = \kappa x_i \text{ for } i \in [\![1,N-1]\!],\\
&\lambda_{N+i}(x) = \gamma x_i \text{ for } i \in [\![1,N]\!].&&
\end{aligned}
\end{equation}

The production of the first species $\mathbf{S_1}$ is inhibited by the last species $\mathbf{S_N}$ through a repressing Hill function. Each species $\mathbf{S_i}$ for $i \in [\![1,N-1]\!]$ gets converted to the next species $\mathbf{S_{i+1}}$ at rate $\kappa$. All species are degraded at rate $\gamma$. This network is nonlinear and we choose $N=10$. We take: $b=1$, $\alpha=100$, $K=10$, $n=1$, $\kappa=5$ and $\gamma=1$.

\section{Representation of expected outputs with an interpretable neural network architecture}
\label{supp_section:sdnet}

In this section, we introduce the \emph{Spectral Decomposition-based network}~(SDnet)\label{first:sd} architecture for representing expected outputs understood as functions: 
\begin{equation}
\begin{aligned}
U : \mathbb{N}^N \times \mathbb{R}_+  & \longrightarrow \mathbb{R} \\
(x,t) & \longmapsto U(x,f,t),
\end{aligned}    
\end{equation}

where $U(x, f, t) = \mathbb{E}_{x}[f(X(t))] = \mathbb{E}[f(X(t)) | X(0)=x]$, as defined in section~\ref{supp_section:srns} of the supplementary material.\newline

SDnet can be trained using various approaches. The simplest is a Supervised Learning~(SL)\label{first:sl} method based, for example, on Monte Carlo estimates of the expected outputs. In this work, however, we employ an extension of the Reinforcement Learning~(RL)\label{first:rl} training strategy introduced in ref.~\cite{gupta2021deepcme}~(see section~\ref{supp_section:training} of the supplementary material).\newline

Expected outputs can be alternatively approximated using existing neural representations. These architectures, however, lack interpretability. The most direct approach employs a feedforward neural network that takes the initial state~$x$ and the current time $t$ as inputs. As an alternative, ref.~\cite{gupta2021deepcme} introduced an architecture which relies on a set of temporal features dependent on learnable parameter triplets.

\subsection{The Spectral Decomposition-based network}

We now introduce in detail SDnet, which is based on the spectral decomposition of $U$ described below.\newline

\textbf{The spectral decomposition of expected outputs.} Assume that the process $(X(t))$ is exponentially ergodic with stationary distribution $\pi$\label{first:pi}, and that the~(complex) eigenvalues of $\mathbb{A}$ are distinct. Arrange them in descending order according to their~(nonpositive) real parts. The sequence $(\tau_{\ell})_{\ell\in\mathbb{N}}$ of eigenvalues starts at 0 and diverges to~$-\infty$~(see refs.~\cite{gupta2022frequency,gupta2025sparse} and references therein). The function $\phi_{\ell} : \mathbb{N}^N \xrightarrow[]{} \mathbb{C}$\label{first:phi} is an eigenfunction associated with $\tau_{\ell}$. The eigenfunction corresponding to $\tau_{0} = 0$ is $\phi_{0} \equiv 1$. Let $L_{ \phi}^2(\pi)$ denote the $L^2$ space spanned by $(\phi_{\ell})_{\ell}$. As in ref.~\cite{gupta2025sparse}, assume that the output function~$f$ lies in  $L_{ \phi}^2(\pi)$. The scalar $\gamma_{\ell}(f) \in \mathbb{C}$\label{first:gamma} is the $\ell$-th coordinate of $f$ in the basis $(\phi_{\ell})_{\ell}$. With this, we can expand $f$ in the basis $(\phi_{\ell})_{\ell}$ as~\cite{gupta2025sparse}:
\begin{equation}
\label{eq:spectral_f}
f = \sum_{\ell=0}^{\infty}\gamma_{\ell}(f)\phi_{\ell} = \gamma_{0}(f) + \sum_{\ell=1}^{\infty}\gamma_{\ell}(f)\phi_{\ell}.
\end{equation}

Next, we  introduce the transition semigroup $\mathcal{K}_t$, defined as the~(linear) operator:
\begin{equation}
\label{eq:koopman_def}
\big[\mathcal{K}_tf\big](x) \coloneqq U(x, f, t) = \big[e^{\mathbb{A}t}f\big](x),
\end{equation}

for any bounded, real-valued function $f$ on $\mathbb{N}^{N}$.  $\mathbb{A}$ and $\mathcal{K}_t$ share the same eigenfunctions. The eigenfunction $\phi_{\ell}$ corresponds to the eigenvalue $\exp(\tau_{\ell}t)$ of $\mathcal{K}_t$. We define the decay mode $\sigma_{\ell}$\label{first:sigma} as the negative of $\tau_{\ell}$. Using equations~\eqref{eq:spectral_f} and \eqref{eq:koopman_def}, we can express $U$ as~\cite{gupta2025sparse}:
\begin{equation}
\label{eq:complex_decomposition_raw}
U(x,f,t) = \big[\mathcal{K}_{t}f\big](x) =  U(f) + \sum_{\ell=1}^{\infty} e^{-\sigma_{\ell}t}\gamma_{\ell}(f)\phi_{\ell}(x),
\end{equation}

where $U(f)$\label{first:vstat} denotes the average of $f$ with respect to the stationary distribution $\pi$. Express the decay mode $\sigma_{\ell}$ as $\sigma_{\ell} \eqqcolon a_{\ell}+i b_{\ell}$\label{first:ab}, the eigenfunction $\phi_{\ell}$ as $\phi_{\ell}(x) \eqqcolon c_{\ell}(x)+i d_{\ell}(x)$\label{first:cd}, and the function coordinate $\gamma_{\ell}(f)$ as $\gamma_{\ell}(f) \eqqcolon g_{\ell}(f)+i h_{\ell}(f)$\label{first:gh}. We can then rewrite equation~\eqref{eq:complex_decomposition_raw} to represent $U$ as:
\begin{equation}
\label{eq:complex_decomposition_expanded}
U(x,f,t) = U(f) + \sum_{\ell=1}^{\infty} e^{-(a_{\ell}+ib_{\ell}) t}(g_{\ell}(f)+ih_{\ell}(f))(c_{\ell}(x)+id_{\ell}(x)).
\end{equation}

Alternatively, let us re-index the eigenfunctions and eigenvalues of $\mathcal{K}_{t}$ as follows. When an eigenfunction is real-valued, we denote it by $\phi_{m}$. When it is complex-valued, its complex conjugate is also an eigenfunction, and we~(arbitrarily) denote  one element of the pair by $\phi_{m}$. In either case, we write $\sigma_{m}$ for the decay mode associated with $\phi_{m}$. When $\phi_{m}$ is real-valued, we define $\gamma_{m}(f)$ as the scalar such that $2\gamma_{m}(f)$ is the corresponding coordinate of the output function~$f$ in the basis of eigenfunctions. When $\phi_{m}$ is complex-valued, we define $\gamma_{m}(f)$ as the corresponding coordinate of $f$. In this case, the complex conjugate $\bar{\gamma}_{m}(f)$ is the coordinate of $f$ associated with $\bar{\phi}_{m}$. We express the eigenfunction~$\phi_{m}$ as $\phi_{m}(x) \eqqcolon c_{m}(x)+i d_{m}(x)$, the decay mode $\sigma_{m}$ as $\sigma_{m} \eqqcolon a_{m}+i b_{m}$, and $\gamma_{m}(f)$ as $\gamma_{m}(f) \eqqcolon g_{m}(f)+i h_{m}(f)$. We can now reformulate equation~\eqref{eq:complex_decomposition_raw} to write $U$ as:
\begin{equation}
\begin{aligned}
\label{eq:real_decomposition_expanded}
U(x,f,t) =  U(f) + 2 \sum_{m=1}^{\infty} e^{-a_{m} t} \bigg[g_{m} \left( f \right) &\left( c_{m} (x) \cos (b_{m} t) + d_{m} (x) \sin (b_{m} t) \right)\\
&+h_{m} \left( f \right) \left( c_{m} (x) \sin (b_{m} t) - d_{m} (x) \cos (b_{m} t) \right)\bigg].
\end{aligned}
\end{equation}

\textbf{Running example~(continued).} Recall from equation~\eqref{eq:pure_death_second_moment} that for $f(x)=x^2$:
\begin{equation*}
U(x, f, t) = e^{-\theta t}x+e^{-2\theta t}x(x-1).
\end{equation*}

Setting:
\begin{align}
\label{eq:pure_death_spectral_components}
U(f)=0,\quad \begin{cases}
\gamma_{1}(f)&= 1,\\
\sigma_{1}&= \theta,\\
\phi_{1} (x)&= x,
\end{cases}
\quad \text{and} \quad
\begin{cases}
\gamma_{2}(f)&= 1,\\
\sigma_{2}&= 2\theta,\\
\phi_{2}(x)&=  x(x-1),
\end{cases}
\end{align}

we have:
\begin{equation*}
U(x,f,t) = U(f) + e^{-\sigma_{1} t} \gamma_{1}(f) \phi_{1} (x)+ e^{-\sigma_{2} t} \gamma_{2}(f) \phi_{2} (x),
\end{equation*}

which is an instance of the spectral decomposition of $U$ given in equation~\eqref{eq:complex_decomposition_raw}. In addition, setting:
\begin{align}
U(f)=0,\quad \begin{cases}
g_{1}(f)&=  1/2,\\
a_{1} &=  \theta,\\
c_{1}(x) &=  x,
\end{cases}
\quad \text{and} \quad
\begin{cases}
g_{2}(f) &=   1/2,\\
a_{2} &= 2\theta,\\
c_{2}(x) &=  x(x-1),
\end{cases}
\end{align}

we have:
\begin{equation}
U(x,f,t) = U(f) + 2e^{-a_{1} t} g_{1}(f) c_{1} (x)+ 2e^{-a_{2} t} g_{2}(f) c_{2} (x),
\end{equation}

which is an instance of the spectral decomposition of $U$ given in equation~\eqref{eq:real_decomposition_expanded}.\newline

\textbf{The Spectral Decomposition-based network (SDnet).} We use the right-hand sides of equations~\eqref{eq:complex_decomposition_expanded} and~\eqref{eq:real_decomposition_expanded} as blueprints for a class of architectures which we call SDnet. More specifically, we refer to the architecture based on equation~\eqref{eq:complex_decomposition_expanded} as \emph{Complex SDnet}~(CSDnet), and the one based on equation~\eqref{eq:real_decomposition_expanded} as \emph{Matched SDnet}~(MSDnet). The designation MSDnet highlights that the complex conjugates are paired in equation~\eqref{eq:real_decomposition_expanded}. We focus the remainder of the presentation on the CSDnet, as it naturally extends to the MSDnet and is the architecture employed in the numerical experiments.\newline

In equation~\eqref{eq:complex_decomposition_expanded}, the most immediate approach would be to replace the stationary mean $U(f)$ with a trainable variable $\hat{U}(f)$. To avoid this, let us introduce the finite-time average $\mathcal{E}(f,t)$ and the ergodic mean $\mathcal{E}(f)$ as follows:
\begin{equation}
\mathcal{E}(f,t) \coloneqq \frac{1}{t}\int_{0}^{t}f(X(s))ds, \quad \text{and} \quad \mathcal{E}(f) \coloneqq \lim_{t \rightarrow \infty} \mathcal{E}(f,t).
\end{equation}

We use the \emph{Ergodic Mean~\emph{(EM)}\label{first:em} estimator} $\mathcal{E}(f,t)$ with a large time $t$ to obtain an estimate $\hat{U}(f)$ of $U(f)$. The approximation can be made arbitrarily precise since $\mathcal{E}(f) = U(f)$, as established by the ergodic theorem~\cite{asmussen2007stochastic}.  Note that $\hat{U}(f)$ could be treated as a trainable variable beyond its initialisation to the value obtained from $\mathcal{E}(f,t)$. When an analytical expression for $U$ is available, the exact value of $U(f)$ can be used directly instead~\cite{schnoerr2017approximation}. This is in particular the case for Stochastic Reaction Networks~(SRNs) exhibiting maximal Robust Perfect Adaptation~(maxRPA), a key property for the reliable performance of engineered cells in therapeutic applications~\cite{aoki2019universal,gupta2022universal,filo2023biomolecular}. Some expected outputs of this class of systems are robust to all reaction network parameters except two, and these outputs have known expressions.\newline

\begin{figure}[H]
\centering
\captionsetup{labelfont=bf}
\includegraphics[width=\textwidth]{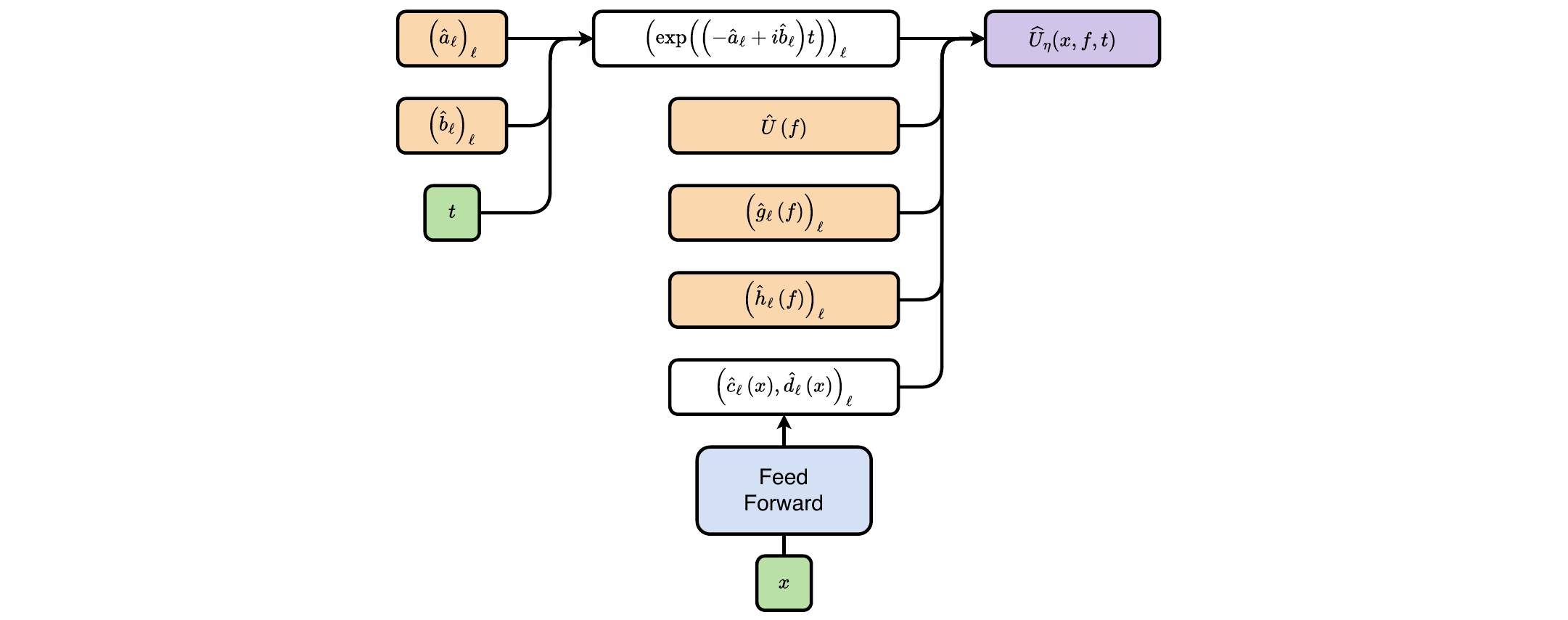}
\caption{\textbf{The Spectral Decomposition-based network (SDnet).} The architecture is motivated by the spectral decomposition of $U$ in equation~\eqref{eq:complex_decomposition_raw}, which is approximated in equation~\eqref{eq:complex_decomposition_expanded_approx}. Specifically, $\hat{\sigma}_{\ell} \coloneqq \hat{a}_{\ell} + i \hat{b}_{\ell}$ serves as an approximation of a decay mode, $\hat{\gamma}_{\ell}(f) \coloneqq \hat{h}_{\ell}(f) + i \hat{g}_{\ell}(f)$ approximates a function coordinate, and $\hat{\phi}_{\ell} \coloneqq \hat{c}_{\ell} + i \hat{d}_{\ell}$ approximates an eigenfunction. Green: inputs. Purple: output. Orange: trainable or fixed variables. Blue: feedforward neural network. White: intermediate quantities.}
\label{supp_figure:sdnet}
\end{figure}

We also truncate in equation~\eqref{eq:complex_decomposition_expanded} the infinite sum to its $r \in\mathbb{N}^{*}$ first elements. Trainable variables $(\hat{g}_{\ell} \left( f \right),\hat{h}_{\ell} \left( f \right))_\ell \in \mathbb{R}^{2r}$ approximate $(g_{\ell} \left( f \right), h_{\ell} \left( f \right))_{\ell\in[\![1,r]\!]}$, while $(\hat{a}_{\ell},\hat{b}_{\ell})_\ell \in (\mathbb{R}_+ \times \mathbb{R})^r$ replace $(a_{\ell},b_{\ell})_{\ell\in[\![1,r]\!]}$. A feedforward neural network $(\hat{c}_{\ell}, \hat{d}_{\ell})_\ell : \mathbb{R}^{N}  \longrightarrow \mathbb{R}^{2r}$ serves as a surrogate for $(c_{\ell}, d_{\ell})_{\ell\in[\![1,r]\!]}$, although alternative neural architectures or even classes of parametric approximation could be employed as well. Pooling all parameters~(trainable variables, feedforward weights and biases) into a single variable $\eta$, the resulting approximation $\widehat{U}_{\eta}$ of $U$ is given by~(see Supp. Fig.~\ref{supp_figure:sdnet} for a visual illustration):
\begin{equation}
\label{eq:complex_decomposition_expanded_approx}
\widehat{U}_{\eta}(x,f,t) = \hat{U}(f) + \sum_{\ell=1}^{r} e^{-(\hat{a}_{\ell}+i\hat{b}_{\ell})t}(\hat{g}_{\ell}(f)+i\hat{h}_{\ell}(f))(\hat{c}_{\ell}(x)+i\hat{d}_{\ell}(x)).
\end{equation}

Observe that $U(x,f,t)$ is a real number, including in equation~\eqref{eq:complex_decomposition_expanded}, whereas $\widehat{U}_{\eta}(x,f,t)$, as defined in equation~\eqref{eq:complex_decomposition_expanded_approx}, may be complex. Most training approaches, including the RL method described in section~\ref{supp_section:training} of the supplementary material, can handle complex values and simply penalise any non-zero imaginary part. However, when using the trained CSDnet in practice, it is often preferable to obtain a real-valued output, in which case we simply take the real part of $\widehat{U}_{\eta}(x,f,t)$. This adjustment is unnecessary for the MSDnet, which always produces real outputs by design.

\textbf{Benefits of the Spectral Decomposition-based network.} The main benefit of SDnet lies in its mathematically interpretable structure --- that is, its components carry clear mathematical meaning. This translates into several advantages that we outline below.\newline

To begin with, this interpretability links the architecture to a well-established body of theoretical results and numerical methods from which it can draw. This strength has already been demonstrated through the use of the EM estimator $\mathcal{E}(f,t)$ to estimate $U(f)$. As a further example, the values of the decay modes and the number of terms $r$ to include could be obtained from the Stochastic Koopman Approximation (SKA) introduced in ref.~\cite{gupta2025sparse}, in which case $\hat{\sigma}_{\ell}$ may either be fixed or treated as a trainable variable beyond its initialisation with the SKA value. A related example, concerning the evaluation of SDnet, is presented below~(see ``Evaluation of of the Spectral Decomposition-based network'').\newline

When representing $U$ even for a single output function $f$, neural networks based on earlier architectures must be sufficiently deep or wide to capture the combined influence of the state $x$ and the time $t$. In the newly introduced architecture, these contributions are explicitly separated, allowing the neural network approximating $(c_{\ell}, d_{\ell})_\ell$ to potentially be much smaller~(see the dependency on $x$ and $t$ in equation~\eqref{eq:complex_decomposition_expanded_approx}).\newline

It is often desirable to represent $U$ for multiple functions simultaneously~(or equivalently, for a vector-valued function $f = (f_i)_{i\in[\![1,n]\!]}$ with $n \in \mathbb{N}^{*}$). In such cases, the earlier architectures are expected to require an increasing number of weights to remain expressive enough. In contrast, SDnet assigns distinct coordinates $\gamma_{\ell}(f_i)$ to each component $f_i$,  while sharing the remainder of the expression~(see the dependency on $f$ in equation~\eqref{eq:complex_decomposition_expanded_approx}). As a result, learning $U$ for multiple~(and possibly many) output functions can be achieved by simply adding $2r$ scalars per function, while keeping the size of the neural network approximating $(c_{\ell},d_{\ell})_\ell$ fixed.\newline

Once the decay modes $\sigma_{\ell}$ and eigenfunctions $\phi_{\ell}$ have been learned for a given set of output functions within SDnet, they can be reused to approximate $U$ for any new output function $f$ of interest. In this case, only the new function coordinates $\gamma_{\ell}(f)$, which correspond to a set of $2r$ scalars, need to be learned~(again refer to the dependency on $f$ in equation~\eqref{eq:complex_decomposition_expanded_approx}). This can be achieved within the training procedure described in section~\ref{supp_section:training} of the supplementary material by allowing only the new function coordinates to be trainable, or by performing regression against the approximation of the eigenfunctions $\phi_{\ell}$. In contrast, with earlier architectures, training the entire neural network must be restarted from scratch.\newline

When an analytical expression for $U$ is desired, symbolic regression can be used to regress $\widehat{U}_{\eta}$ onto a library of functions~\cite{schmidt2009distilling,dubvcakova2011eureqa,udrescu2020ai,cranmer2023interpretable}. For this, the neural network can be evaluated efficiently once trained, allowing the generation of a large number of pairs $((x,t), \widehat{U}_{\eta}(x,f,t))$ to perform the regression. An alternative would be the~(potentially) time-consuming simulation of $(X(t))$ across a range of initial states~$x$ and times~$t$ to obtain precise estimates of $U(x,f,t)$. Moreover, the architecture based on equation~\eqref{eq:complex_decomposition_expanded_approx} is a sum of decaying exponentials, which tends to yield smooth approximations of $U$~(see the results in the main text). In contrast, Monte Carlo approximations typically produce noisy training data for symbolic regression. Pairs $(x, \hat{c}_{\ell}(x))$ and $(x, \hat{d}_{\ell}(x))$ can themselves be used for regression, which is not feasible from Monte Carlo simulations of $(X(t))$ alone. Each of these regression problems is one dimension lower than that for $\widehat{U}_{\eta}$, since the pairs $(x, \hat{c}_{\ell}(x))$ and $(x, \hat{d}_{\ell}(x))$ do not depend on time $t$. This approach breaks down the initial regression task into $2r$ subproblems, which are expected to be simpler.\newline

Another advantage of SDnet is that it enforces accuracy of the approximation $\widehat{U}_{\eta}$ in the large-time limit, as:
\begin{equation}
\lim_{t \rightarrow \infty} \widehat{U}_{\eta}(x,f,t) = \hat{U}(f),
\end{equation}

where $\hat{U}(f)$ is either computed from the ergodic average or known a priori. SDnet therefore ensures physically meaningful steady-state behaviour, which is not possible, for instance, with earlier architectures.\newline

\textbf{Evaluation of the Spectral Decomposition-based network.} Let us define $\hat{\gamma}_{\ell}(f)$, $\hat{\sigma}_{\ell}$, and $\hat{\phi}_{\ell}$ as follows:
\begin{equation}
\begin{cases}
\hat{\gamma}_{\ell} (f) &\coloneqq \hat{g}_{\ell}(f) + i \hat{h}_{\ell}(f),\\
\hat{\sigma}_{\ell} &\coloneqq \hat{a}_{\ell}+i \hat{b}_{\ell},\\
\hat{\phi}_{\ell} &\coloneqq \hat{c}_{\ell}+i \hat{d}_{\ell}.
\end{cases}   
\end{equation}

Their mathematical meaning can be used to evaluate SDnet. Recall from equation~\eqref{eq:spectral_f} that:
\begin{equation}
\label{eq:output_consistency}
f(x) = \gamma_0(f) + \sum_{\ell=1}^{\infty}\gamma_{\ell}(f)\phi_{\ell}(x).
\end{equation}

The right-hand side can be estimated using quantities from SDnet. The difference from the left-hand side indicates how well $\hat{\gamma}_{\ell}(f)$ and $\hat{\phi}_{\ell}$ satisfy the properties of function coordinates and eigenfunctions. Since $\exp(-\sigma_{\ell}t)$ is an eigenvalue of $\mathcal{K}_t$ and $\phi_{\ell}$ an eigenfunction, we also have that:  
\begin{equation}
\mathcal{K}_t \phi_{\ell}(x) = E_x \left[ \phi_{\ell} (X(t)) \right] = e^{-\sigma_{\ell} t}\phi_{\ell} (x),
\end{equation}

which can be decomposed into:
\begin{align}
\label{eq:koopman_consistency}
\begin{cases}
E_x \left[c_{\ell} (X(t)) \right] &= e^{-a_{\ell} t} (c_{\ell}(x)\cos(b_{\ell} t) + d_{\ell}(x)\sin(b_{\ell} t)),\\
E_x \left[d_{\ell} (X(t)) \right] &= e^{-a_{\ell} t}(d_{\ell}(x)\cos(b_{\ell} t) - c_{\ell}(x)\sin(b_{\ell} t)).
\end{cases}
\end{align}

The left-hand sides in equation~\eqref{eq:koopman_consistency} can be estimated from simulations of the process $(X(t))$, while the right-hand sides can be computed directly from SDnet. Comparing these values provides an indication of how well $\hat{\sigma}_{\ell} = \hat{a}_{\ell}+i \hat{b}_{\ell}$ and $\hat{\phi}_{\ell}=\hat{c}_{\ell}+i \hat{d}_{\ell}$ satisfy the properties of decay modes and eigenfunctions.\newline

\textbf{Running example~(continued).} Recall from equation~\eqref{eq:pure_death_spectral_components} that for $f(x)=x^2$:
\begin{align*}
U(f) = \gamma_0(f) = 0,\quad \begin{cases}
\gamma_{1}(f) &=  1,\\
\sigma_{1} &=  \theta,\\
\phi_{1}(x) &=  x,
\end{cases}
\quad \text{and} \quad
\begin{cases}
\gamma_{2}(f) &=   1,\\
\sigma_{2} &=  2\theta,\\
\phi_{2}(x) &=  x(x-1).
\end{cases}
\end{align*}

It is immediate that:
\begin{equation}
x^2 = \gamma_{0}(f) + \gamma_{1}(f)\phi_{1}(x)+\gamma_{2}(f)\phi_{2}(x),
\end{equation}

which is an instance of the spectral decomposition of $f$ given in equation~\eqref{eq:output_consistency}, and that:
\begin{equation}
E_x \left[ c_{1} (X(t)) \right] = e^{-a_{1} t}c_{1} (x),\quad \text{and} \quad E_x \left[ c_{2} (X(t)) \right] = e^{-a_{2} t}c_{2} (x),
\end{equation}

which is an instance of the eigenvalue equation given in equation~\eqref{eq:koopman_consistency}.

\subsection{Extensions of the Spectral Decomposition-based network}

\textbf{The Spectral Decomposition-based network for the solution to the Poisson equation (P-SDnet).}   When analysing the steady-state properties of the process $(X(t))$, it is often useful to introduce a real-valued function $F$ on $\mathbb{N}^{N}$ that solves the Poisson equation given by~\cite{milias2014fast,durrenberger2019finite}:
\begin{equation}
\label{eq:poisson_eq}
\mathbb{A}F = U(f) - f.
\end{equation}

Solutions to the Poisson equation are unique up to an additive constant function (see ref.~\cite{durrenberger2019finite} and references therein). Choosing $F$ such that $U(F)= 0$, it can be expressed as~(see section~\ref{app:proof} of the supplementary material for a proof):
\begin{equation}
\label{eq:poisson_spectral}
F(x) = \sum_{\ell=1}^{\infty}\frac{\gamma_{\ell}(f)}{\sigma_{\ell}}\phi_{\ell}(x),
\end{equation}

which motivates introducing the approximation $\widehat{F}_{\eta}$ of $F$ defined as:
\begin{equation}
\label{eq:poisson_spectral_approx}
\widehat{F}_{\eta}(x) \coloneqq \sum_{\ell=1}^{r}\frac{\hat{\gamma}_{\ell}(f)}{\hat{\sigma}_{\ell}}\hat{\phi}_{\ell}(x).
\end{equation}

\begin{figure}[H]
\centering
\captionsetup{labelfont=bf}
\includegraphics[width=\textwidth]{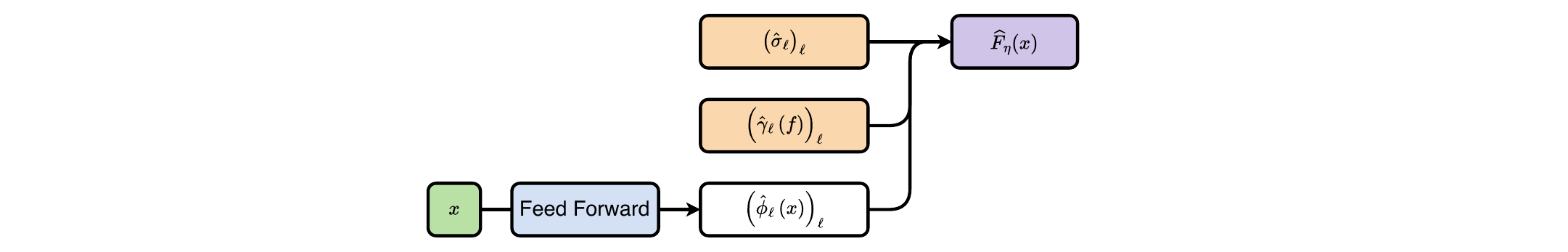}
\caption{\textbf{The Poisson Spectral Decomposition-based network (P-SDnet).} The architecture is motivated by the spectral decomposition of $F$ in equation~\eqref{eq:poisson_spectral}, which is approximated in equation~\eqref{eq:poisson_spectral_approx}. Green: inputs. Purple: output. Orange: fixed or trainable variables. Blue: feedforward neural network. White: intermediate quantities.}
\label{supp_figure:psdnet}
\end{figure}

The quantities in equation~\eqref{eq:poisson_spectral_approx} are defined analogously to those in SDnet. We refer to the architecture defined by equation~\eqref{eq:poisson_spectral_approx} as the \emph{Poisson Spectral Decomposition-based network}~(P-SDnet)~(see Supp. Fig.~\ref{supp_figure:psdnet} for a visual illustration).\newline

\textbf{Running example~(continued).} Recall from equation~\eqref{eq:pure_death_spectral_components} that for $f(x)=x^2$:
\begin{align*}
U(f)= \gamma_{0}(f) = 0,\quad \begin{cases}
\gamma_{1}(f)&= 1,\\
\sigma_{1}&= \theta,\\
\phi_{1} (x)&= x,
\end{cases}
\quad \text{and} \quad
\begin{cases}
\gamma_{2}(f)&= 1,\\
\sigma_{2}&= 2\theta,\\
\phi_{2}(x)&= x(x-1).
\end{cases}
\end{align*}

It is immediate that:
\begin{equation}
\theta x\Bigg[\frac{\gamma_{1}(f)}{\sigma_{1}}\big(\phi_{1}(x-1)- \phi_{1}(x)\big)+\frac{\gamma_{2}(f)}{\sigma_{2}}\big(\phi_{2}(x-1)- \phi_{2}(x)\big)\Bigg]=- x^2,
\end{equation}

which is an instance of equation~\eqref{eq:poisson_eq} where $F$ has been replaced by its expression given in equation~\eqref{eq:poisson_spectral}.

\section{Training of neural approximations of expected outputs for any time and initial state}
\label{supp_section:training}

In this section, we generalise the training procedure from ref.~\cite{gupta2021deepcme} into a strategy to learn parametric approximations of expected outputs for any time and initial state, \emph{i.e.} to learn them as functions: 
\begin{align}
U : \mathbb{N}^N \times \mathbb{R}_+  & \longrightarrow \mathbb{R} \\
(x, t) & \longmapsto U(x, f, t).
\end{align}

The approach can be used to train any parametrised approximation of $U$, including, for example, a parametrised Padé approximation~\cite{gupta2022pade,marano1991multipoint,gupta2025sparse}. In this work, we focus exclusively on~(deep) neural networks, specifically employing the Spectral Decomposition-based network~(SDnet)~(see section~\ref{supp_section:sdnet} of the supplementary material).

\subsection{Training method}

The training method outlined below is presented for scalar-valued output functions $f$ and extends naturally to the vector-valued case.\newline

\textbf{Almost sure relationship for the training method.} Let us introduce a vector-valued compensated Poisson  process $(\tilde{R}(t))_{t \in \mathbb{R}_{+}}$, which we define as $\tilde{R}(t)\coloneqq(\tilde{R}_k(t))_{k \in [\![1, M]\!]}$\label{first:tilder} with:
\begin{equation}
\label{eq:compensated_poisson}
\tilde{R}_k(t) \coloneqq R_k(t) - \int_{0}^{t} \lambda_{k}(X(s))ds.
\end{equation}

We also introduce $\Delta U$ as:
\begin{equation}
\Delta U(x, f, t) \coloneqq (\Delta_{\zeta_1} U(x, f, t), \dots, \Delta_{\zeta_M} U(x,f, t)) = (U(x+\zeta_1, f, t) -  U(x, f, t), \dots, U(x+\zeta_M,f, t) - U(x,f, t)).
\end{equation}

Let $\Gamma \coloneqq \{t_0,\dots, t_{J}\}$ be a time grid with $t_0 = 0$ and $t_J = T$. Select two points $t_\alpha$ and $t_\beta$ on the grid $\Gamma$, independently of $(X(t))$, such that $t_\alpha < t_\beta$. The starting point of the training methodology is the following almost sure relationship satisfied by $U$~(see section~\ref{app:proof} of the supplementary material for a proof):
\begin{equation}
\label{eq:fc_int_mean}
f(X(t_\beta)) = U(X(t_{\alpha}), f, t_{\beta} - t_{\alpha}) + \int_{t_{\alpha}}^{t_{\beta}}\sum_{k=1}^{M} \Delta_{\zeta_k} U(X(s), f, t_{\beta}-s)d\tilde{R}_k(s).
\end{equation}

The integration against $(\tilde{R}_k(t))$ can be understood both from the perspective of stochastic calculus and Riemann–Stieltjes integration.\newline

\textbf{Running example~(continued).} Consider our running example with $\theta=1$. From equation~\eqref{eq:pure_death_first_moment}, we know that for $f(x)=x$:
\begin{equation}
U(x, f, t_{\beta})=  x \exp(-t_{\beta}),\quad \text{and} \quad \Delta_{\zeta} U(x, f, t_{\beta}-t) = U(x-1, f, t_{\beta}-t) -  U(x, f, t_{\beta}-t) = -\exp(-(t_{\beta}-t)). 
\end{equation}

Assume that the initial state is $10$ and that the reaction occurs twice over the time interval $[0, 1]$, the first time at time $0.2$ and the second one at time $0.7$. Pick $t_{\alpha}= 0.15$ and $t_{\beta}=0.4$. In that case, $X(t_{\beta}) = 9$, and the right-hand side of equation~\eqref{eq:fc_int_mean} can be rewritten as:
\begin{align}
10 e^{-0.25} &- \int_{0.15}^{0.4}e^{-(0.4-s)}dR(s) + \int_{0.15}^{0.4}e^{-(0.4-s)}X(s)ds\nonumber\\   
&= 10 e^{-0.25} -e^{-0.2} + 10 \int_{0.15}^{0.2}e^{-(0.4-s)}ds + 9 \int_{0.2}^{0.4}e^{-(0.4-s)}ds\nonumber\\
&= 10 e^{-0.25} -e^{-0.2} + 10(e^{-0.2}-e^{-0.25}) + 9(1-e^{-0.2})\nonumber\\
&=9, \text{ as expected.}
\end{align}

\textbf{Penalty for the training method.} Choose any time point~$t_{\bar{\alpha}}$ on the grid such that $t_{\alpha} \leq t_{\bar{\alpha}} < t_{\beta}$. Over the interval $[t_{\bar{\alpha}}, t_{\beta}]$, a relationship analogous to equation~\eqref{eq:fc_int_mean} holds, with $t_{\alpha}$ simply replaced by $t_{\bar{\alpha}}$.\newline

Define $\widehat{U}_{\eta} : \mathbb{N}^N \times \mathbb{R}_+ \xrightarrow[]{} \mathbb{C}$ as an approximation of $U$, parametrised by $\eta$. Let $\mathcal{Y}_{\bar{\alpha}, \beta}^{x}$ denote a trajectory of the process $(X^{x}(t), \tilde{R}(t))$ between times $t_{\bar{\alpha}}$ and $t_\beta$ when $(X^{x}(t))$ starts from state $x$ at time $t=0$, and define $\mathcal{F}_{\bar{\alpha}, \beta}$ as:
\begin{equation}
\label{eq:predictor_fc_int_mean}
\mathcal{F}_{\bar{\alpha}, \beta}(\mathcal{Y}_{\bar{\alpha}, \beta}^{x}, \eta) = \widehat{U}_{\eta}(X^{x}(t_{\bar{\alpha}}), f, t_{\beta} - t_{\bar{\alpha}}) + \int_{t_{\bar{\alpha}}}^{t_{\beta}}\sum_{k=1}^{M} \Delta_{\zeta_k} \widehat{U}_{\eta}(X^{x}(s), f, t_{\beta}-s)d\tilde{R}_k(s).
\end{equation}

\begin{figure}[H]
\centering
\captionsetup{labelfont=bf}
\includegraphics[width=\textwidth]{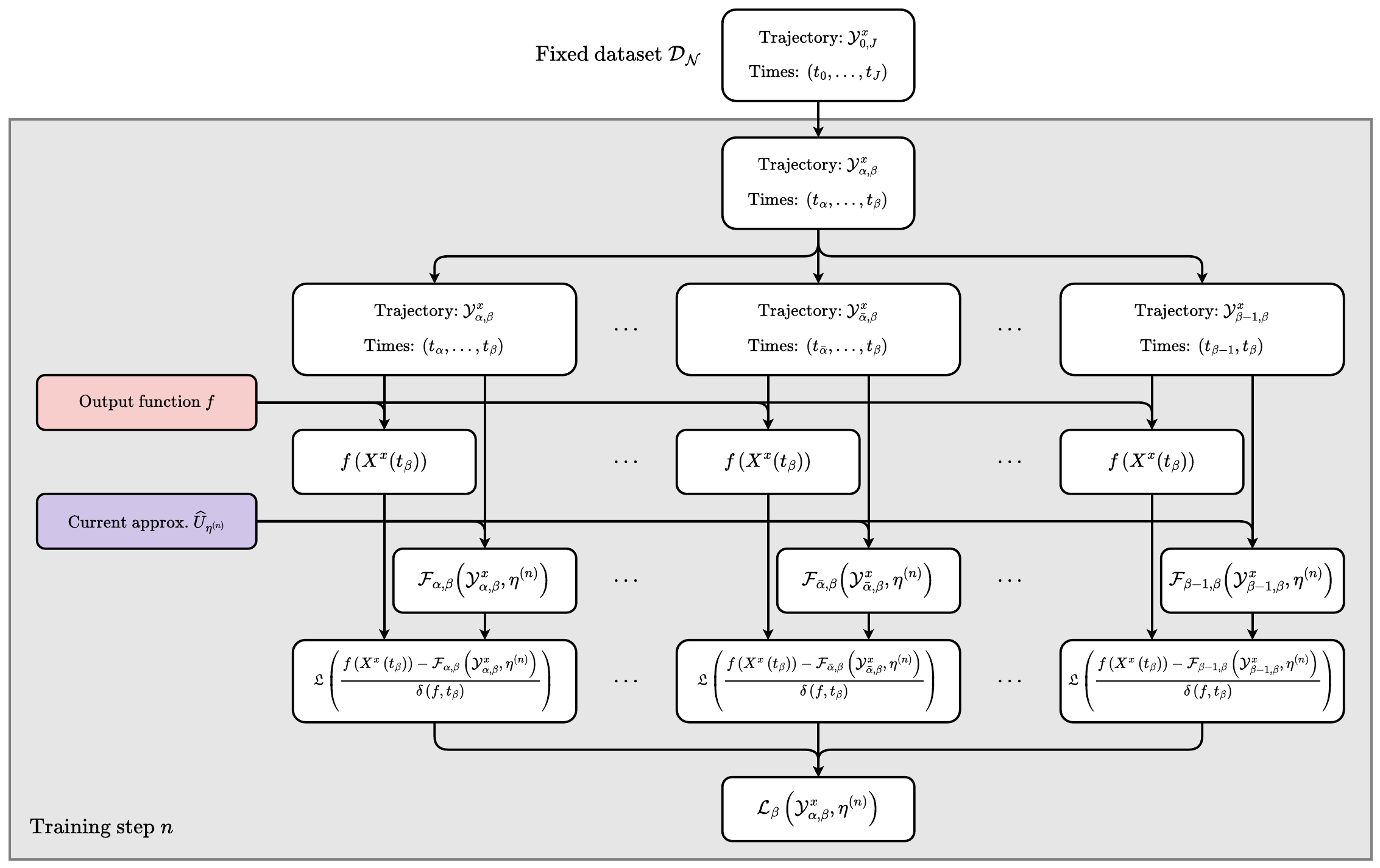}
\caption{\textbf{Computation of the penalty $\mathcal{L}_{\beta}$ in the training method.} The penalty is motivated by the almost sure relationship in equation~\eqref{eq:fc_int_mean}. The trajectory $\mathcal{Y}_{\alpha, \beta}^{x}$  between time $t_{\alpha}$ and $t_{\beta}$ is created from the pool of trajectories $\mathcal{Y}_{0,J}^{x}$  between time $t_{0}$ and $t_{J}$ in the fixed dataset $\mathcal{D}_{\mathcal{N}}$. At the training step $n$, the parameter $\eta^{(n)}$ is used to compute $\mathcal{L}_{\beta}$.}
\label{supp_figure:penalty_computation}
\end{figure}

Motivated by equation~\eqref{eq:fc_int_mean}, the methodology aims to find $\eta$ such that:
\begin{equation}
\label{eq:training_motivation}
f(X^{x}(t_{\beta})) = \mathcal{F}_{\bar{\alpha}, \beta}(\mathcal{Y}_{\bar{\alpha}, \beta}^{x}, \eta)
\end{equation}

for every possible simulated trajectory, or at least so that the values of the left- and right-hand sides of equation~\eqref{eq:training_motivation} are close. To formalise the requirement of closeness, we introduce the penalty $\mathcal{L}_{\beta}$ as follows~(see Supp. Fig.~\ref{supp_figure:penalty_computation} for a visual illustration):
\begin{equation}
\label{eq:training_penalty}
\mathcal{L}_{\beta}(\mathcal{Y}_{\alpha, \beta}^{x}, \eta) \coloneqq \sum_{\bar{\alpha}=\alpha}^{\beta-1} \mathfrak{L}\bigg(\frac{ f(X^{x}(t_{\beta})) - \mathcal{F}_{\bar{\alpha}, \beta}(\mathcal{Y}_{\bar{\alpha}, \beta}^{x}, \eta)}{\delta(f,t_{\beta})}\bigg),
\end{equation}

where we leave the scaling factor $\delta(f,t_{\beta})\in\mathbb{R}$ unspecified for now, and where:
\begin{equation}
\mathfrak{L}(x) = \begin{cases} 
x^{2} & \text{if }  |x| <1 \\
2|x|-1 & \text{otherwise}. 
\end{cases}
\end{equation}

Note that $\mathcal{L}_{\beta}$ depends on $\mathcal{F}_{\bar{\alpha}, \beta}$ for $\bar{\alpha} \neq \alpha$. This construction ensures that the pathwise relationship in equation~\eqref{eq:fc_int_mean} is satisfied by $\widehat{U}_{\eta}$ over the time interval $[t_{\alpha}, t_\beta]$, without allowing errors to cancel each other out in the summation of equation~\eqref{eq:predictor_fc_int_mean}. The penalty could be regularised to encourage sparsity in the weights of the neural network approximating $U$. Alternative functional forms of $\mathfrak{L}$ may also be employed.\newline

\textbf{Computational considerations.} Although the integration against $(\tilde{R}_k(t))$ can in principle be performed exactly on a path-by-path basis in equation~\eqref{eq:predictor_fc_int_mean}, it is instead carried out on the grid $\Gamma$ to improve numerical efficiency. Accordingly, $\mathcal{F}_{\bar{\alpha}, \beta}(\mathcal{Y}_{\bar{\alpha}, \beta}^{x}, \eta)$ is computed as:
\begin{equation}
\label{eq:predictor_fc_int_mean_approx}
\widehat{U}_{\eta}(X^{x}(t_{\bar{\alpha}}), f, t_{\beta} - t_{\bar{\alpha}}) + \sum_{q=0}^{\beta - \bar{\alpha} -1}\sum_{k=1}^{M} \Delta_{\zeta_k}  \widehat{U}_{\eta}(X^{x}(t_{\bar{\alpha}+q}), f, t_{\beta} - t_{\bar{\alpha} + q})(\tilde{R}_k(t_{\bar{\alpha} + q+1})-\tilde{R}_k(t_{\bar{\alpha} + q})).
\end{equation}

In addition, this formulation permits storing $\mathcal{Y}_{\bar{\alpha}, \beta}^{x}$ as the values $ \{(X^{x}(t_{\bar{\alpha}}), \tilde{R}(t_{\bar{\alpha}})), \dots, (X^{x}(t_{\beta}), \tilde{R}(t_{\beta}))\}$ of the process $(X^{x}(t), \tilde{R}(t))$ sampled on $\Gamma$ between times $t_{\bar{\alpha}}$ and $t_\beta$.  A natural question is how fine the grid $\Gamma$ should be to ensure that the increments $\tilde{R}_k(t_{\bar{\alpha} + q+1})-\tilde{R}_k(t_{\bar{\alpha} + q})$ provide a sufficiently accurate approximation of the integral with respect to $(\tilde{R}_k(t))$. In most practical scenarios, $(\tilde{R}_k(t))$ is a martingale with zero mean, so the sum of these increments is expected to be close to zero. This sum can be computed alongside  equation~\eqref{eq:predictor_fc_int_mean_approx}, providing a useful diagnostic for assessing whether the resolution of the grid should be increased.\newline

When using $\widehat{U}_{\eta}$ to approximate $\Delta U_{\eta}$ in the computation of $\mathcal{F}_{\bar{\alpha}, \beta}$, $M+1$ evaluations are required at each time point selected on the grid $\Gamma$. To address this, we use two separate approximations in computing $\mathcal{F}_{\bar{\alpha}, \beta}$ from equation~\eqref{eq:predictor_fc_int_mean_approx}: $\widehat{U}_{\eta_1}$, parametrised by $\eta_1$, for approximating $U$, and  $\widehat{\Delta} U_{\eta_2}$, parametrised by $\eta_2$, for approximating $\Delta U$. When SDnet is employed, both $\widehat{U}_{\eta_1}$ and $\widehat{\Delta} U_{\eta_2}$ share the approximations $\hat{\sigma}_{\ell}$ of the decay modes and $\hat{\gamma}_{\ell}(f)$ of the function coordinates. For clarity of presentation, we will continue to write $\Delta \widehat{U}_{\eta}$ even when $\widehat{\Delta} U_{\eta_2}$ is used to approximate $\Delta U$.

\textbf{Training method.} Let $\mathcal{C} \subset \mathbb{N}^{N}$ be a hypercube. We seek $\eta^{*}$ as the solution to:
\begin{equation}
\label{eq:training_objective}
\eta^{*} =  \argmin_{\eta} \Upsilon(\eta) \coloneqq \argmin_{\eta} \mathbb{E}_{\substack{x \sim \mathcal{U}(\mathcal{C}), \omega \sim \mathcal{U}(1, J)\\ \alpha \sim \mathcal{U}(0, J-\omega)}}\big[\mathcal{L}_{\beta}(\mathcal{Y}_{\alpha, \alpha + \omega}^{x}, \eta)\big],
\end{equation}

and $\widehat{U}_{\eta^{*}}$ gives an approximation of $U$ for every initial state $x$ and time $t$.\newline

We obtain $\eta^{*}$ by performing the stochastic gradient descent algorithm Adam using trajectories stored in a fixed training dataset $\mathcal{D}_{\mathcal{N}} \coloneqq \{\mathcal{Y}_i\}_{i \in [\![1, \mathcal{N}]\!]}$, where $\mathcal{Y}_{i}$ is a realisation of the trajectory $\mathcal{Y}_{0,J}^{x}$ and $\mathcal{N}$ corresponds to dataset size. This dataset is generated by simulating the process $(X(t))$ over the time interval $[0,T]$ using a standard Stochastic Simulation Algorithm~(SSA) for Stochastic Reaction Networks~(SRNs)~\cite{gillespie1976general,gillespie1977exact,anderson2007modified}. The modified next reaction method, based on the RTC from equations~\eqref{eq:reaction_count_process}–\eqref{eq:rtc}, keeps track of the integral in equation~\eqref{eq:compensated_poisson}~\cite{anderson2007modified}. This makes the algorithm efficient at jointly generating trajectories of $X(t)$ and $\tilde{R}(t)$.\newline

As featured in equation~\eqref{eq:training_objective}, randomising the initial state helps promote greater diversity in the states visited by the process $(X(t))$ over the interval $[0, T]$. In particular, this mechanism guarantees that the states within the hypercube are visited at least at time $t=0$.\newline

We also exploit the time homogeneity of the Markov process  by randomly selecting segments of length $T_{\alpha, \beta}$ at each step $n$ from the trajectories generated over the interval $[0, T]$, following equation~\eqref{eq:fc_int_mean}. This allows us to enforce a greater number of almost sure relationships without incurring the cost of generating new trajectories. In practice, we randomly choose a number $\omega$ of intervals to consider on the grid $\Gamma$, along with a starting time point $t_{\alpha}$~(which then determines the terminal time $t_\beta \coloneqq t_{\alpha + \omega}$), such that both are compatible with $\Gamma$. We then randomly select $\mathcal{N}_{\omega}$ trajectories from $\mathcal{D}_{\mathcal{N}}$ on which to evaluate equation~\eqref{eq:training_penalty}, ensuring that the product $\omega\times\mathcal{N}_{\omega}$ remains constant throughout training. This two-step procedure is repeated multiple times at each step $n$, with gradients accumulated before performing a gradient update to stabilise the training process. If desired, $t_\alpha$ and $t_\beta$ could of course still be chosen~(deterministically) as $0$ and $T$.\newline

From the $\mathcal{N}_{\omega}$ selected trajectories spanning $t_{\alpha}$ to $t_{\beta}$, we can compute the sample mean $\widehat{\mathcal{M}}(f,t_{\beta},\mathcal{N}_{\omega})$ and standard deviation $\widehat{\mathcal{S}}(f,t_{\beta},\mathcal{N}_{\omega})$. These quantities are then used to define the scaling factor $\delta(f,t_{\beta})$ as $\delta(f,t_{\beta}) \coloneqq 1 + |\widehat{\mathcal{M}}(f,t_{\beta},\mathcal{N}_{\omega})| + 2\widehat{\mathcal{S}}(f,t_{\beta},\mathcal{N}_{\omega})$.  Given the number of trajectories $\mathcal{N}_{\omega}$ used, these estimates are typically rough but sufficiently accurate for scaling purposes. Alternative choices for the scaling factor include, for example, the Ergodic Mean~(EM).\newline

\textbf{Interpretation of the training method as a reinforcement learning approach.} As observed in ref.~\cite{gupta2021deepcme}, the objective defined in equation~\eqref{eq:training_objective} belongs to the domain of continuous-time Reinforcement Learning~(RL) and can be viewed as a stochastic target reaching task. To illustrate this, consider an agent whose state over the time interval $[t_{\bar{\alpha}}, t_{\beta}]$ is represented by $\mathcal{X}(t) \in \mathbb{R}$, and whose dynamics are given by:
\begin{equation}
\label{eq:rl_connection}
\mathcal{X}(t) \coloneqq \widehat{U}_{\eta}(X^x(t_{\bar{\alpha}}), f, t_{\beta} - t_{\bar{\alpha}}) + \int_{t_{\bar{\alpha}}}^{t}\sum_{k=1}^{M} \Delta_{\zeta_k} \widehat{U}_{\eta}(X^x(s), f, t_{\beta}-s)d\tilde{R}_k(s).
\end{equation}

\begin{figure}[H]
\centering
\captionsetup{labelfont=bf}
\includegraphics[width=\textwidth]{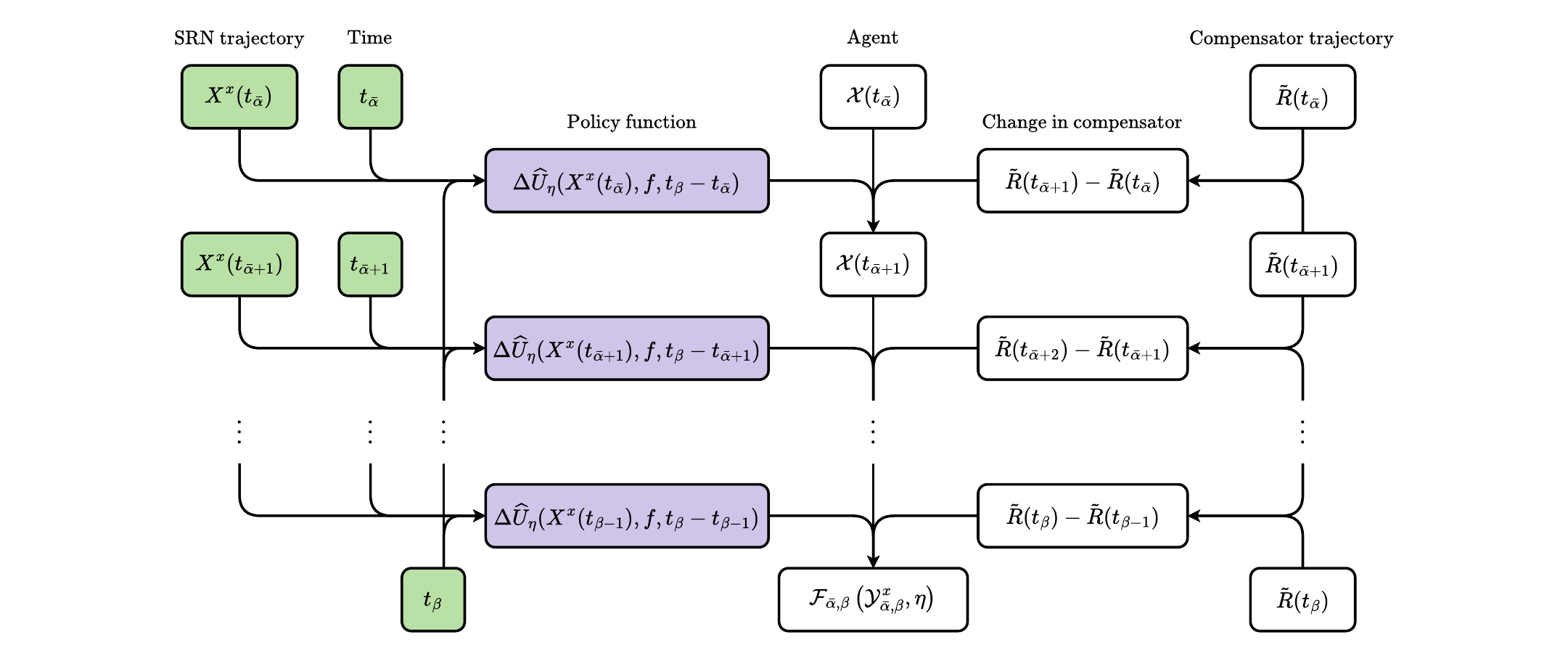}
\caption{\textbf{Interpretation of the training method as a reinforcement learning approach.} The evolution of the agent $(\mathcal{X}(t))$  is dictated by a policy function $\widehat{U}_{\eta}$ and random fluctuations $(\tilde{R}(t))$. At the end of the time interval $[t_{\bar{\alpha}}, t_{\beta}]$, the agent is in the state  $\mathcal{F}_{\bar{\alpha}, \beta}(\mathcal{Y}_{\bar{\alpha}, \beta}^{x}, \eta)$ while its target is $f(X^x(t_{\beta}))$.}
\label{supp_figure:rl_interpretation}
\end{figure}

The evolution described in equation~\eqref{eq:rl_connection} is driven by the deterministic policy function $\widehat{U}_{\eta}$ together with the random fluctuations of the process $(\tilde{R}(t))$. The policy $\widehat{U}_{\eta}$ can depend on the current state of the SRN, $X^x(t)$, as well as on the remaining time horizon $t_{\beta}-t$, with $t \in [t_{\bar{\alpha}}, t_{\beta}]$. At the terminal time $t_{\beta}$, the agent is required to reach the stochastic target $f(X^x(t_{\beta}))$.\newline

The dynamics in equation~\eqref{eq:predictor_fc_int_mean_approx} approximate those in equation~\eqref{eq:rl_connection}, with the final state $\mathcal{X}(t_{\beta})$ of the agent obtained from
$\mathcal{F}_{\bar{\alpha}, \beta}(\mathcal{Y}_{\bar{\alpha}, \beta}^{x}, \eta)$~(see Supp. Fig.~\ref{supp_figure:rl_interpretation} for a visual illustration). The penalty incurred by the agent depends on its final distance from the target $f(X^x(t_{\beta}))$, as captured by equation~\eqref{eq:training_penalty}. Unlike Supervised Learning~(SL), no explicit $((x,t), U(x,f,t))$ pairs are used.\newline

\textbf{Benefits of the training method.} The most straightforward alternative to the training method introduced above is a SL strategy, where $U$ is learned using Monte Carlo estimates of expected outputs~(see Supplementary Table~\ref{supp_table:training}). However, this approach can be very time-consuming, since the estimates must be sufficiently accurate to serve as reliable training data, while the convergence rate of Monte Carlo methods is notoriously slow~\cite{asmussen2007stochastic}. This challenge is especially pronounced here because precise estimates are required for a \emph{large} number of initial states $x$, \emph{each} of which demands its own dedicated set of simulations.\newline

In contrast, the RL training method completely eliminates the need for Monte Carlo estimates by exploiting the relationship in equation~\eqref{eq:fc_int_mean} that holds path-by-path. In the enhanced training method introduced here, randomisation of the initial state is only optional, as the trajectories truncated over the time intervals $[t_{\alpha},t_{\beta}]$ already begin from diverse starting points.

\begin{table}[H]
\centering
\captionsetup{labelfont=bf}
\begin{tabular}{c|c|c|c|}
    \multicolumn{2}{c|}{} & \multicolumn{2}{c|}{Requires estimates $\dots$} 
    \\  \cline{2-4}
    & Requires estimates of $U(x,f,t)$  & $\dots$ which are precise &  $\dots$ for various initial states $x$\\ \hline
    \multicolumn{1}{l|}{RL method}  &{\textcolor{red}{\ding{55}}}
    & {\textcolor{red}{\ding{55}}} & {\textcolor{red}{\ding{55}}}\\ \hline
    \multicolumn{1}{l|}{SL method} & {\textcolor{ForestGreen}{\ding{51}}}
    & {\textcolor{ForestGreen}{\ding{51}}}  & {\textcolor{ForestGreen}{\ding{51}}} \\ \hline
\end{tabular}
\caption{\textbf{Main characteristics of the reinforcement learning training method.} The Reinforcement Learning~(RL) method does not require any estimate of $U(x,f,t)$. In contrast, a Supervised Learning~(SL) method depends on precise estimates of $U(x,f,t)$ for a range of initial states $x$.}
\label{supp_table:training}
\end{table}

\textbf{Alternative sampling strategies.} Trajectories of $(X(t))$ and $(\tilde{R}(t))$ for the dataset $\mathcal{D}_{\mathcal{N}}$ could equivalently be sampled  under an alternative probability distribution obtained by modifying the propensities while keeping the stoichiometry vectors unchanged, provided that the new distribution is absolutely continuous with respect to the original one. This can be achieved, for example, using SSA with Deep Importance Sampling~(SSA\label{first:is} with DeepIS)~(see section~\ref{supp_section:dlmc} of the supplementary material). Under the assumption of absolute continuity, the almost sure relationship~\eqref{eq:fc_int_mean} remains valid under the alternative distribution with the same $U$.\newline

\textbf{Alternative almost sure relationships.} The almost sure relationship~\eqref{eq:fc_int_mean} can be substituted with alternative equations for training, such as~(see equation~\eqref{eq:intermediate_pinn} and above in section~\ref{app:proof} of the supplementary material for a proof):
\begin{equation}
\label{eq:fc_pde_mean}
\begin{dcases}
\int_{t_{\alpha}}^{t_{\beta}}\frac{\partial U}{\partial t}(X(s), f, t_{\beta} - s)ds &= \int_{t_{\alpha}}^{t_{\beta}} \sum_{k=1}^{M} \lambda_k(X(s))\Delta_{\zeta_k}U(X(s), f, t_{\beta}-s)ds,\\
U(x,f, 0)  &=  f(x).
\end{dcases}
\end{equation}

A training strategy based on this equation constitutes a stochastic analogue to conventional Deep Galerkin Methods~(DGMs) and Physics-Informed Neural Networks~(PINNs), wherein neural networks are trained to approximate solutions of differential equations by enforcing that their time derivative matches the right-hand side of the differential equation of interest~\cite{sirignano2018dgm,raissi2019physics,lu2021deepxde}. Observe that for SDnet, an analytical expression for the time derivative in equation~\eqref{eq:fc_pde_mean} follows directly from equation~\eqref{eq:complex_decomposition_expanded_approx} and can be computed at the same cost as evaluating $\widehat{U}_{\eta}$ itself, by leveraging the fact that:
\begin{equation}
\frac{\partial \widehat{U}_{\eta}}{\partial t}(x,f,t) = -\sum_{\ell=1}^{r} e^{-\hat{\sigma}_{\ell}t}\hat{\sigma}_{\ell}\hat{\gamma}_{\ell}(f)\hat{\phi}_{\ell}(x).
\end{equation}

A deterministic counterpart to equation~\eqref{eq:fc_pde_mean}, closer in spirit to the standard DGM/PINN framework, can be derived from the Kolmogorov backward equation~\eqref{eq:kb_intro}. The corresponding strategy is simulation-free but requires truncation of the state space.\newline

When the output function $f$ is positive, it holds similarly to equations~\eqref{eq:fc_int_mean} and \eqref{eq:fc_pde_mean} that~(see section~\ref{app:proof} of the supplementary material for a proof):
\begin{align}
\label{eq:fc_int_log_mean}
\begin{split}
\log(f(X(t_{\beta}))) &=\log(U(X(t_{\alpha}), f, t_{\beta} - t_{\alpha})) +  \sum_{k=1}^{M}\int_{t_{\alpha}}^{t_{\beta}}  \lambda_k(X(s))\left(1 - \Pi_{\zeta_k}U(X(s),f,t_{\beta}-s)\right)ds\\
&+\sum_{k=1}^{M}  \int_{t_{\alpha}}^{t_{\beta}}\Delta_{\zeta_k}\log\left(U(X(s), f, t_{\beta}-s)\right) dR_k(s),  
\end{split}
\end{align}

where $\Pi_{\zeta_k}f(x) \coloneqq f(x+\zeta_k) / f(x)$. From this equation, it is possible to learn, for example, $U$,  $\Pi_{\zeta_k}U$, $\log U$, and  $\Delta_{\zeta_k}\log U$. Importantly, the output activation layers of a neural network approximating $U$ in equation~\eqref{eq:fc_int_log_mean} must enforce positivity. To handle cases where $f$ is non-negative, it suffices to apply equation~\eqref{eq:fc_int_log_mean} with $\tilde{f} = f + \epsilon$, where $\epsilon \in \mathbb{R}_{+}^{*}$ is a small positive scalar. This amounts to approximating $E_{x}[f(X(t))]+\epsilon$ instead of $E_{x}[f(X(t))]$. Approximations of $\Pi_{\zeta_k}U$ and $\Delta_{\zeta_k}\log U$ can be particularly useful for SSA with DeepIS (see section~\ref{supp_section:dlmc} of the supplementary material), since they avoid numerical instabilities that may otherwise occur when evaluating $\widehat{U}_{\eta}$ in the denominator.\newline

Under the same positivity assumption for $f$, we define a family of reaction counting processes $(R_{k}^{\text{IS}^{*}}(f,t,T))_{t \in [0,T]}$ and a state process $(X^{\text{IS}^*}(f,t,T))_{t \in [0,T]}$ over the time interval~$[0,T]$. These are defined analogously to $(R_{k}(t))$ and $(X(t))$ in equations~\eqref{eq:reaction_count_process} and~\eqref{eq:rtc} but using modified propensities $\lambda_{k}^{\text{IS}^{*}}$ given by:
\begin{equation}
\lambda_k^{\text{IS}^{*}}(x, f, t) \coloneqq \Pi_{\zeta_k}U(x, f, t) \lambda_k(x).
\end{equation}

As detailed in section~\ref{supp_section:dlmc} of the supplementary material, the following almost sure relationship holds over any interval~$[t_{\alpha},t_{\beta}] \subset [0,T]$ (see equation~\eqref{eq:is_unbiased}): 
\begin{equation}
\label{eq:is_for_training}
\begin{aligned}
 U(X^{\text{IS}^{*}}(f, t_{\alpha}, t_{\beta}), f, t_{\beta} - t_{\alpha}) = f(X^{\text{IS}^{*}}(f,t_{\beta},t_{\beta}))  \exp&\Bigg(
    \sum_{k=1}^{M}
    \int_{t_{\alpha}}^{t_{\beta}} \log \Bigg(\frac{\lambda_k(X^{\text{IS}^{*}}(f,s,t_{\beta}))}{\lambda_k^{\text{IS}^{*}}(X^{\text{IS}^{*}}(f,s,t_{\beta}), f, t_{\beta}-s)} \Bigg)dR_{k}^{\text{IS}^{*}}(f,s,t_{\beta})\\ &- \sum_{k=1}^{M} \int_{t_{\alpha}}^{t_{\beta}}\big(\lambda_k(X^{\text{IS}^{*}}(f,s,t_{\beta}))
   - \lambda_k^{\text{IS}^{*}}(X^{\text{IS}^{*}}(f,s,t_{\beta}), f, t_{\beta}-s)\big) ds \Bigg).
\end{aligned}
\end{equation}

Here too, the formulation enables the learning of $U$,  $\Pi_{\zeta_k}U$, $\log U$, and $\Delta_{\zeta_k}\log U$.

\subsection{Extensions of the training method}

\textbf{Training method for the solution to the Poisson equation.} 
Training SDnet using the RL method introduced above yields approximations of the function coordinates $\gamma_{\ell}$, decay modes $\sigma_{\ell}$, and eigenfunctions $\phi_{\ell}$, which can then be used to compute other quantities beyond $U$~(see ref.~\cite{gupta2025sparse} for several such examples). This is particularly the case for the solution to the Poisson equation when approximated with P-SDnet~(see section~\ref{supp_section:sdnet} of the supplementary material).\newline

\textbf{Training method for the variance.} The almost sure relationships presented so far have been introduced for training approximations of:
\begin{equation*}
U(x, f, t) \coloneqq \mathbb{E}[f(X(t)) | X(0)=x].
\end{equation*}

It is also common to be interested in computing the expectation of integral output functions, specifically to evaluate $W$ defined as:
\begin{equation}
\label{eq:def_int_out}
W(x, f, t) = \mathbb{E}\Bigg[\int_{0}^{t}f(X(s),s)ds\Big| X(0) = x\Bigg].
\end{equation}

We consider one quantity of this form, the variance $\mathcal{W}$, which is defined as:
\begin{equation}
\label{eq:variance_def}
\mathcal{W}(x,f, t) \coloneqq \text{Var}_{x}\Big(f(X(t))\Big) \coloneqq \mathbb{E}_{x}\Bigg[\Big(f(X(t)) - U(x,f,t)\Big)^2\Bigg].
\end{equation}

It can be shown that the variance admits the following integral representation~(see section~\ref{app:proof} of the supplementary material for a proof):
\begin{equation}
\label{eq:variance_int_rep}
\mathcal{W}(x,f, t) = \mathbb{E}_x\Bigg[\sum_{k=1}^{M}\int_{0}^{t} \lambda_{k}(X(s))\Big(\Delta_{\zeta_k}U(X(s), f, t-s)\Big)^2ds\Bigg],
\end{equation}

which is an instance of equation~\eqref{eq:def_int_out}. As with equation~\eqref{eq:fc_int_mean}, we have that~(see section~\ref{app:proof} of the supplementary material for a proof):
\begin{equation}
\label{eq:fc_int_var}
\begin{split}
 \sum_{k=1}^{M} \int_{t_{\alpha}}^{t_{\beta}} \lambda_{k}(X(s))\Big(\Delta_{\zeta_k}U(X(s), f, t_{\beta}-s)\Big)^2ds &= \mathcal{W}(X(t_{\alpha}), f, t_{\beta} - t_{\alpha}) \\
 &+ \sum_{k=1}^{M} \int_{t_{\alpha}}^{t_{\beta}}  \Delta_{\zeta_k} \mathcal{W}(X(s), f, t_{\beta}-s)d\tilde{R}_k(s).
 \end{split}
\end{equation}

\section{Deep Learning/Monte Carlo estimators of expected outputs with guaranteed reliability}
\label{supp_section:dlmc}

In this section, we introduce two \emph{Deep Learning/Monte Carlo~\emph{(DLMC)}~\label{first:dlmc} estimators }of expected outputs $U(x,f,t)$. These hybrid methods retain the same reliability guarantees as the crude Monte Carlo estimator, which we refer to as the \emph{Stochastic Simulation Algorithm~\emph{(SSA)} estimator}. They leverage a neural network approximation of $U$ to achieve variance reduction~(see ref.~\cite{asmussen2007stochastic} for a general overview of variance-reduction techniques, and refs.~\cite{anderson2012multilevel,anderson2022conditional} for examples tailored to estimating $U$ for Stochastic Reaction Networks~(SRNs)). Additionally, we propose uncertainty metrics for the neural approximation of $U$. To stay consistent with the previous sections, we continue to denote the approximation of $U$ used in the DLMC methods by $\widehat{U}_{\eta}$.\newline

The variance-reduction schemes can also be applied with neural architectures other than the Spectral Decomposition-based network~(SDnet), or even alternative types of approximations. For instance, $U$ could be obtained via a moment closure approach~(see, for example, the mass fluctuation kinetics method in ref.~\cite{gomez2007mass}, the separable derivative matching method in ref.~\cite{singh2010approximate}, and an overview in ref.~\cite{sukys2022momentclosure}). That said, once trained, neural networks offer the advantage of fast evaluation for any initial state $x$ and time $t$, whereas moment closure methods require solving an Ordinary Differential Equation~(ODE) for each $(x,t)$ pair.

\subsection{The SSA with Deep Importance Sampling and SSA with Deep Control Variates estimators}

In what follows, the notations are expanded relative to the main text to facilitate the subsequent ``Comparison between the Deep Learning/Monte Carlo estimators''.\newline

\textbf{The SSA with Deep Importance Sampling estimator (SSA with DeepIS).} When the output function $f$ is positive, we can define a family of reaction counting processes as follows:
\begin{equation}
\label{eq:reaction_counting_is}
R_{k}^{\text{IS}}(f, t,T) \coloneqq Y_{k}\bigg(\int_0^t \lambda_k^{\text{IS}}(X(s), f, T-s) ds\bigg) \quad \forall t \in [0,T],
\end{equation}

where the processes $(Y_k(t))$ are independent unit-rate Poisson processes as earlier, and:
\begin{equation}
\label{eq:prop_is}
\lambda_k^{\text{IS}}(x, f, t) \coloneqq \Pi_{\zeta_k}\widehat{U}_{\eta}(x, f, t) \lambda_k(x) \quad \forall t \in [0,T],
\end{equation}

with $\Pi_{\zeta_k}f(x) \coloneqq f(x+\zeta_k) / f(x)$. Whenever $x+\zeta_k \succcurlyeq 0$, note that $\widehat{U}_{\eta}(x,f,t)$ must be non-zero for $\lambda_k^{\text{IS}}$ to be well-defined in equation~\eqref{eq:prop_is}, and that $\Pi_{\zeta_k}\widehat{U}_{\eta}(x, f, t)$ must be non-negative for $\lambda_k^{\text{IS}}$ to represent a valid propensity. If any element of  $x+\zeta_k$ is negative, then $\lambda_k(x) =0$, which suffices to ensure that $\lambda_k^{\text{IS}}$ is valid.\newline

We use the reaction counting processes $(R_{k}^{\text{IS}}(f, t,T))_{t \in [0,T]}$ to define the process  $(X^{\text{IS}}_{\eta}(f,t,T))_{t \in [0,T]}$ as:
\begin{equation}
X_{\eta}^{\text{IS}}(f,t,T) \coloneqq x + \sum_{k=1}^{M} \zeta_k R^{\text{IS}}_{k}(f, t,T),
\end{equation}

as well as the likelihood ratio $Z_{\eta}^{\text{IS}}(f, t,T)$ over the time interval $[0,T]$:
\begin{equation}
\label{eq:lr_is}
\begin{aligned}
Z_{\eta}^{\text{IS}}(f, t,T) \coloneqq \exp\Bigg(
    \sum_{k=1}^{M}
    \int_{0}^{t} &\log \Bigg(\frac{\lambda_k(X_{\eta}^{\text{IS}}(f,s,T))}{\lambda_k^{\text{IS}}(X_{\eta}^{\text{IS}}(f,s,T), f, T-s)} \Bigg)dR_{k}^{\text{IS}}(f,s,T)\\ &- \sum_{k=1}^{M} \int_{0}^{t}\big(\lambda_k(X_{\eta}^{\text{IS}}(f,s,T))
   - \lambda_k^{\text{IS}}(X_{\eta}^{\text{IS}}(f,s,T), f, T-s)\big) ds \Bigg).
\end{aligned}
\end{equation}

\begin{figure}[H]
\captionsetup{labelfont=bf}
\centering
\includegraphics[width=\textwidth]{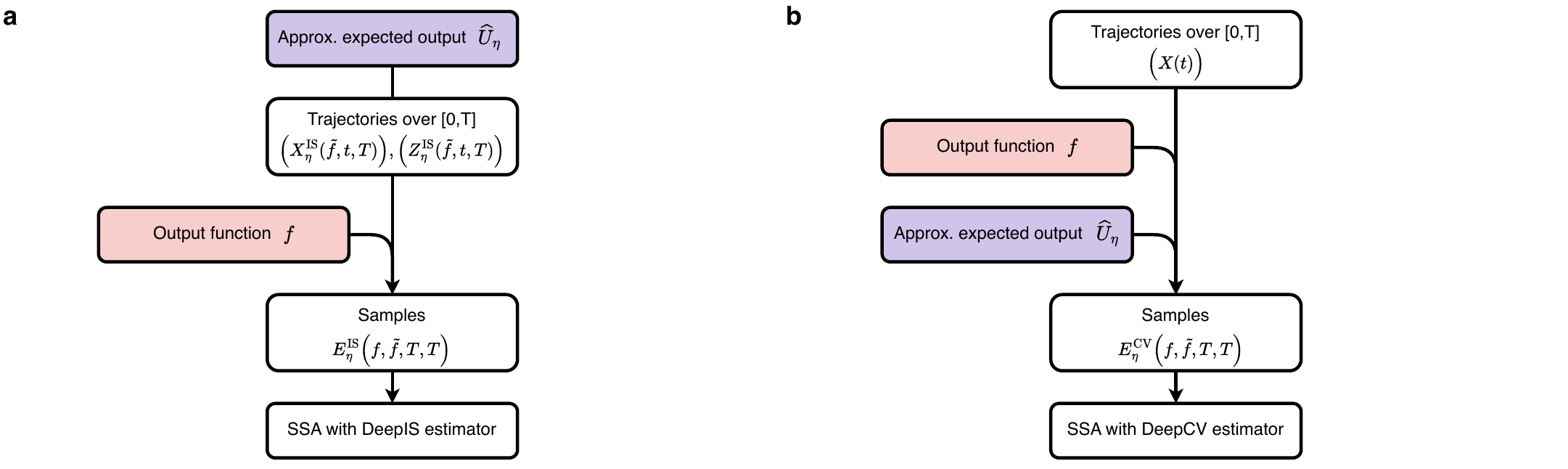}
\caption{\textbf{Deep Learning/Monte Carlo estimators of expected outputs.} \textbf{a,} The SSA with Deep Importance Sampling~(SSA with DeepIS) estimator. The approximate expected output $\widehat{U}_{\eta}$ is used to compute the propensities $\lambda_k^{\text{IS}}$ required for simulating the processes $(X^{\text{IS}}_{\eta}(\tilde{f}, t, T))$ and $(Z_{\eta}^{\text{IS}}(\tilde{f}, t, T))$. These simulated trajectories are then used to generate samples of $E_{\eta}^{\text{IS}}(f,\tilde{f},T,T)$. \textbf{b,} The SSA with Deep Control Variates~(SSA with DeepCV) estimator. The approximate expected output $\widehat{U}_{\eta}$ is used to construct the control variate $(Z_{\eta}^{\text{CV}}(\tilde{f}, t, T))$ based on trajectories of $(X(t))$. The control variate is then used to generate samples of $E_{\eta}^{\text{CV}}(f,\tilde{f},T,T)$.}
\label{supp_figure:deep_is_cv}
\end{figure}

Define $E_{\eta}^{\text{IS}}$\label{first:eis} as:
\begin{equation}
\label{eq:def_eis}
E_{\eta}^{\text{IS}}(f,\tilde{f}, t,T) \coloneqq f(X^{\text{IS}}_{\eta}(\tilde{f}, t, T))Z_{\eta}^{\text{IS}}(\tilde{f}, t, T),    
\end{equation}

for two functions $f$ and $\tilde{f}$, and a time $t \leq T$. By the Girsanov theorem, we know that  $E_{\eta}^{\text{IS}}(f,\tilde{f},T,T)$ has the same expectation as $f(X(T))$~\cite{wilkinson2018stochastic}. We refer to the corresponding unbiased estimator as the~\emph{SSA with Deep Importance Sampling~\emph{(SSA with DeepIS)} estimator}~(see Supp. Fig.~\ref{supp_figure:deep_is_cv} for a visual illustration). If $\widehat{U}_{\eta}$ exactly matches $U$, then we have:
\begin{equation}
\label{eq:is_unbiased}
E_{\eta}^{\text{IS}}(f,f,T,T) = \mathbb{E}_x[f(X(T))] \quad \text{almost surely},
\end{equation}

which implies that $E_{\eta}^{\text{IS}}(f,f,T,T)$ has zero variance. This conclusion follows from the classical theory of Doob’s $h$-transform (see section 6 of chapters 3 and 4 in ref.~\cite{rogers2000diffusions} for general background, and section 3 of ref.~\cite{corstanje2025guided} for the corresponding statement in the case of SRNs). Although importance sampling is typically applied in rare-event estimation, this result holds without any additional assumptions on the specific form of 
$f$, aside from its positivity.

Observe in equation~\eqref{eq:reaction_counting_is} that the propensities are time-dependent, which necessitates the use of specialised SSA algorithms~(see, for example, the Extrande algorithm in ref.~\cite{voliotis2016stochastic}). In this work, we replace $\lambda_k^{\text{IS}}$ with a surrogate that is piecewise-constant in time. The resulting estimator remains unbiased but no longer has zero variance, even when the approximation of $U$ is exact. Note that, although making the surrogate piecewise-constant over shorter time intervals may bring it ``closer'' to the propensity $\lambda_k^{\text{IS}}$, it also increases the simulation time per sample. We refer to the simulation procedure for $(X^{\text{IS}}_{\eta}(f, t,T))$ described above as the \emph{DeepSSA}, to reflect the incorporation of a neural network in the stochastic simulations.\newline

\textbf{The SSA with Deep Control Variates estimator (SSA with DeepCV).} Let us now introduce the process $(Z^{\text{CV}}_{\eta}(f, t, T))_{t \in [0,T]}$ as follows:
\begin{equation}
\label{eq:covariate_mean}
Z^{\text{CV}}_{\eta}(f, t, T) \coloneqq \sum_{k=1}^{M}\int_{0}^{t} \Delta_{\zeta_k} \widehat{U}_{\eta}(X(s), f, T-s)d\tilde{R}_k(s).
\end{equation}

Note that if any component of $X(s)+\zeta_k$ is negative, then $\lambda_k(X(s)) = 0$. This ensures, in particular, that the integrand in equation~\eqref{eq:covariate_mean} is well-defined.\newline

Define $E_{\eta}^{\text{CV}}$\label{first:ecv} as:
\begin{equation}
E_{\eta}^{\text{CV}}(f,\tilde{f},t,T) \coloneqq f(X( t)) - \rho Z_{\eta}^{\text{CV}}(\tilde{f}, t, T),
\end{equation}

for two functions $f$ and $\tilde{f}$, a time $t \leq T$, and a coefficient $\rho \in \mathbb{R}$. The process $(Z^{\text{CV}}_{\eta}(\tilde{f}, t, T))$ is correlated with the process $(X(t))$ and, in most practical scenarios, is a martingale with zero mean. Consequently, $E_{\eta}^{\text{CV}}(f,\tilde{f},T,T)$ shares the same expectation as $f(X(T))$, and $Z^{\text{CV}}_{\eta}(\tilde{f},T,T)$ serves as a control variate. We refer to the resulting unbiased estimator as the~\emph{SSA with Deep Control Variates~\emph{(SSA with DeepCV)}\label{first:cv} estimator}~(see Supp. Fig.~\ref{supp_figure:deep_is_cv} for a visual illustration).\newline

It is well-known that the variance of $E_{\eta}^{\text{CV}}(f,f,T,T)$ is~\cite{glasserman2004monte,asmussen2007stochastic}:
\begin{equation}
\label{eq:var_cv_estimator}
\text{Var}_{x}\Big(E_{\eta}^{\text{CV}}(f,f,T,T)\Big) = \text{Var}_{x}(f(X( T))) + \rho^2  \text{Var}_{x}\Big(Z_{\eta}^{\text{CV}}(f, T, T)\Big) - 2 \rho \text{Cov}_{x}\Big(f(X(T)), \, Z_{\eta}^{\text{CV}}(f, T, T)\Big),
\end{equation}

and that the coefficient $\rho^*$ which minimises the variance of $E_{\eta}^{\text{CV}}(f,f,T,T)$ is~\cite{glasserman2004monte,asmussen2007stochastic}:
\begin{equation}
\label{eq:optimal_rho}
\rho^* = \frac{\text{Cov}_{x}\Big(f(X(T)), \, Z_{\eta}^{\text{CV}}(f, T, T)\Big)}{\text{Var}_{x}\Big(Z_{\eta}^{\text{CV}}(f, T, T)\Big)},
\end{equation}

where $\text{Cov}$ denotes the covariance.\newline

In practice, the  coefficient $\rho^*$ that is optimal when using $Z_{\eta}^{\text{CV}}(f, T, T)$ as a covariate can be estimated by substituting the numerator and denominator in equation~\eqref{eq:optimal_rho} with their empirical counterparts. This procedure ensures a reduction in the asymptotic variance of $E_{\eta}^{\text{CV}}(f,f,T,T)$, even in the unfavourable case where the correlation between $f(X(T))$ and $Z_{\eta}^{\text{CV}}(f, T, T)$ is weak. However, estimation of $\rho^*$ complicates the construction of confidence intervals by inducing dependence across samples~(see section 4 in ref.~\cite{glasserman2004monte}).\newline

Using the fact that $Z^{\text{CV}}_{\eta}(f, T, T)$ has zero mean together with the isometry property of stochastic integrals~(see chapter~3 in ref.~\cite{bremaud1981point}), we obtain that:
\begin{equation}
\label{eq:var_z}
\text{Var}_{x}\Big(Z_{\eta}^{\text{CV}}(f, T, T)\Big) 
= \mathbb{E}_{x}\Bigg[\int_{0}^{T}\sum_{k=1}^{M} \Big(\Delta_{\zeta_k} \widehat{U}_{\eta}(X(s), f, T-s\Big)^2\lambda_{k}(X(s))ds\Bigg].
\end{equation}

Using the same observations together with the almost sure relationship~\eqref{eq:fc_int_mean}, we deduce that:
\begin{equation}
\label{eq:covar_x_z}
\begin{aligned}
\text{Cov}_{x}\Big(f(X(T)), \, Z_{\eta}^{\text{CV}}(f, T, T)\Big)
= \mathbb{E}_{x}\Bigg[\int_{0}^{T}\sum_{k=1}^{M} \Big(\Delta_{\zeta_k} U(X(s), f, T-s)\Big)\Big(\Delta_{\zeta_k} \widehat{U}_{\eta}(X(s), f, T-s)\Big)\lambda_{k}(X(s))ds\Bigg].
\end{aligned}
\end{equation}

If $\widehat{U}_{\eta}$ is exactly equal to $U$, we obtain from equations~\eqref{eq:var_z} and \eqref{eq:covar_x_z} that:
\begin{equation}
\text{Var}_{x}\Big(Z_{\eta}^{\text{CV}}(f, T, T)\Big) = \text{Cov}_{x}\Big(f(X(T)), \, Z_{\eta}^{\text{CV}}(f, T, T)\Big),
\end{equation}

and from equation~\eqref{eq:variance_def} that:
\begin{equation}
\text{Var}_{x}\Big(Z_{\eta}^{\text{CV}}(f, T, T)\Big) = \text{Var}_{x}(f(X( T))).
\end{equation}

Given these results, equation~\eqref{eq:optimal_rho} directly implies that $\rho^{*}=1$. Consequently, equation~\eqref{eq:var_cv_estimator} simplifies to:
\begin{equation}
\label{eq:zero_variance}
\text{Var}_{x}\Big(E_{\eta}^{\text{CV}}(f,f,T,T)\Big) = 0,
\end{equation}

indicating that the variance of $Z_{\eta}^{\text{CV}}(f, T, T)$ and its correlation with $X(T)$ eliminate all variability in $f(X(T))$. Equation~\eqref{eq:zero_variance} could have been deduced immediately from the almost sure relationship in equation~\eqref{eq:fc_int_mean}~(see the paragraph ``Hedges as controls'' in section 4 of ref.~\cite{glasserman2004monte} for a related observation in the context of Stochastic Differential Equations~(SDEs)). Henceforth, we take $\rho=1$ and omit the coefficient for brevity.\newline

To understand how the variance of $E_{\eta}^{\text{CV}}(f,f,T,T)$ is affected when $\widehat{U}_{\eta}$ is only an approximate version of $U$, we introduce the error function $\varepsilon$ defined as $\varepsilon(x, f, t) \coloneqq U(x, f, t) - \widehat{U}_{\eta}(x, f, t)$. It holds that~(see section~\ref{app:proof} of the supplementary material for a proof):
\begin{equation}
\label{eq:variance_cv}
\text{Var}_{x}\Big(E_{\eta}^{\text{CV}}(f,f,T,T)\Big) = \mathbb{E}_x\Bigg[\sum_{k=1}^{M}\int_{0}^{T} \Big(\Delta_{\zeta_k}\varepsilon(X(s), f, T-s)\Big)^2\lambda_{k}(X(s))ds\Bigg].
\end{equation}

Equation~\eqref{eq:variance_cv} implies, in particular, that the variance of $E_{\eta}^{\text{CV}}(f,f,T,T)$ depends ``smoothly'' on the error $\varepsilon$.\newline

As noted above, performing exact pathwise integration in equation~\eqref{eq:covariate_mean} yields an unbiased estimator. When the integration is instead carried out on a deterministic grid for numerical efficiency, as in equation~\eqref{eq:predictor_fc_int_mean_approx}, a numerical integration error is introduced, leading to a potentially non-zero mean of the integral term. The resulting bias can be made arbitrarily small by refining the time grid used for integration.\newline

\textbf{Comparison between the Deep Learning/Monte Carlo estimators.} The SSA with DeepIS and the SSA with DeepCV, based on $E_{\eta}^{\text{IS}}(f,f,T,T)$ and $E_{\eta}^{\text{CV}}(f,f,T,T)$ respectively, are both unbiased estimators of $\mathbb{E}_x[f(X(T))]$~(see Supplementary Table~\ref{supp_table:dlmc}). They have the potential to be sample-optimal, in the sense that a single trajectory suffices to estimate the expectation at the final time $T$ when the exact function $U$ is used in their computation.

\begin{table}[H]
\centering
\captionsetup{labelfont=bf}
\begin{tabular}{c|c|c|c|}
    & Is unbiased  & Uses $\widehat{U}_{\eta}$ \\ \hline
    \multicolumn{1}{l|}{SSA with DeepIS estimator} & {\textcolor{ForestGreen}{\ding{51}}}
    & {\textcolor{ForestGreen}{\ding{51}}}  \\ \hline
    \multicolumn{1}{l|}{SSA with DeepCV estimator} & {\textcolor{myorange}{\ding{51}}}
    & {\textcolor{ForestGreen}{\ding{51}}}  \\ \hline
    \multicolumn{1}{l|}{SSA estimator} & {\textcolor{ForestGreen}{\ding{51}}} & {\textcolor{red}{\ding{55}}}  \\ \hline
\end{tabular}
\caption{\textbf{Main characteristics of the Deep Learning/Monte Carlo estimators of expected outputs.} The SSA with Deep Importance Sampling (SSA with DeepIS) estimator and the SSA with Deep Control Variates (SSA with DeepCV) estimator are unbiased. When the integration in equation~\eqref{eq:covariate_mean} is performed on a deterministic grid, a numerical integration error is introduced, which can result in a non-zero mean of the integral term. The resulting bias, however, can be made arbitrarily small by refining the time grid used for integration~(hence the yellow tick). Both estimators leverage $\widehat{U}_{\eta}$ for variance reduction.}
\label{supp_table:dlmc}
\end{table}

In practice, the variance of SSA with DeepCV can always be directly compared to that of the standard SSA estimator, since both are computed from the same sampled trajectories. This ensures that, even in the unfavourable scenario where SSA with DeepCV increases variance, the standard SSA estimator can still be recovered at no additional computational cost. Furthermore, in such cases, the optimal value $\rho^*$ can be estimated and used to set the coefficient $\rho$ to a value other than $1$, thereby guaranteeing that the variance of the SSA with DeepCV estimator is reduced relative to the crude SSA estimator.\newline

It is often of interest to estimate $\mathbb{E}_x[f(X(t_i))]$ at multiple times $t_i \in \mathbb{R}_+$. This can be achieved in two ways using the DLMC estimators. The first approach involves simulating the process $(E^{\text{IS}}_{\eta}(f,f,t,t_{i^*}))_{t \in [0, t_{i^*}]}$ or $(E^{\text{CV}}_{\eta}(f,f,t,t_{i^*}))_{t \in [0, t_{i^*}]}$, where the largest time point $t_{i^*}$ is taken as the terminal time. For any earlier time $t_i < t_{i^{*}}$, samples of the random variable $E^{\text{IS}}_{\eta}(f, f,t_i,t_{i^*})$ or $E^{\text{CV}}_{\eta}(f,f,t_i,t_{i^*})$ can be obtained at no additional cost, and these associated estimators remain unbiased. However, they are generally not sample-optimal, even when the exact function $U$ is used. Alternatively, separate processes $(E^{\text{IS}}_{\eta}(f,f,t,t_i))_{t\in[0,t_i]}$ or $(E^{\text{CV}}_{\eta}(f,f,t,t_i))_{t\in[0,t_i]}$ can be simulated, each with its own terminal time $t_i$. Importantly, the SSA with DeepCV estimator can leverage the same trajectories of $(X(t))$ to generate the process $(E^{\text{CV}}_{\eta}(f,f,t,t_i))$ for multiple terminal times $t_i$. In this case, only the covariate process $(Z^{\text{CV}}_{\eta}(f,t,t_i))_{t\in[0,t_i]}$ needs to be adjusted for each terminal time $t_i$. This adjustment involves only evaluating the neural network $\widehat{U}_{\eta}$, without any additional stochastic simulation. Therefore, the SSA with DeepCV estimator is generally expected to be more time-efficient when many times $t_i$ are of interest.\newline

It is also common to estimate $\mathbb{E}_x[f_i(X(T))]$ for multiple output functions $f_i : \mathbb{N}^N \xrightarrow[]{} \mathbb{R}$. Again, this can be done in two ways. The first approach consists of simulating the process $(E^{\text{IS}}_{\eta}(f_{i^*},f_{i^*},t,T))$ or $(E^{\text{CV}}_{\eta}(f_{i^*},f_{i^*},t,T))$, where $f_{i^*}$ is one of the functions of interest. For any other function $f_i \neq f_{i^*}$, the samples of the random variable $E^{\text{IS}}_{\eta}(f_i,f_{i^*},t,T)$ or $E^{\text{CV}}_{\eta}(f_i,f_{i^*},t,T)$ can be computed at no additional cost, and these corresponding estimators remain unbiased. However, they are generally not sample-optimal even when the exact function $U$ is used. Alternatively, distinct processes $(E^{\text{IS}}_{\eta}(f_i,f_i,t,T))$ or $(E^{\text{CV}}_{\eta}(f_i,f_i,t,T))$ can be simulated, each using a different output function $f_i$ as the second argument. As before, the SSA with DeepCV estimator can utilise the same trajectories of $(X(t))$ to generate the process $(E_{\eta}^{\text{CV}}(f_i,f_i,t,T))$ for multiple output functions $f_i$. Consequently, the SSA with DeepCV estimator is typically more time-efficient when many output functions $f_i$ are of interest.\newline

As mentioned earlier, SSA with DeepIS is restricted to positive output functions $f$. This restriction does not apply to SSA with DeepCV.\newline

Finally, the presence of a ratio in equation~\eqref{eq:prop_is} and of a logarithm in equation~\eqref{eq:lr_is} can cause numerical instabilities when $\widehat{U}_{\eta}$ takes very small values. On the other hand, numerical integration of equation~\eqref{eq:covariate_mean} over a deterministic grid may require evaluating $\Delta_{\zeta_k} \widehat{U}_{\eta}$ at negative values of $X(t)+\zeta_k$, where $U$ is not defined. This issue typically arises when the integration against $(\tilde{R}(t))$ is performed on a grid which is too coarse and therefore needs refinement.\newline

\textbf{Evaluation of SDnet through the Deep Learning/Monte Carlo estimators.} The variance and coefficient of variation of $E_{\eta}^{\text{IS}}(f,f,T,T)$ and $E_{\eta}^{\text{CV}}(f,f,T,T)$ reflect the accuracy of $\widehat{U}_{\eta}$ in terms of values directly related to the quantity of interest. In particular, their coefficient of variation can serve as a principled stopping criterion during training.

\subsection{Extensions of the Deep Learning/Monte Carlo estimators}

\textbf{The SSA with Deep Control Variates estimator for the transition kernel.} The control variates scheme introduced above can be applied, as a special case, to the unbiased estimation of the transition kernel $K(x,x',t) \coloneqq \mathbb{P}(X(t)=x'|X(0)=x)$. Suppose that we have obtained an approximation $\widehat{K}$ of $K$, for example, using the methodologies presented in refs.~\cite{sukys2022approximating,liu2024distilling,tang2023neural}, and that we are interested in the transition kernel of $X(t)$ for a~(potentially large) range of values of $x'$. We can compute:
\begin{align}
\delta_{x'}(X(t)) -\sum_{k=1}^{M} \int_{0}^{t} \Delta_{\zeta_k}\widehat{K}(X(s),x',t-s)d\tilde{R}_{k}(s) 
&= \delta_{x'}(X(t)) -\sum_{k=1}^{M} \int_{0}^{t} \Delta_{\zeta_k}\widehat{U}(X(s), \delta_{x'}, t-s)d\tilde{R}_{k}(s)\nonumber\\
&= \delta_{x'}(X(t)) - Z_{\eta}^{\text{CV}}(\delta_{x'}, t, t)
\end{align}

for all values $x'$ of interest, using the same trajectories of $(X(t))$, where $\delta_{x'}$ denotes the Kronecker delta, defined such that $\delta_{x'}(x) = 1$ if $x = x'$, and $\delta_{x'}(x) = 0$ otherwise. This yields a family of unbiased DLMC estimators of the transition kernel at time $t$, simultaneously covering all values of $x'$.\newline

\textbf{Deep Learning/Monte Carlo estimator for the steady-state mean.} We now present a deep control variates scheme for estimating steady-state outputs, following ideas introduced in refs.~\cite{henderson1997variance,milias2014fast}. To this end, we define the finite-time average $\mathcal{E}^{\text{CV}}_{\eta}(f,t)$ as:
\begin{equation}
\label{eq:poisson_cv}
\mathcal{E}^{\text{CV}}_{\eta}(f,t) \coloneqq \frac{1}{t}\int_{0}^{t}[f+\mathbb{A}\widehat{F}_{\eta}](X(s))ds,
\end{equation}

where $\widehat{F}_{\eta}$ employs P-SDnet, and $\mathbb{A}\widehat{F}_{\eta}$ is computed using equations~\eqref{eq:def_generator} and \eqref{eq:poisson_spectral_approx}, as follows:
\begin{equation}
\label{eq:operator_on_poisson}
\mathbb{A}\widehat{F}_{\eta}(x) = \sum_{k=1}^{M} \lambda_k(x)\Bigg[\sum_{\ell=1}^{r}\frac{\hat{\gamma}_{\ell}(f)}{\hat{\sigma}_{\ell}}\Delta_{\zeta_k}\hat{\phi}_{\ell}(x)\Bigg].
\end{equation}

As earlier,  in equation~\eqref{eq:operator_on_poisson} $\hat{\gamma}_{\ell}(f)$,  $\hat{\sigma}_{\ell}$ and $\hat{\phi}_{\ell}$ are approximations of the function coordinates, decay modes and eigenfunctions. From the ergodic theorem, we know  that~\cite{asmussen2007stochastic}:
\begin{equation}
\label{eq:first_part_shadow}
\lim_{t\to\infty}\frac{1}{t}\int_{0}^{t}f(X(s))ds = U(f).
\end{equation}

At stationarity, the Chemical Master Equation~(CME) gives $\mathbb{Q}\pi = 0$~(see equation~\eqref{eq:cme}). Combined with the ergodic theorem, this implies that:
\begin{equation}
\label{eq:second_part_shadow}
\lim_{t\to\infty}\frac{1}{t}\int_{0}^{t}[\mathbb{A}\widehat{F}_{\eta}](X(s))ds = U(\mathbb{A}\widehat{F}_{\eta})=\langle \mathbb{A}\widehat{F}_{\eta}, \pi \rangle = \langle \widehat{F}_{\eta}, \mathbb{A}^{*}\pi \rangle = \langle \widehat{F}_{\eta}, \mathbb{Q}\pi \rangle = 0.
\end{equation}

\begin{figure}[H]
\centering
\captionsetup{labelfont=bf}
\includegraphics[width=\textwidth]{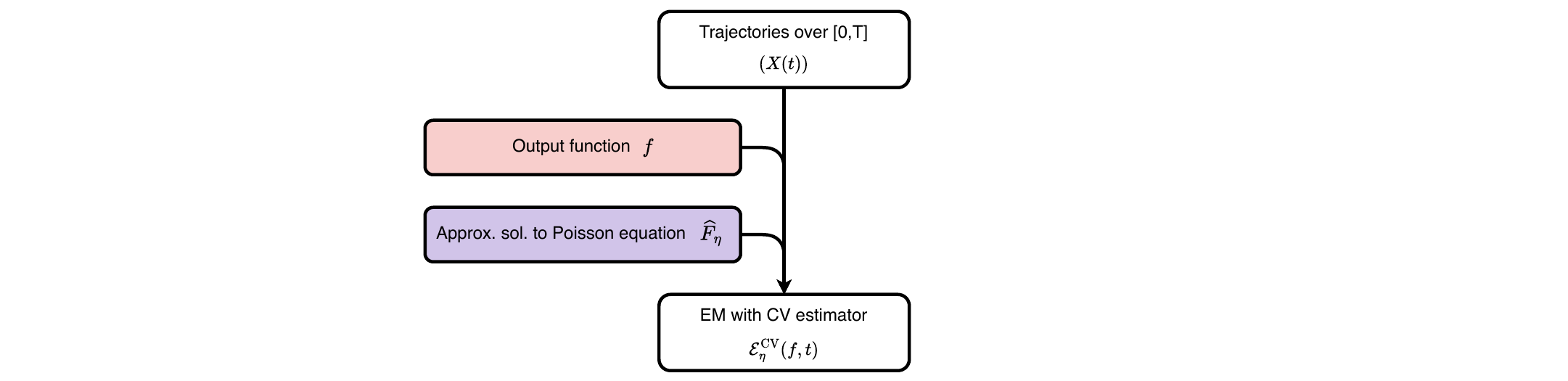}
\caption{\textbf{The Ergodic Mean with Deep Control Variates estimator (EM with DeepCV).} The approximate solution $\widehat{F}_{\eta}$ to the Poisson equation~\eqref{eq:poisson_eq} is used to compute the control variate of the Ergodic Mean with Deep Control Variates (EM with DeepCV) estimator based on the trajectories of $(X(t))$.}
\label{supp_figure:em_deepcv}
\end{figure}

The fact that $U(\mathbb{A}\widehat{F}_{\eta}) = 0$ can be established rigorously using Echeverria’s theorem~(see ref.~\cite{ethier2009markov} and its application in ref.~\cite{gupta2022frequency}). Equations~\eqref{eq:first_part_shadow} and \eqref{eq:second_part_shadow} together imply that $\mathcal{E}^{\text{CV}}_{\eta}(f,t)$ is an exact DLMC estimator of $U(f)$ in the large-time limit. Following ref.~\cite{milias2014fast}, we refer to $\mathbb{A}\widehat{F}_{\eta}$ as a \emph{shadow function} in equation~\eqref{eq:poisson_cv}. We also refer to $\mathcal{E}^{\text{CV}}_{\eta}(f,t)$ as the \emph{Ergodic Mean with Deep Control Variates~\emph{(EM with DeepCV)} estimator}~(see Supp. Fig.~\ref{supp_figure:em_deepcv} for a visual illustration). If $\widehat{F}_{\eta}$ is an exact solution to the Poisson equation~\eqref{eq:poisson_eq}, then it follows immediately from equation~\eqref{eq:poisson_cv} that:
\begin{equation}
\mathcal{E}^{\text{CV}}_{\eta}(f,t) = U(f),
\end{equation}

which implies that $\mathcal{E}^{\text{CV}}_{\eta}(f,t)$ is an exact estimator of the stationary mean $U(f)$, with zero variance for any finite time~$t$.

\begin{figure}[H]
\centering
\captionsetup{labelfont=bf}
\includegraphics[width=\textwidth]{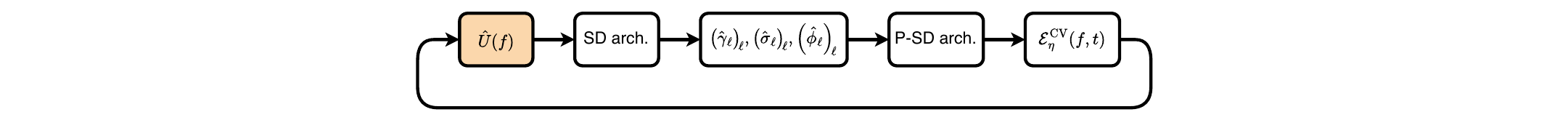}
\caption{\textbf{Iterative loop associated to the framework.}}
\label{supp_figure:iterative_loop}
\end{figure}

The connection between SDnet, P-SDnet, and EM with DeepCV suggests the following iterative loop~(see Supp. Fig.~\ref{supp_figure:iterative_loop} for a visual illustration). Given an initial estimate $\hat{U}(f)$ of $U(f)$, SDnet is trained~(\emph{e.g.}, using the Reinforcement Learning~(RL) method of section~\ref{supp_section:training}), yielding approximations of the function coordinates $(\gamma_{\ell})_\ell$, decay modes $(\sigma_{\ell})_\ell$, and eigenfunctions $(\phi_{\ell})_\ell$. These quantities are then used to construct P-SDnet, which is employed by the EM with DeepCV estimator $\mathcal{E}^{\text{CV}}_{\eta}(f,t)$ to produce an updated estimate $\hat{U}(f)$ of $U(f)$, thus completing the loop.

\textbf{Deep Learning/Monte Carlo estimator for the variance.} Recall from equation~\eqref{eq:variance_int_rep} that the variance $\mathcal{W}$ can be represented as:
\begin{equation*}
\text{Var}_{x}\Big(f(X(t))\Big) = \mathbb{E}_x\Bigg[\sum_{k=1}^{M}\int_{0}^{t} \lambda_{k}(X(s))\Big(\Delta_{\zeta_k}U(X(s), f, t-s)\Big)^2ds\Bigg].
\end{equation*}

\begin{figure}[H]
\captionsetup{labelfont=bf}
\centering
\includegraphics[width=\textwidth]{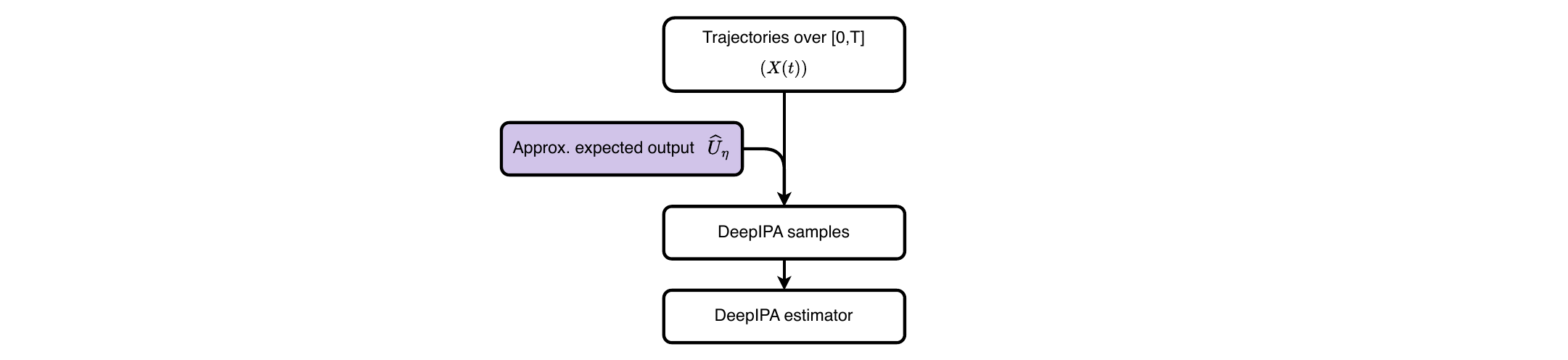}
\caption{\textbf{The Deep Integral Path Algorithm estimator (DeepIPA).} The approximate expected output $\widehat{U}_{\eta}$ is used to evaluate the pathwise integral  shown in equation~\eqref{eq:variance_int_rep}, which is then employed to compute the Deep Integral Path Algorithm~(DeepIPA) estimator.}
\label{supp_figure:deepipa}
\end{figure}

$\widehat{U}_{\eta}$ can be used to approximate $\Delta U$ in this formula, producing what we call the \emph{Deep Integral Path Algorithm \emph{(DeepIPA)} estimator}~(see Supp. Fig.~\ref{supp_figure:deepipa}). Following the approach used to construct the SSA with DeepCV estimator, we can also define a \emph{DeepIPA with Deep Control Variates~\emph{(DeepIPA with DeepCV)} estimator} from equation~\eqref{eq:fc_int_var}. These two DLMC estimators are biased whenever $\Delta \widehat{U}_{\eta}$ is not exact, but they offer the potential for substantially reduced variance.

\textbf{Deep Learning/Monte Carlo estimator for the time average variance constant.} Introduce $\hat{\mu}_{n}(x,f,t)$ as the average of $n$ samples of the random variable $X(t)$, where the process starts from state $x$ at time $t=0$. It is well-known from the Central Limit Theorem~(CLT) that~\cite{asmussen2007stochastic}:
\begin{equation}
\sqrt{n}(\hat{\mu}_{n}(x,f,t) - U(x, f, t)) \xrightarrow[]{\text{dist.}} \mathcal{N}(0, \sigma_{\text{MC}}^2),
\end{equation}

which means that the random variable $\sqrt{n}(\hat{\mu}_{n}(x,f,t) - U(x, f, t))$ converges in distribution to a centred normal distribution with variance $\sigma_{\text{MC}}^2$. Furthermore, it is known that $\sigma_{\text{MC}}^2 = \text{Var}_{x}(f(X(t)))$, which means that $\sigma_{\text{MC}}^2$ can be readily estimated and subsequently used to construct confidence intervals.

\begin{figure}[H]
\centering
\captionsetup{labelfont=bf}
\includegraphics[width=\textwidth]{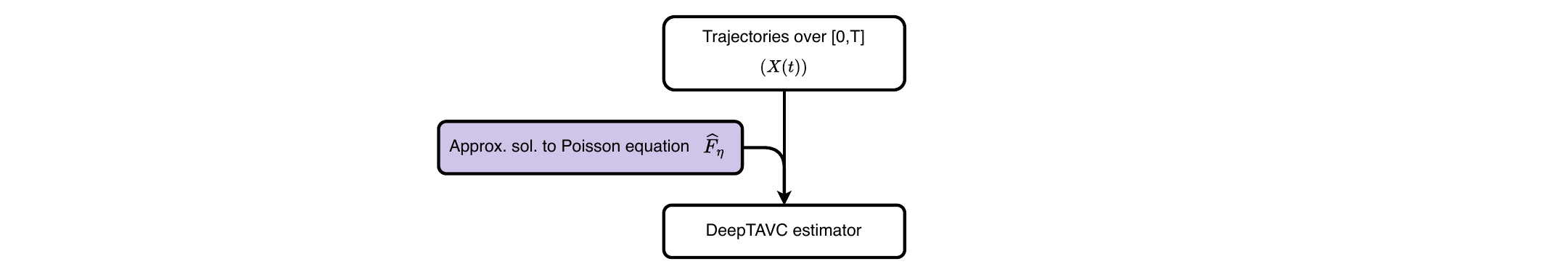}
\caption{\textbf{The Deep Time Average Variance Constant estimator (DeepTAVC).} The approximate solution $\widehat{F}_{\eta}$ to the Poisson equation~\eqref{eq:poisson_eq} is used to evaluate the pathwise integral  shown in equation~\eqref{eq:tav_em}.}
\label{supp_figure:deeptavc}
\end{figure}

An analogous CLT holds for the ergodic mean, which states that~\cite{asmussen2007stochastic}:
\begin{equation}
\sqrt{t}(\mathcal{E}(f,t) - U(f)) \xrightarrow[]{\text{dist.}} \mathcal{N}(0, \sigma_{\text{TAVC}}^{2}(f)),
\end{equation}

where $\sigma_{\text{TAVC}}^{2}(f)$ is known as the Time Average Variance Constant~(TAVC). In general, estimating $\sigma_{\text{TAVC}}^{2}(f)$ is a non-trivial problem~\cite{asmussen2007stochastic}. It can be shown that~(see section~\ref{app:proof} of the supplementary material for a proof):
\begin{equation}
\label{eq:tavc}
\text{TAVC}(f) = \mathbb{E}_{\pi}\left[\sum_{k=1}^{M}\lambda_{k}(\cdot)\Big(\Delta_{\zeta_k}F\Big)^2\right] = \sum_{x\in\mathbb{N}^N}\sum_{k=1}^{M}\lambda_{k}(x)\Big(\Delta_{\zeta_k}F(x)\Big)^2\pi(x),
\end{equation}

which implies that the TAVC can be estimated through the following ergodic average:
\begin{equation}
\frac{1}{t}\int_{0}^{t}\sum_{k=1}^{M}\lambda_{k}(X(s))\Big(\Delta_{\zeta_k}\widehat{F}_{\eta}(X(s))\Big)^2ds.
\label{eq:tav_em}
\end{equation}

We refer to the corresponding estimator as the \emph{Deep Time Average Variance Constant~\emph{(DeepTAVC)} estimator}, which is biased whenever $\widehat{F}_{\eta}$ is not exact~(see Supp. Fig.~\ref{supp_figure:deeptavc} for a visual illustration).

\section{Solving times}
\label{supp_section:solving_times}

\begin{table}[H]
\centering
\captionsetup{labelfont=bf}
\begin{tabularx}{\textwidth}{l l|S[table-format=2.2, table-space-text-post=hours]|X|X|X}
\hline
\rowcolor{gray!20}
& \textbf{Name} & {\textbf{Solving time}} & \textbf{\% for training} & \textbf{\% for SSA} & \textbf{\% for EM} \\
\hline
1. & Const. gene expression & 48.55~min & 95.4 & 0.7 & 3.9 \\
\hline
\rowcolor{gray!20}
2. & Self-reg. gene expression & 52.42~min & 99.8 & 0.2 & 0.1 \\
\hline
3. & Toggle switch & 1.16~h & 98.4 & 0.3 & 1.3 \\
\hline
\rowcolor{gray!20}
4. & SIR & 1.07~h & 99.8 & 0.2 & $<0.1$ \\
\hline
5. & reference-based AIC & 1.23~h & 95.6 & 0.7 & 3.7 \\
\hline
\rowcolor{gray!20}
6. & sensor-based AIC & 1.12~h & 95.3 & 0.8 & 3.9 \\
\hline
7. & Repressilator & 3.73~h & 99.0 & 0.1 & 0.9 \\
\hline
\rowcolor{gray!20}
8. & NCC & 4.59~h & 97.5 & 0.4 & 2.2 \\
\hline
9. & LCF & 4.49~h & 97.6 & 0.4 & 2.0 \\
\hline
\end{tabularx}
\caption{\textbf{Solving times for the stochastic reaction networks studied.}}
\end{table}

\section{Numerical results for additional nonlinear stochastic reaction networks}
\label{supp_section:nonlinear}

\subsection{Self-regulatory gene expression network}

\begin{figure}[H]
\centering
\captionsetup{labelfont=bf}
\includegraphics[width=\textwidth]{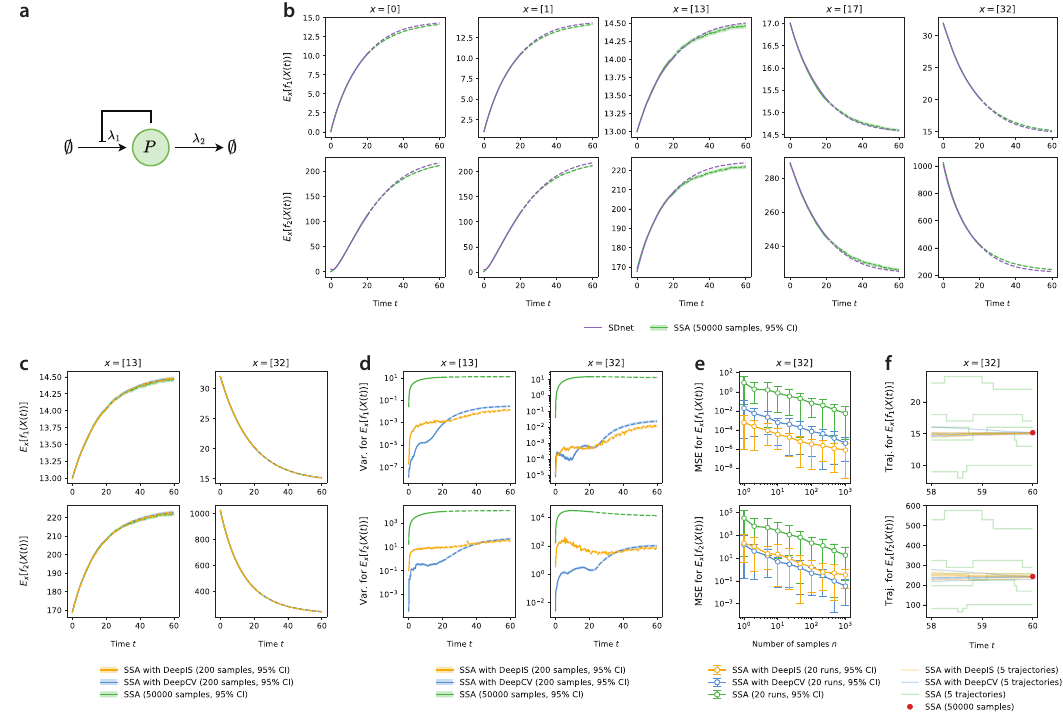}
\caption{\textbf{Results for the self-regulatory gene expression.}  \textbf{a}, Reaction graph of the system. The self-regulatory gene expression network consists of a single protein, $\mathbf{P}$, that represses its own production through a Hill-type inhibition and is degraded according to mass-action kinetics. \textbf{b}, Mean dynamics for different initial states and output functions obtained with the Spectral Decomposition-based network~(SDnet) and the Stochastic Simulation Algorithm~(SSA). Each plot shows the temporal evolution of the mean from time~0 to~60 for a given initial state and output function. Purple lines indicate SDnet predictions, and green bands show the mean and 95\% confidence interval~(CI) computed from 50 000 SSA samples. As in all other plots, solid lines correspond to the training interval $[0,20]$, and dashed lines indicate times beyond the training interval. Initial states vary horizontally from left to right, output functions vary vertically from top to bottom. The first output function corresponds to the first moment of protein $\mathbf{P}$~($f_1(x) = x$), and the second to the second moment~($f_2(x) = x^2$).
\textbf{c}, Mean dynamics under different initial states and output functions obtained with SSA with Deep Importance Sampling~(SSA with DeepIS), SSA with Deep Control Variates~(SSA with DeepCV), and SSA. The layout follows panel~b. Curves for the Deep Learning/Monte Carlo (DLMC) estimators are computed from 200 samples and displayed with 95\% CIs.
\textbf{d}, Temporal evolution of the variance of the estimators for different initial states and output functions. The panel layout mirrors panels~b and~c.
\textbf{e}, Mean-squared error~(MSE) of the estimators at time $t=5$ for the initial state $32$ as a function of sample size. Reference values are computed from SSA with DeepCV using 100 000 samples. The squared error is averaged over 20 independent runs.
\textbf{f}, Five sample paths obtained with SSA with DeepIS, SSA with DeepCV, and SSA. The red dot indicates the reference mean abundance at $t=60$, computed from 50 000 SSA samples.}
\label{supp_figure:slf}
\end{figure}

\subsection{Susceptible–infected–recovered network}

\begin{figure}[H]
\centering
\captionsetup{labelfont=bf}
\includegraphics[width=\textwidth]{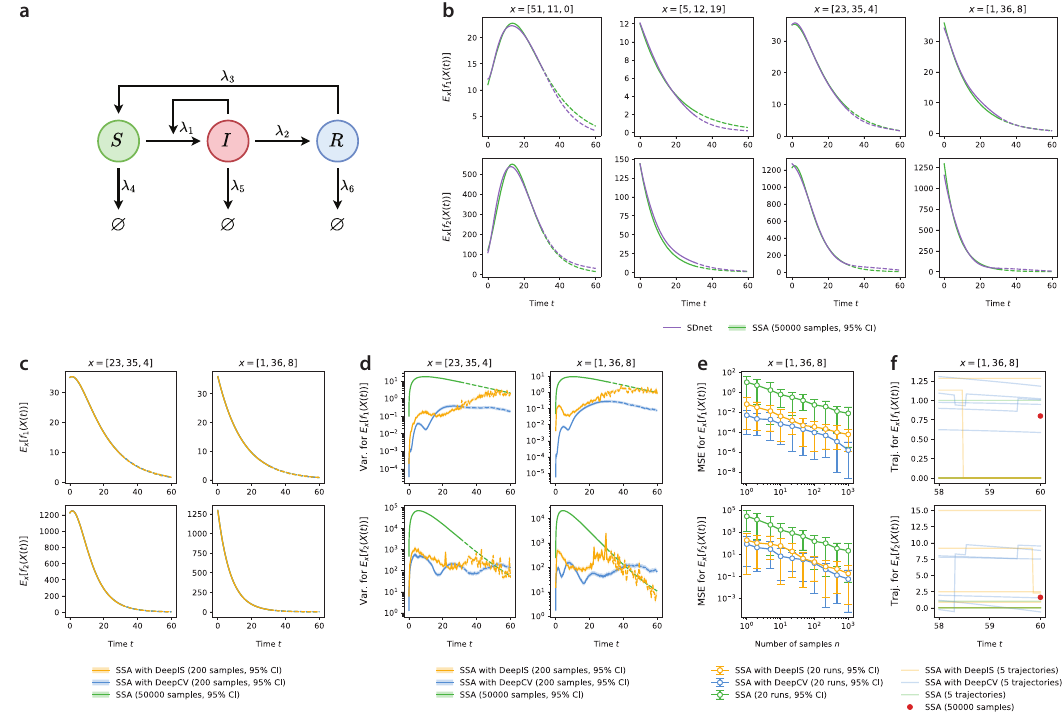}
\caption{\textbf{Results for the susceptible–infected–recovered network.} The conventions are the same as in Fig.~\ref{supp_figure:slf}. In this system, three populations interact: susceptible individuals $\mathbf{S}$, infected individuals $\mathbf{I}$, and recovered individuals $\mathbf{R}$. Infection occurs when a susceptible individual interacts with an infected one. Infected individuals may recover, while recovered individuals lose immunity and return to the susceptible class. Each population undergoes natural death. The reaction graph of this system is displayed in panel~\textbf{a}, the mean dynamics for different initial states and output functions obtained with the Spectral Decomposition-based network~(SDnet) and the Stochastic Simulation Algorithm~(SSA) in panel~\textbf{b}, 
the mean dynamics under different initial states and output functions obtained with SSA with Deep Importance Sampling~(SSA with DeepIS), SSA with Deep Control Variates~(SSA with DeepCV), and SSA in panel~\textbf{c}, 
the temporal evolution of the variance of the estimators for different initial states and output functions in panel~\textbf{d}, the mean-squared error~(MSE) of the estimators at time $t=5$ for the initial state $[1, 36, 8]$ as a function of sample size in panel~\textbf{e}, 
and five sample paths obtained with SSA with DeepIS, SSA with DeepCV, and SSA in panel~\textbf{f}. The first output function corresponds to the first moment of infected individuals $\mathbf{I}$~($f_1(x) = x_2$), and the second to the second moment~($f_2(x) = (x_2)^2$).}
\end{figure}

\subsection{Sensor-based antithetic integral control of gene expression network}

\begin{figure}[H]
\centering
\captionsetup{labelfont=bf}
\includegraphics[width=\textwidth]{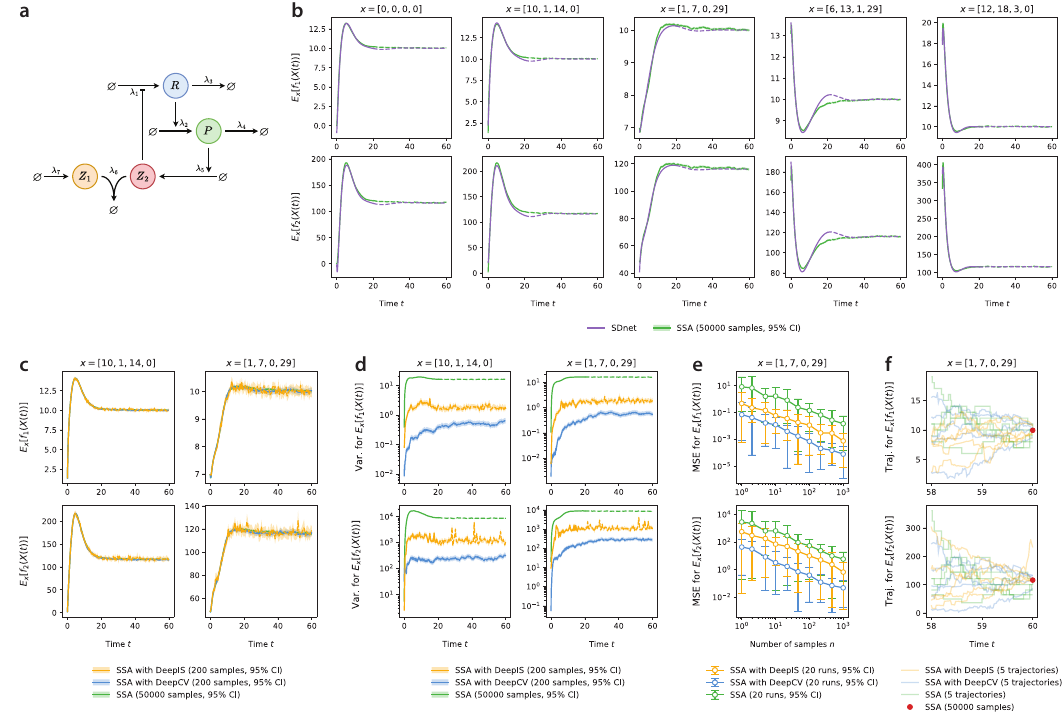}
\caption{\textbf{Results for the sensor-based antithetic integral control of gene expression.} The conventions are the same as in Fig.~\ref{supp_figure:slf}. In this system, the protein $\mathbf{P}$ is regulated by two controller species, $\mathbf{Z_1}$ and $\mathbf{Z_2}$. The species $\mathbf{Z_2}$ is produced in proportion to $\mathbf{P}$~(sensing) and inhibits the transcription of the mRNA $\mathbf{R}$~(actuation). Translation of $\mathbf{R}$ yields $\mathbf{P}$, both $\mathbf{R}$ and $\mathbf{P}$ undergo degradation, and the controller species $\mathbf{Z_1}$ and $\mathbf{Z_2}$ are jointly removed through an annihilation reaction. The reaction graph of this system is displayed in panel~\textbf{a}, the mean dynamics for different initial states and output functions obtained with the Spectral Decomposition-based network~(SDnet) and the Stochastic Simulation Algorithm~(SSA) in panel~\textbf{b}, 
the mean dynamics under different initial states and output functions obtained with SSA with Deep Importance Sampling~(SSA with DeepIS), SSA with Deep Control Variates~(SSA with DeepCV), and SSA in panel~\textbf{c}, 
the temporal evolution of the variance of the estimators for different initial states and output functions in panel~\textbf{d}, the mean-squared error~(MSE) of the estimators at time $t=5$ for the initial state $[1, 7, 0, 29]$ as a function of sample size in panel~\textbf{e}, 
and five sample paths obtained with SSA with DeepIS, SSA with DeepCV, and SSA in panel~\textbf{f}. The first output function corresponds to the first moment of protein $\mathbf{P}$~($f_1(x) = x_2$), and the second to the second moment~($f_2(x) = (x_2)^2$).}
\end{figure}

\subsection{Nonlinear conversion cascade network}

\begin{figure}[H]
\centering
\captionsetup{labelfont=bf}
\includegraphics[width=\textwidth]{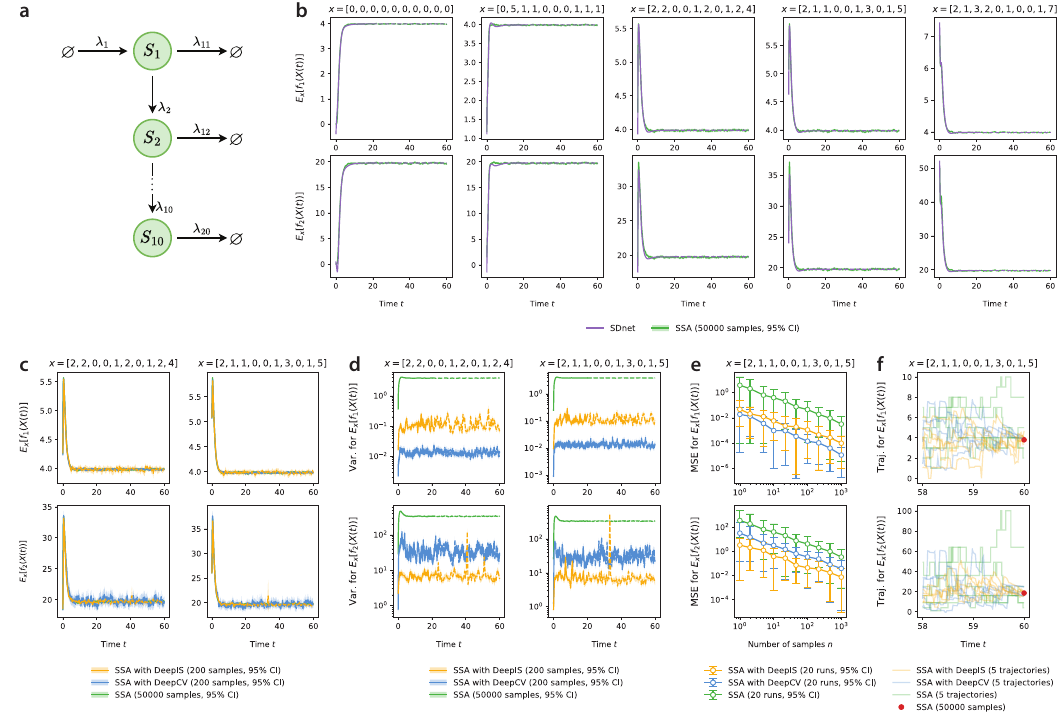}
\caption{\textbf{Results for the nonlinear conversion cascade.}  The conventions are the same as in Fig.~\ref{supp_figure:slf}. In this system, the first species $\mathbf{S_1}$ is produced constitutively. Each species $\mathbf{S_1}, \dots, \mathbf{S_9}$ is sequentially converted into $\mathbf{S_2}, \dots, \mathbf{S_{10}}$ through a Hill-type activation. All species undergo degradation. The reaction graph of this system is displayed in panel~\textbf{a}, the mean dynamics for different initial states and output functions obtained with the Spectral Decomposition-based network~(SDnet) and the Stochastic Simulation Algorithm~(SSA) in panel~\textbf{b}, 
the mean dynamics under different initial states and output functions obtained with SSA with Deep Importance Sampling~(SSA with DeepIS), SSA with Deep Control Variates~(SSA with DeepCV), and SSA in panel~\textbf{c}, 
the temporal evolution of the variance of the estimators for different initial states and output functions in panel~\textbf{d}, the mean-squared error~(MSE) of the estimators at time $t=5$ for the initial state $[2, 1, 1, 0, 0, 1, 3, 0, 1, 5]$ as a function of sample size in panel~\textbf{e}, 
and five sample paths obtained with SSA with DeepIS, SSA with DeepCV, and SSA in panel~\textbf{f}. The first output function corresponds to the first moment of species $\mathbf{S_{10}}$~($f_1(x) = x_{10}$), and the second to the second moment~($f_2(x) = (x_{10})^2$).}
\end{figure}

\section{Numerical results for a linear stochastic reaction network}
\label{supp_section:linear}

\subsection{Constitutive gene expression network}

\begin{figure}[H]
\centering
\captionsetup{labelfont=bf}
\includegraphics[width=\textwidth]{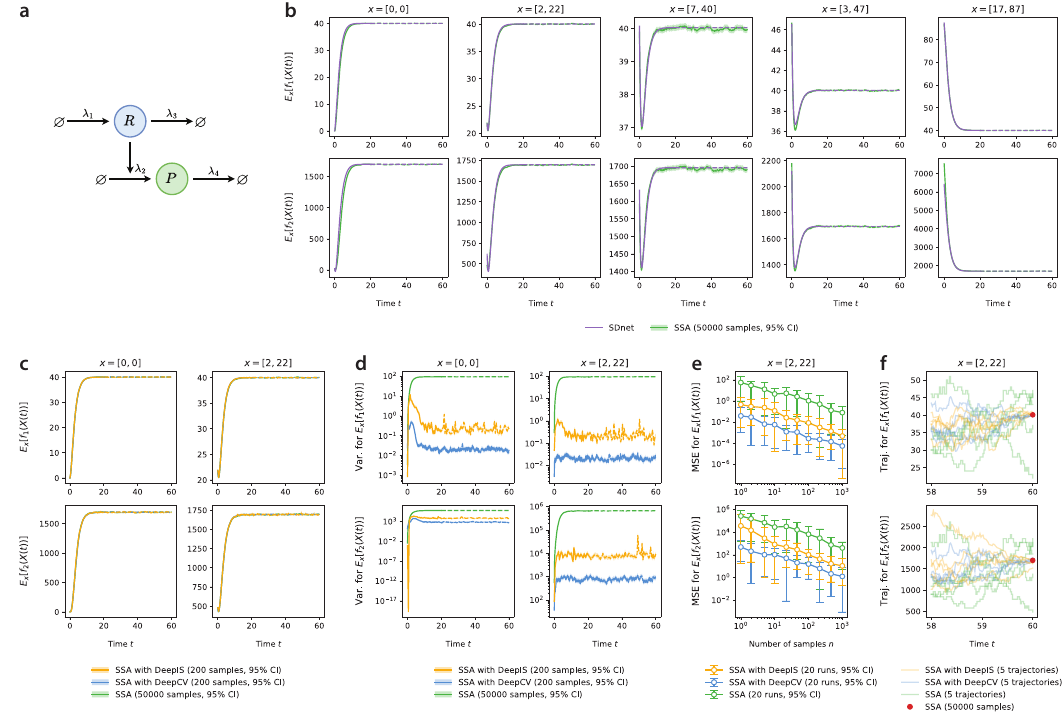}
\caption{\textbf{Results for the constitutive gene expression.}   \textbf{a}, Reaction graph of the system. The constitutive gene expression network consists of an mRNA, $\mathbf{R}$, which is constitutively transcribed, and a protein, $\mathbf{P}$, produced by translation of $\mathbf{R}$. Both $\mathbf{R}$ and $\mathbf{P}$ are subject to degradation. \textbf{b}, Mean dynamics for different initial states and output functions obtained with the Spectral Decomposition-based network~(SDnet) and the Stochastic Simulation Algorithm~(SSA). Each plot shows the temporal evolution of the mean from time~0 to~60 for a given initial state and output function. Purple lines indicate SDnet predictions, and green bands show the mean and 95\% confidence interval~(CI) computed from 50 000 SSA samples. As in all other plots, solid lines correspond to the training interval $[0,20]$, and dashed lines indicate times beyond the training interval. Initial states vary horizontally from left to right, output functions vary vertically from top to bottom. The first output function corresponds to the first moment of protein $\mathbf{P}$~($f_1(x) = x_2$), and the second to the second moment~($f_2(x) = (x_2)^2$).
\textbf{c}, Mean dynamics under different initial states and output functions obtained with SSA with Deep Importance Sampling~(SSA with DeepIS), SSA with Deep Control Variates~(SSA with DeepCV), and SSA. The layout follows panel~b. Curves for the Deep Learning/Monte Carlo (DLMC) estimators are computed from 200 samples and displayed with 95\% CIs.
\textbf{d}, Temporal evolution of the variance of the estimators for different initial states and output functions. The panel layout mirrors panels~b and~c.
\textbf{e}, Mean-squared error~(MSE) of the estimators at time $t=5$ for the initial state $[2, 22]$ as a function of sample size. Reference values are computed from the exact expressions, which are available in this case. The squared error is averaged over 20 independent runs.
\textbf{f}, Five sample paths obtained with SSA with DeepIS, SSA with DeepCV, and SSA. The red dot indicates the reference mean abundance at $t=60$, computed from 50 000 SSA samples.}
\label{supp_figure:cge}
\end{figure}

\clearpage
\section{Numerical results for the Deep Learning/Monte Carlo estimator of the steady-state mean}
\addtocontents{toc}{\protect\setcounter{tocdepth}{1}}
\label{supp_section:ergodic_mean}

In this section, we present a systematic comparison of the Ergodic Mean with Deep Control Variates~(EM with DeepCV)  estimator against the  Ergodic
Mean~(EM) baseline. A comparison of the blue and green bars in panel~\textbf{a}~(\emph{i.e.} of EM with DeepCV and EM over an extended time horizon)  indicates that EM with DeepCV attains the accuracy guaranteed by its theoretical construction. At the same time, the Deep Learning/Monte Carlo~(DLMC) estimator exhibits a pronounced advantage over EM~(in panel~\textbf{b}), achieving a variance reduction of several orders of magnitude relative to EM when evaluated over the same short time horizon (brown bar), and a variance comparable to that of EM computed over a time horizon 100 times longer (green bar).

\subsection{Constitutive gene expression network}

\begin{figure}[H]
\centering
\captionsetup{labelfont=bf}
\includegraphics[width=0.94\textwidth]{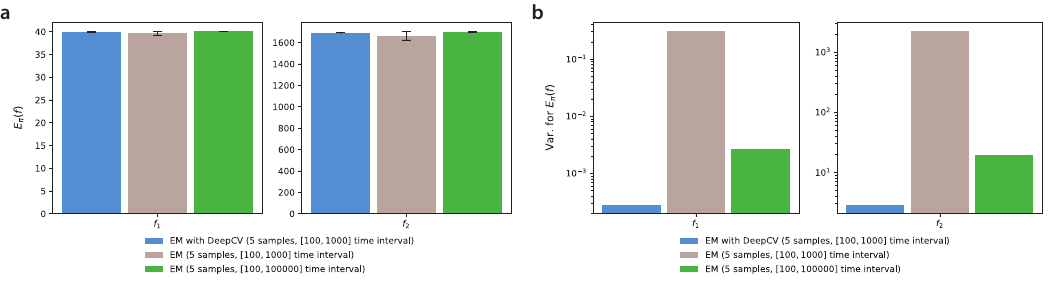}
\caption{\textbf{Steady-state mean for the constitutive gene expression.} \textbf{a}, Steady-state means under different output functions obtained with the Ergodic Mean with Deep Control Variates~(EM with DeepCV) and the Ergodic Mean~(EM). \textbf{b}, Variance of the estimators under different output functions. For both panels, each bar is computed from five independent trajectories generated by the Stochastic Simulation Algorithm~(SSA) from the initial state $[2,22]$. EM with DeepCV was obtained over the time interval $[100, 1000]$, whereas EM was obtained either over the interval $[100, 1000]$ or $[1000, 100000]$. The first output function corresponds to the first moment of protein $\mathbf{P}$~($f_1(x) = x_2$), and the second to the second moment~($f_2(x) = (x_2)^2$).}
\end{figure}

\subsection{Self-regulatory gene expression network}

\begin{figure}[H]
\centering
\captionsetup{labelfont=bf}
\includegraphics[width=0.94\textwidth]{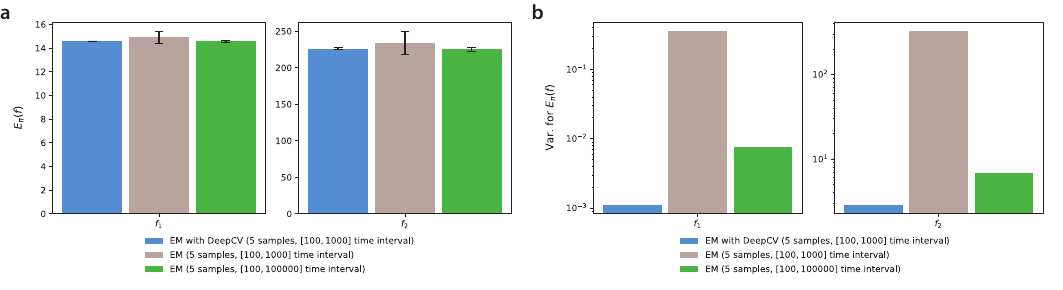}
\caption{\textbf{Steady-state mean for the self-regulatory gene expression.} \textbf{a}, Steady-state means under different output functions obtained with the Ergodic Mean with Deep Control Variates~(EM with DeepCV) and the Ergodic Mean~(EM). \textbf{b}, Variance of the estimators under different output functions. For both panels, each bar is computed from five independent trajectories generated by the Stochastic Simulation Algorithm~(SSA) from the initial state $32$. EM with DeepCV was obtained over the time interval $[100, 1000]$, whereas EM was obtained either over the interval $[100, 1000]$ or $[1000, 100000]$. The first output function corresponds to the first moment of protein $\mathbf{P}$~($f_1(x) = x$), and the second to the second moment~($f_2(x) = x^2$).}
\end{figure}

\subsection{Genetic toggle switch network}

\begin{figure}[H]
\centering
\captionsetup{labelfont=bf}
\includegraphics[width=\textwidth]{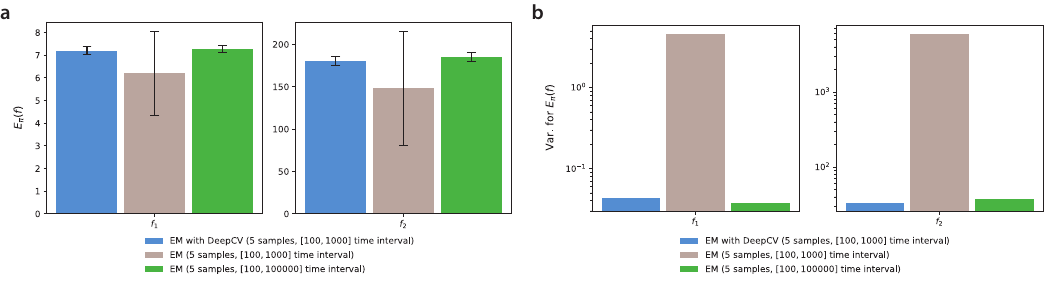}
\caption{\textbf{Steady-state mean for the toggle switch.} \textbf{a}, Steady-state means under different output functions obtained with the Ergodic Mean with Deep Control Variates~(EM with DeepCV) and the Ergodic Mean~(EM). \textbf{b}, Variance of the estimators under different output functions. For both panels, each bar is computed from five independent trajectories generated by the Stochastic Simulation Algorithm~(SSA) from the initial state $[5, 1]$. EM with DeepCV was obtained over the time interval $[100, 1000]$, whereas EM was obtained either over the interval $[100, 1000]$ or $[1000, 100000]$. The first output function corresponds to the first moment of protein $\mathbf{P_1}$~($f_1(x) = x_1$), and the second to the second moment~($f_2(x) = (x_1)^2$).}
\end{figure}

\subsection{Susceptible–infected–recovered network}

\begin{figure}[H]
\centering
\captionsetup{labelfont=bf}
\includegraphics[width=0.9\textwidth]{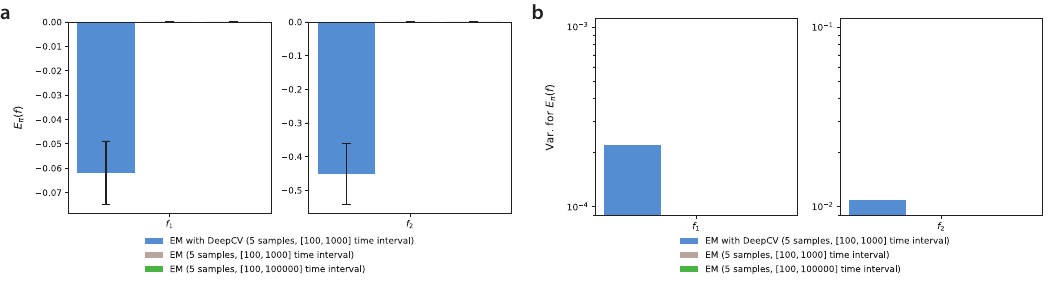}
\caption{\textbf{Steady-state mean for the susceptible–infected–recovered network.} \textbf{a}, Steady-state means under different output functions obtained with the Ergodic Mean with Deep Control Variates~(EM with DeepCV) and the Ergodic Mean~(EM). \textbf{b}, Variance of the estimators under different output functions. For both panels, each bar is computed from five independent trajectories generated by the Stochastic Simulation Algorithm~(SSA) from the initial state $[1, 36, 8]$. EM with DeepCV was obtained over the time interval $[100, 1000]$, whereas EM was obtained either over the interval $[100, 1000]$ or $[1000, 100000]$. The first output function corresponds to the first moment of infected individuals $\mathbf{I}$~($f_1(x) = x_2$), and the second to the second moment~($f_2(x) = (x_2)^2$).}
\end{figure}

\subsection{Reference-based antithetic integral control of gene expression network}

\begin{figure}[H]
\centering
\captionsetup{labelfont=bf}
\includegraphics[width=\textwidth]{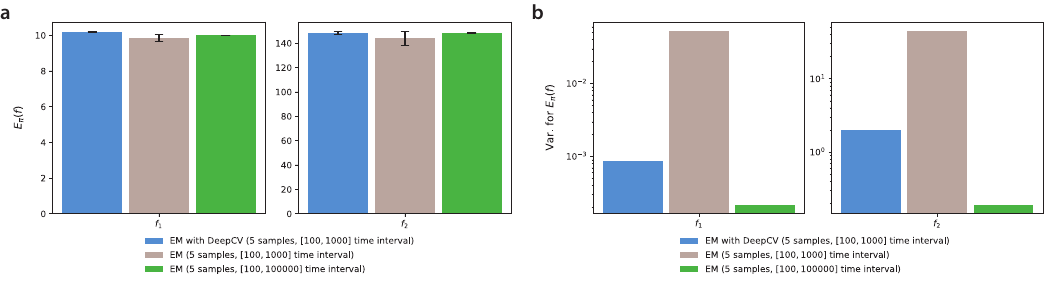}
\caption{\textbf{Steady-state mean for the reference-based antithetic integral control of gene expression.} \textbf{a}, Steady-state means under different output functions obtained with the Ergodic Mean with Deep Control Variates~(EM with DeepCV) and the Ergodic Mean~(EM). \textbf{b}, Variance of the estimators under different output functions. For both panels, each bar is computed from five independent trajectories generated by the Stochastic Simulation Algorithm~(SSA) from the initial state $[0, 6, 0, 22]$. EM with DeepCV was obtained over the time interval $[100, 1000]$, whereas EM was obtained either over the interval $[100, 1000]$ or $[1000, 100000]$. The first output function corresponds to the first moment of protein $\mathbf{P}$~($f_1(x) = x_2$), and the second to the second moment~($f_2(x) = (x_2)^2$).}
\end{figure}

\subsection{Sensor-based antithetic integral control of gene expression network}

\begin{figure}[H]
\centering
\captionsetup{labelfont=bf}
\includegraphics[width=\textwidth]{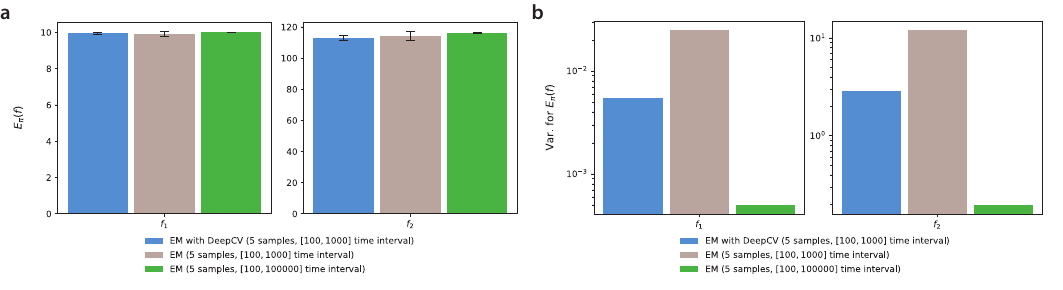}
\caption{\textbf{Steady-state mean for the sensor-based antithetic integral control of gene expression.} \textbf{a}, Steady-state means under different output functions obtained with the Ergodic Mean with Deep Control Variates~(EM with DeepCV) and the Ergodic Mean~(EM). \textbf{b}, Variance of the estimators under different output functions. For both panels, each bar is computed from five independent trajectories generated by the Stochastic Simulation Algorithm~(SSA) from the initial state $[1, 7, 0, 29]$. EM with DeepCV was obtained over the time interval $[100, 1000]$, whereas EM was obtained either over the interval $[100, 1000]$ or $[1000, 100000]$. The first output function corresponds to the first moment of protein $\mathbf{P}$~($f_1(x) = x_2$), and the second to the second moment~($f_2(x) = (x_2)^2$).}
\end{figure}

\subsection{Repressilator network}

\begin{figure}[H]
\centering
\captionsetup{labelfont=bf}
\includegraphics[width=\textwidth]{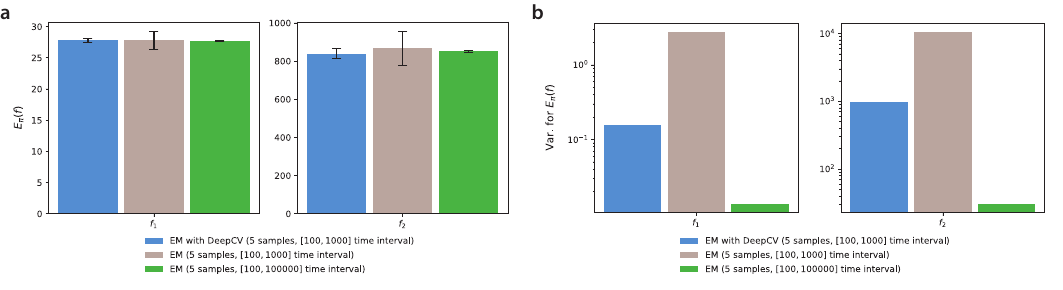}
\caption{\textbf{Steady-state mean for the repressilator.} \textbf{a}, Steady-state means under different output functions obtained with the Ergodic Mean with Deep Control Variates~(EM with DeepCV) and the Ergodic Mean~(EM). \textbf{b}, Variance of the estimators under different output functions. For both panels, each bar is computed from five independent trajectories generated by the Stochastic Simulation Algorithm~(SSA) from the initial state $[1, 1, 1, 37, 40, 11]$. EM with DeepCV was obtained over the time interval $[100, 1000]$, whereas EM was obtained either over the interval $[100, 1000]$ or $[1000, 100000]$. The first output function corresponds to the first moment of protein $\mathbf{P_1}$~($f_1(x) = x_4$), and the second to the second moment~($f_2(x) = (x_4)^2$).}
\end{figure}

\subsection{Nonlinear conversion cascade network}

\begin{figure}[H]
\centering
\captionsetup{labelfont=bf}
\includegraphics[width=\textwidth]{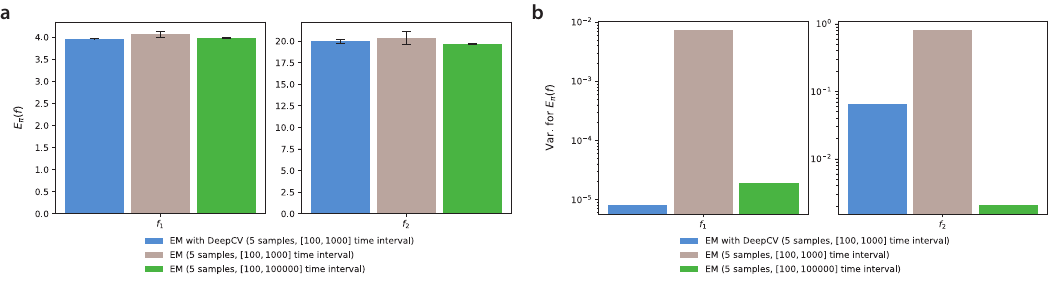}
\caption{\textbf{Steady-state mean for the nonlinear conversion cascade.} \textbf{a}, Steady-state means under different output functions obtained with the Ergodic Mean with Deep Control Variates~(EM with DeepCV) and the Ergodic Mean~(EM). \textbf{b}, Variance of the estimators under different output functions. For both panels, each bar is computed from five independent trajectories generated by the Stochastic Simulation Algorithm~(SSA) from the initial state $[2, 1, 1, 0, 0, 1, 3, 0, 1, 5]$. EM with DeepCV was obtained over the time interval $[100, 1000]$, whereas EM was obtained either over the interval $[100, 1000]$ or $[1000, 100000]$. The first output function corresponds to the first moment of species $\mathbf{S_{10}}$~($f_1(x) = x_{10}$), and the second to the second moment~($f_2(x) = (x_{10})^2$).}
\end{figure}

\subsection{Linear conversion cascade with feedback network}

\begin{figure}[H]
\centering
\captionsetup{labelfont=bf}
\includegraphics[width=\textwidth]{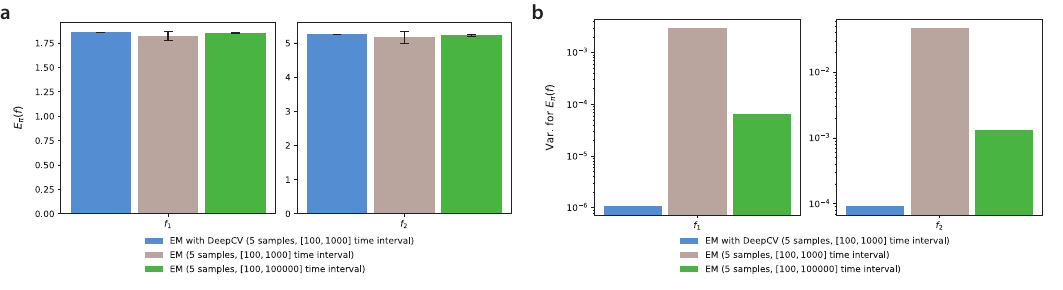}
\caption{\textbf{Steady-state mean for the linear conversion cascade with feedback.} \textbf{a}, Steady-state means under different output functions obtained with the Ergodic Mean with Deep Control Variates~(EM with DeepCV) and the Ergodic Mean~(EM). \textbf{b}, Variance of the estimators under different output functions. For both panels, each bar is computed from five independent trajectories generated by the Stochastic Simulation Algorithm~(SSA) from the initial state $[0, 1, 1, 2, 0, 4, 0, 1, 0, 3]$. EM with DeepCV was obtained over the time interval $[100, 1000]$, whereas EM was obtained either over the interval $[100, 1000]$ or $[1000, 100000]$. The first output function corresponds to the first moment of species $\mathbf{S_{10}}$~($f_1(x) = x_{10}$), and the second to the second moment~($f_2(x) = (x_{10})^2$).}
\end{figure}

\clearpage
\section{Numerical results for the Deep Learning/Monte Carlo estimator of the variance}
\label{supp_section:variance}

In this section, we evaluate the accuracy of the Deep Integral Path Algorithm (DeepIPA) for variance estimation~(in beige), relative to the classical sample variance baseline computed from SSA (in green). The results indicate that, although DeepIPA is generally biased, it maintains strong accuracy and attains high precision even when applied with a limited number of samples.

\subsection{Constitutive gene expression network}

\begin{figure}[H]
\centering
\captionsetup{labelfont=bf}
\includegraphics[width=0.93\textwidth]{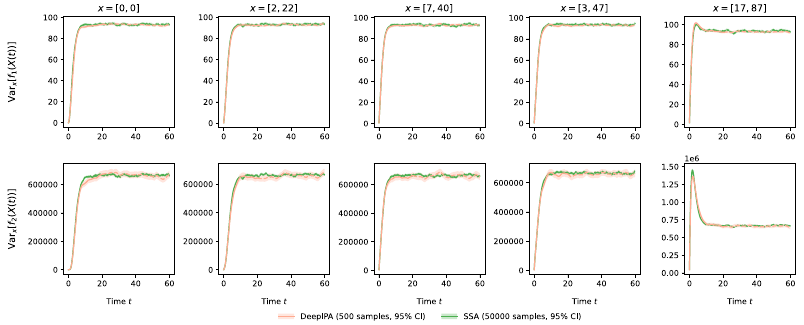}
\caption{\textbf{Variance for the constitutive gene expression.} Variance dynamics under different initial states and output functions obtained with SSA and with the Deep Integral Path Algorithm (DeepIPA). The first output function corresponds to the first moment of protein $\mathbf{P}$~($f_1(x) = x_2$), and the second to the second moment~($f_2(x) = (x_2)^2$).}
\end{figure}

\subsection{Self-regulatory gene expression network}

\begin{figure}[H]
\centering
\captionsetup{labelfont=bf}
\includegraphics[width=0.93\textwidth]{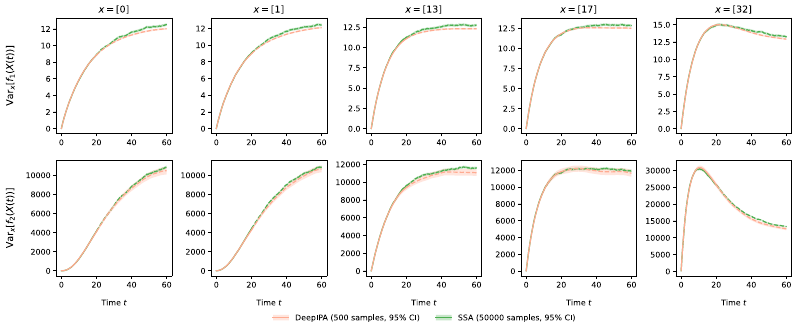}
\caption{\textbf{Variance for the self-regulatory gene expression.} Variance dynamics under different initial states and output functions obtained with SSA and with the Deep Integral Path Algorithm (DeepIPA). The first output function corresponds to the first moment of protein $\mathbf{P}$~($f_1(x) = x$), and the second to the second moment~($f_2(x) = x^2$).}
\end{figure}

\subsection{Genetic toggle switch network}

\begin{figure}[H]
\centering
\captionsetup{labelfont=bf}
\includegraphics[width=\textwidth]{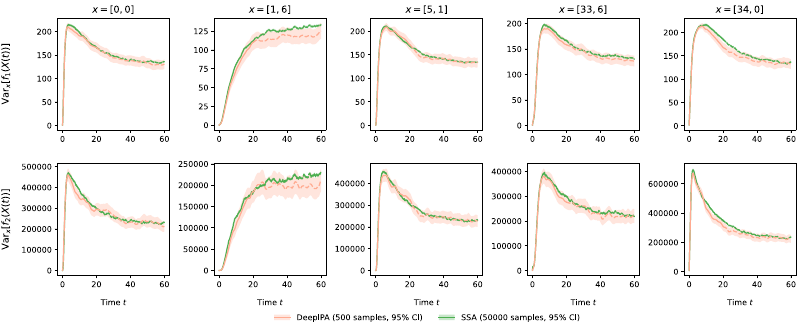}
\caption{\textbf{Variance for the toggle switch.} Variance dynamics under different initial states and output functions obtained with SSA and with the Deep Integral Path Algorithm (DeepIPA). The first output function corresponds to the first moment of protein $\mathbf{P_1}$~($f_1(x) = x_1$), and the second to the second moment~($f_2(x) = (x_1)^2$).}
\end{figure}

\subsection{Susceptible–infected–recovered network}

\begin{figure}[H]
\centering
\captionsetup{labelfont=bf}
\includegraphics[width=\textwidth]{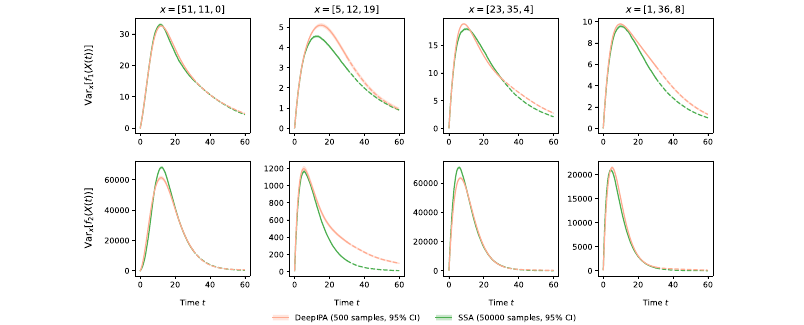}
\caption{\textbf{Variance for the susceptible–infected–recovered network.} Variance dynamics under different initial states and output functions obtained with SSA and with the Deep Integral Path Algorithm (DeepIPA). The first output function corresponds to the first moment of infected individuals $\mathbf{I}$~($f_1(x) = x_2$), and the second to the second moment~($f_2(x) = (x_2)^2$).}
\end{figure}

\subsection{Reference-based antithetic integral control of gene expression network}

\begin{figure}[H]
\centering
\captionsetup{labelfont=bf}
\includegraphics[width=0.93\textwidth]{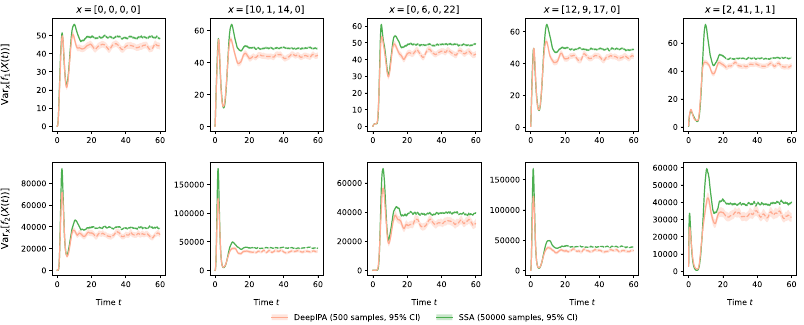}
\caption{\textbf{Variance for the reference-based antithetic integral control of gene expression.} Variance dynamics under different initial states and output functions obtained with SSA and with the Deep Integral Path Algorithm~(DeepIPA). The first output function corresponds to the first moment of protein $\mathbf{P}$~($f_1(x) = x_2$), and the second to the second moment~($f_2(x) = (x_2)^2$).}
\end{figure}

\subsection{Sensor-based antithetic integral control of gene expression network}

\begin{figure}[H]
\centering
\captionsetup{labelfont=bf}
\includegraphics[width=0.93\textwidth]{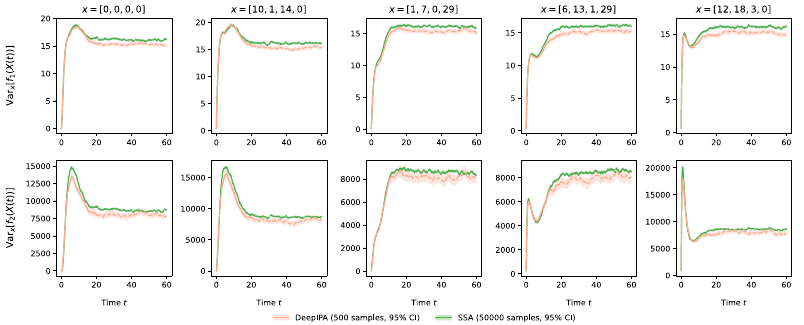}
\caption{\textbf{Variance for the sensor-based antithetic integral control of gene expression.} Variance dynamics under different initial states and output functions obtained with SSA and with the Deep Integral Path Algorithm~(DeepIPA). The first output function corresponds to the first moment of protein $\mathbf{P}$~($f_1(x) = x_2$), and the second to the second moment~($f_2(x) = (x_2)^2$).}
\end{figure}

\subsection{Repressilator network}

\begin{figure}[H]
\centering
\captionsetup{labelfont=bf}
\includegraphics[width=0.93\textwidth]{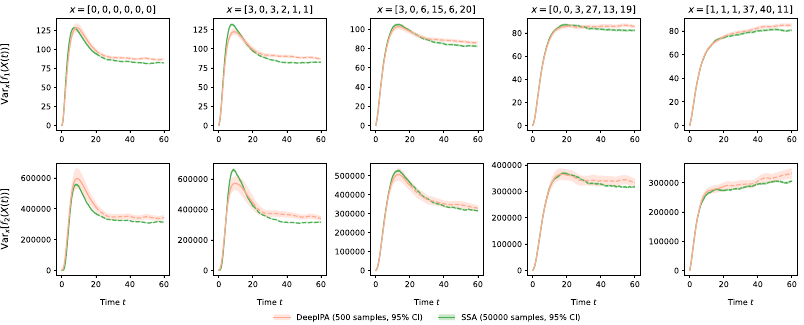}
\caption{\textbf{Variance for the repressilator.} Variance dynamics under different initial states and output functions obtained with SSA and with the Deep Integral Path Algorithm (DeepIPA). The first output function corresponds to the first moment of protein $\mathbf{P_1}$~($f_1(x) = x_4$), and the second to the second moment~($f_2(x) = (x_4)^2$).}
\end{figure}

\subsection{Nonlinear conversion cascade network}

\begin{figure}[H]
\centering
\captionsetup{labelfont=bf}
\includegraphics[width=0.93\textwidth]{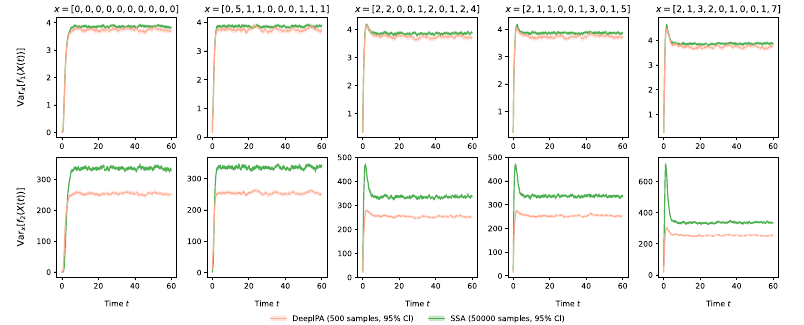}
\caption{\textbf{Variance for the nonlinear conversion cascade.} Variance dynamics under different initial states and output functions obtained with SSA and with the Deep Integral Path Algorithm (DeepIPA). The first output function corresponds to the first moment of species $\mathbf{S_{10}}$~($f_1(x) = x_{10}$), and the second to the second moment~($f_2(x) = (x_{10})^2$).}
\end{figure}

\subsection{Linear conversion cascade with feedback network}

\begin{figure}[H]
\centering
\captionsetup{labelfont=bf}
\includegraphics[width=0.93\textwidth]{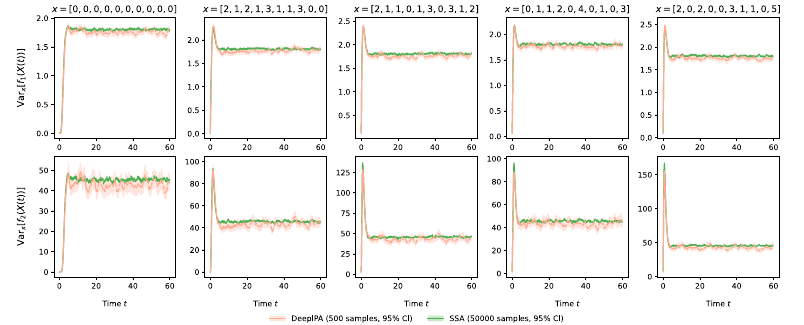}
\caption{\textbf{Variance for the linear conversion cascade with feedback.} Variance dynamics under different initial states and output functions obtained with SSA and with the Deep Integral Path Algorithm (DeepIPA). The first output function corresponds to the first moment of species $\mathbf{S_{10}}$~($f_1(x) = x_{10}$), and the second to the second moment~($f_2(x) = (x_{10})^2$).}
\end{figure}

\clearpage
\addtocontents{toc}{\protect\setcounter{tocdepth}{2}}
\section{Proofs}
\label{app:proof}

From this point onward, we omit the dependency of $U$, $\mathcal{W}$, and $E_{\eta}^{\text{CV}}$ on $f$ whenever it is clear from the context. Throughout the proofs, we make use of two time points, $t_{\min}$ and $t_{\max}$, with $0\leq t_{\min} \leq t_{\max}$. Without loss of generality, we assume that $f$ is scalar-valued.

\subsection{Proofs for section~\ref{supp_section:sdnet}}

The upcoming proof utilises the Poisson equation, Dynkin's formula, and the spectral decomposition of expected outputs in Stochastic Reaction Networks~(SRNs).

\begin{proof}[Proof of equation~\eqref{eq:poisson_spectral}] \textbf{Step 1.} Recall that $F$ is a solution to the Poisson equation given by~(see equation~\eqref{eq:poisson_eq}):
\begin{equation*}
\mathbb{A}F = U(f) - f.
\end{equation*}

Integrating it along trajectories of the process $(X(t))$ over the time interval $[0,t]$, we then have:
\begin{equation}
\int_{0}^{t}\mathbb{A}F(X(s))ds = \int_{0}^{t}\Big(U(f) - f(X(s))\Big)ds.
\end{equation}

Taking the expectation leads to:
\begin{equation}
\label{eq:integrated_poisson_finite}
\mathbb{E}_x\left[\int_{0}^{t}\mathbb{A}F(X(s))ds\right] = \int_{0}^{t}\Big(U(f) - U(x,f,s)\Big)ds.
\end{equation}

\textbf{Step 2.} From Dynkin's formula, we know that~\cite{anderson2015stochastic}:
\begin{equation}
U(x, F, t) = F(x) + \mathbb{E}_x\left[\int_{0}^{t}\mathbb{A}F(X(s))ds\right] \iff \mathbb{E}_x\left[\int_{0}^{t}\mathbb{A}F(X(s))ds\right] = U(x, F, t) - F(x) .
\end{equation}

In the large-time limit, this means that:
\begin{equation}
\mathbb{E}_x\left[\int_{0}^{\infty}\mathbb{A}F(X(s))ds\right] = U(F) - F(x).
\end{equation}

We have chosen $F$ to have zero steady-state mean~(\emph{i.e.} $U(F) = 0$), which implies that:
\begin{equation}
\label{eq:integrated_poisson_lhs}
\mathbb{E}_x\left[\int_{0}^{\infty}\mathbb{A}F(X(s))ds\right] = -F(x).
\end{equation}

\textbf{Step 3.} Recall from the spectral decomposition of expected outputs in equation~\eqref{eq:complex_decomposition_raw} that:
\begin{equation*}
U(x,f,t) =  U(f) + \sum_{\ell=1}^{\infty} e^{-\sigma_{\ell}t}\gamma_{\ell}(f)\phi_{\ell}(x) \iff U(f) - U(x,f,t) = - \sum_{\ell=1}^{\infty} e^{-\sigma_{\ell}t}\gamma_{\ell}(f)\phi_{\ell}(x).
\end{equation*}

Integrating it over the time interval $[0,t]$, we then have:
\begin{align}
\int_{0}^{t}\Big(U(f) - U(x,f,s)\Big)ds &= - \sum_{\ell=1}^{\infty}\gamma_{\ell}(f)\phi_{\ell}(x)\int_{0}^{t} e^{-\sigma_{\ell}s}ds\nonumber\\
&= -\sum_{\ell=1}^{\infty}\frac{\gamma_{\ell}(f) \phi_{\ell}(x)}{\sigma_{\ell}}(1 - e^{-\sigma_{\ell}t})\\
\end{align}

In the large-time limit, this means that:
\begin{equation}
\label{eq:integrated_poisson_rhs}
\int_{0}^{\infty}\Big(U(f) - U(x,f,s)\Big)ds = - \sum_{\ell=1}^{\infty}\frac{\gamma_{\ell}(f)\phi_{\ell}(x)}{\sigma_{\ell}}
\end{equation}

\textbf{Step 4.} In the large-time limit, equation~\eqref{eq:integrated_poisson_finite} yields:
\begin{equation}
\label{eq:integrated_poisson_infinite}
\mathbb{E}_x\left[\int_{0}^{\infty}\mathbb{A}F(X(s))ds\right] =\int_{0}^{\infty}\Big(U(f) - U(x,f,s)\Big)ds.
\end{equation}

Using equations~\eqref{eq:integrated_poisson_lhs} and \eqref{eq:integrated_poisson_rhs} in equation~\eqref{eq:integrated_poisson_infinite} allows us to conclude that:
\begin{equation}
F(x) = \sum_{\ell=1}^{\infty}\frac{\gamma_{\ell}(f)\phi_{\ell}(x)}{\sigma_{\ell}}.
\end{equation}
\end{proof}

\subsection{Proofs for section~\ref{supp_section:training}}

The following proofs build upon the Kolmogorov backward equation and Itô's formula for SRNs. For convenience, the derivations in this subsection will focus on the expected output $V$ defined as:
\begin{equation*}
V(x,f,t,T) = \mathbb{E}[f(X(T)) | X(t)=x].
\end{equation*}

As mentioned earlier, by the time homogeneity of the process $(X(t))$, we have that: $V(x, f, t, T) = U(x, f, T-t)$.

\subsubsection{Proofs for expected outputs}

We begin with derivations that establish two almost sure relationships satisfied by the expected output $V$.

\begin{proof}[Proof of equation~\eqref{eq:fc_int_mean}]
\textbf{Step 1.} Recall that $V$ is the solution of the Kolmogorov backward equation given by~(see equation~\eqref{eq:kb_intro}):
\begin{equation}
0 = \mathbb{A}V(\cdot, s, t_{\max}) + \frac{\partial V}{\partial s}(\cdot, s, t_{\max}),
\end{equation}

with the terminal condition $V(x, t_{\max}, t_{\max}) = f(x)$. Integrating it along trajectories of $(X(t))$ over the time interval $[t_{\min},t_{\max}]$, we then have:
\begin{equation}
\label{eq:intermediate_pinn}
0 = \int_{t_{\min}}^{t_{\max}} \mathbb{A}V(X(s), s, t_{\max}) ds + \int_{t_{\min}}^{t_{\max}} \frac{\partial V}{\partial s} (X(s), s, t_{\max})ds.
\end{equation}

\textbf{Step 2.} Itô’s formula for SRNs states that:
\begin{equation}
\label{eq:ito}
\begin{aligned}
V(X(t_{\max}), t_{\max}, t_{\max}) &= V(X(t_{\min}), t_{\min}, t_{\max})+ \int_{t_{\min}}^{t_{\max}} \frac{\partial V}{\partial s} (X(s), s, t_{\max})ds + \int_{t_{\min}}^{t_{\max}} \mathbb{A}V(X(s), s, t_{\max}) ds\\
&+ \sum_{k=1}^{M}  \int_{t_{\min}}^{t_{\max}}\Delta_{\zeta_k} V(X(s), s, t_{\max}) d\tilde{R}_k(s).
\end{aligned}
\end{equation}

\textbf{Step 3.} Combining equations~\eqref{eq:intermediate_pinn} and~\eqref{eq:ito}, we get:
\begin{equation}
\begin{aligned}
\label{eq:fc_int_mean_proof}
V(X(t_{\max}), t_{\max}, t_{\max}) &= V(X(t_{\min}), t_{\min}, t_{\max}) + \sum_{k=1}^{M}  \int_{t_{\min}}^{t_{\max}}\Delta_{\zeta_k} V(X(s), s, t_{\max}) d\tilde{R}_k(s)\\
\iff f(X(t_{\max})) &= V(X(t_{\min}), t_{\min}, t_{\max}) + \sum_{k=1}^{M}  \int_{t_{\min}}^{t_{\max}}\Delta_{\zeta_k} V(X(s), s, t_{\max}) d\tilde{R}_k(s). 
\end{aligned}
\end{equation}

\end{proof}

\begin{proof}[Proof of equation~\eqref{eq:fc_int_log_mean}] \textbf{Step 1.} Introduce $\mathcal{V}$ as:
\begin{equation}
\mathcal{V}(x,t,T)=-\log\left(V(x,t,T)\right).
\end{equation}

From the chain rule and the Kolmogorov backward equation~(see equation~\eqref{eq:kb_intro}), we know that:
\begin{align}
\frac{\partial \mathcal{V}}{\partial s}(x, s, t_{\max}) &= -\frac{\partial V}{\partial s}(x, s, t_{\max})\frac{1}{V(x, s, t_{\max})}\nonumber\\
&= \frac{\mathbb{A}V(x, s, t_{\max})}{V(x, s, t_{\max})}\nonumber\\
&= \sum_{k=1}^{M} \lambda_k(x)\frac{\Delta_{\zeta_k}V(x,s,t_{\max})}{V(x, s, t_{\max})}\nonumber\\
&= \sum_{k=1}^{M} \lambda_k(x)\left(\Pi_{\zeta_k}V(x,s,t_{\max})-1\right),
\end{align}

with the terminal condition $\mathcal{V}(x, t_{\max}, t_{\max}) = -\log(f(x))$. Integrating it along trajectories of $(X(t))$ over $[t_{\min},t_{\max}]$, we then have:
\begin{equation}
\label{eq:intermediate_log}
\int_{t_{\min}}^{t_{\max}} \frac{\partial \mathcal{V}}{\partial s}(X(s), s, t_{\max})ds = \sum_{k=1}^{M} \int_{t_{\min}}^{t_{\max}} \lambda_k(X(s))\left(\Pi_{\zeta_k}V(X(s),s,t_{\max})-1\right)ds.
\end{equation}

\textbf{Step 2.} Itô’s formula (see for example equation~\eqref{eq:ito}) can be expressed as:
\begin{equation}
\label{eq:ito_bis}
\mathcal{V}(X(t_{\max}), t_{\max}, t_{\max}) = \mathcal{V}(X(t_{\min}), t_{\min}, t_{\max})+ \int_{t_{\min}}^{t_{\max}} \frac{\partial \mathcal{V}}{\partial s}~(X(s), s, t_{\max})ds + \sum_{k=1}^{M}  \int_{t_{\min}}^{t_{\max}}\Delta_{\zeta_k} \mathcal{V}(X(s), s, t_{\max}) dR_k(s).
\end{equation}

\textbf{Step 3.} Combining equations~\eqref{eq:intermediate_log} and~\eqref{eq:ito_bis}, we get:
\begin{equation}
\begin{aligned}
\mathcal{V}(X(t_{\max}), t_{\max}, t_{\max}) &= \mathcal{V}(X(t_{\min}), t_{\min}, t_{\max}) +  \sum_{k=1}^{M}\int_{t_{\min}}^{t_{\max}}  \lambda_k(X(s))\left(\Pi_{\zeta_k}V(X(s),s,t_{\max})-1\right)ds\\
&+ \sum_{k=1}^{M}  \int_{t_{\min}}^{t_{\max}}\Delta_{\zeta_k} \mathcal{V}(X(s), s, t_{\max}) dR_k(s)\\
\iff \log(f(X(t_{\max}))) &= \log(V(X(t_{\min}), t_{\min}, t_{\max}))+  \sum_{k=1}^{M}\int_{t_{\min}}^{t_{\max}}  \lambda_k(X(s))\left(1 -  \Pi_{\zeta_k}V(X(s),s,t_{\max})\right)ds\\
& + \sum_{k=1}^{M}  \int_{t_{\min}}^{t_{\max}}\Delta_{\zeta_k}\log\left(V(X(s),s,t_{\max})\right) dR_k(s). 
\end{aligned}
\end{equation}

\end{proof}

\subsubsection{Proof for the variance}

In the proof of equation~\eqref{eq:fc_int_var}, we make use of the fact that if $W$ is the solution of the following Partial Differential Equation~(PDE):
\begin{equation}
\label{eq:fc_starting_point_implication}
0 = f(x,s,t_{\max}) + \mathbb{A}W(x, s,t_{\max}) + \frac{\partial}{\partial s}W(x, s,t_{\max}),
\end{equation}

then:
\begin{equation}
\label{eq:integral_output}
\int_{t_{\min}}^{t_{\max}} f(X(s),s,t_{\max})ds = W(X(t_{\min}), t_{\min},t_{\max}) + \sum_{k=1}^{M} \int_{t_{\min}}^{t_{\max}}  \Delta_{\zeta_k} W(X(s), s,t_{\max})d\tilde{R}_k(s).
\end{equation}

We establish this implication in the proof below, and then proceed to the derivation of equation~\eqref{eq:fc_int_var}.

\begin{proof}[Proof of the implication \eqref{eq:fc_starting_point_implication} $\!\implies\!$ \eqref{eq:integral_output}] \textbf{Step 1.} Take $W$ to be a solution~(if it exists) of:

\begin{equation}
\label{eq:fc_starting_point}
0 = f(x,s,t_{\max}) + \mathbb{A}W(x, s,t_{\max}) + \frac{\partial}{\partial s}W(x, s,t_{\max}),
\end{equation}

with the terminal condition $W(x, t_{\max},t_{\max}) = 0$. Integrating it along trajectories of $(X(t))$ over $[t_{\min},t_{\max}]$, we then have:
\begin{equation}
0 = \int_{t_{\min}}^{t_{\max}} f(X(s),s,t_{\max})ds +  \int_{t_{\min}}^{t_{\max}}\mathbb{A}W(X(s), s,t_{\max})ds + \int_{t_{\min}}^{t_{\max}} \frac{\partial}{\partial s} W(X(s), s,t_{\max})ds.
\end{equation}

\textbf{Step 2.} Using Itô's formula~(see for example equation~\eqref{eq:ito}), we get:
\begin{equation}
\begin{aligned}
&W(X(t_{\min}), t_{\min},t_{\max}) = W(X(t_{\max}), t_{\max},t_{\max}) + \int_{t_{\min}}^{t_{\max}} f(X(s),s,t_{\max})ds - \sum_{k=1}^{M} \int_{t_{\min}}^{t_{\max}}  \Delta_{\zeta_k} W(X(s), s, t_{\max})d\tilde{R}_k(s)\\
&\iff \int_{t_{\min}}^{t_{\max}} f(X(s),s,t_{\max})ds = W(X(t_{\min}), t_{\min},t_{\max}) + \sum_{k=1}^{M} \int_{t_{\min}}^{t_{\max}}  \Delta_{\zeta_k} W(X(s), s,t_{\max})d\tilde{R}_k(s).
\end{aligned}
\end{equation}

\end{proof}

Having established the implication \eqref{eq:fc_starting_point_implication} $\!\implies\!$ \eqref{eq:integral_output}, we are now ready to prove the almost sure relationship satisfied by the variance $\mathcal{W}$.

\begin{proof}[Proof of equation~\eqref{eq:fc_int_var}] \textbf{Step 1.} Introduce $W$ as $W(x,s,t_{\max}) \coloneqq\mathcal{W}(x,t_{\max}-s)$. We know from the König–Huygens formula that:
\begin{equation}
\label{eq:kh_var}
\begin{aligned}
W(x,s,t_{\max}) &= \mathbb{E}_{x}\left[\left( f(X(t_{\max}-s)) \right)^2\right] - \Big(\mathbb{E}_{x}[f(X(t_{\max}-s))] \Big)^2 = V(x,s,\tilde{f}, t_{\max}) -  \tilde{V}(x,s,f,t_{\max}),\\
\iff &0 =  \tilde{V}(x,s,f,t_{\max}) - V(x,s,\tilde{f}, t_{\max})  + W(x,s,t_{\max}),
\end{aligned}
\end{equation}

where $\tilde{f}$ is defined as $\tilde{f}(x) \coloneqq (f(x))^2$ and $\tilde{V}$ as $\tilde{V}(x,s,f,t_{\max}) \coloneqq (V(x,f,s,t_{\max}))^2$.\newline

\textbf{Step 2.} Differentiating equation~\eqref{eq:kh_var} with respect to the time $s$, we get that:
\begin{equation}
\label{eq:kh_var_diff}
0 =  \frac{\partial\tilde{V}}{\partial s}(x,s,f,t_{\max}) - \frac{\partial V}{\partial s}(x,s,\tilde{f}, t_{\max})  +\frac{\partial W}{\partial s}(x,s,t_{\max}) ,
\end{equation}

Differentiating $\tilde{V}$ with respect to the time $s$, and using the Kolmogorov backward equation~(see equation~\eqref{eq:kb_intro}), we have that:
\begin{align}
 \frac{\partial \tilde{V}}{\partial s}(x,f,s,t_{\max}) &= 2\frac{\partial V}{\partial s}(x,f,s,t_{\max})V(x,f,s,t_{\max})\nonumber\\
 &= - 2\mathbb{A}V(x,f,s,t_{\max}) \times V(x,f,s,t_{\max}).\label{eq:var_part2}   
\end{align}

We also know from the Kolmogorov backward equation that:
\begin{equation}
\label{eq:var_part1}
\frac{\partial V}{\partial s}(\cdot, \tilde{f}, s, t_{\max}) = - \mathbb{A}V(\cdot, s, \tilde{f}, t_{\max}).
\end{equation}

Using equations~\eqref{eq:var_part1} and~\eqref{eq:var_part2} in equation~\eqref{eq:kh_var_diff}, we get that:
\begin{equation}
\label{eq:kh_var_diff_explicit}
0 = -2\mathbb{A}V(x,f,s,t_{\max}) \times V(x,f,s,t_{\max}) + \mathbb{A}V(x,\tilde{f},s,t_{\max}) + \frac{\partial W}{\partial s}(x,s,t_{\max}).
\end{equation}

\textbf{Step 3.} Adding and subtracting $\mathbb{A}\tilde{V}(x,f,s,t_{\max})$ in equation~\eqref{eq:kh_var_diff_explicit}, we then have that:
\begin{equation}
\label{eq:var_intermediate}
\begin{aligned}
&0 =  -2\mathbb{A}V(x,f,s,t_{\max}) \times V(x,f,s,t_{\max}) + \mathbb{A}V(x,\tilde{f},s,t_{\max}) + \mathbb{A}\tilde{V}(x,f,s,t_{\max}) - \mathbb{A}\tilde{V}(x,f,s,t_{\max})+ \frac{\partial W}{\partial s}(x,s,t_{\max})\\
&\iff 0 = \mathbb{A}\tilde{V}(x,f,s,t_{\max}) - 2V(x,f,s,t_{\max})  \times \mathbb{A}V(x,f,s,t_{\max})+ \mathbb{A}V(x,\tilde{f},s,t_{\max}) - \mathbb{A}\tilde{V}(x,f,s,t_{\max})+ \frac{\partial W}{\partial s}(x,s,t_{\max}).
\end{aligned}
\end{equation}

Now also observe that:
\begin{align}
\label{eq:var_part1_bis}
&\mathbb{A}\tilde{V}(x,f,s,t_{\max}) - 2V(x,f,s,t_{\max}) \times \mathbb{A}V(x,f,s,t_{\max})\nonumber \\
&= \sum_{k=1}^M\lambda_{k}(x)(\tilde{V}(x+\zeta_k,s,f,t_{\max})-\tilde{V}(x,s,f,t_{\max}))-2V(x,f,s,t_{\max})\sum_{k=1}^M\lambda_{k}(x) (V(x+\zeta_k,s,f,t_{\max})-V(x,s,f,t_{\max}))\nonumber\\
&= \sum_{k=1}^M\lambda_{k}(x)(\tilde{V}(x+\zeta_k,s,f,t_{\max})-\tilde{V}(x,s,f,t_{\max})-2V(x,f,s,t_{\max}) V(x+\zeta_k,s,f,t_{\max})+ 2V(x,f,s,t_{\max}) V(x,s,f,t_{\max}))\nonumber\\
&= \sum_{k=1}^M\lambda_{k}(x)(\tilde{V}(x+\zeta_k,s,f,t_{\max})-\tilde{V}(x,s,f,t_{\max})-2V(x,f,s,t_{\max}) V(x+\zeta_k,s,f,t_{\max})+ 2\tilde{V}(x,f,s,t_{\max}))\nonumber\\
&= \sum_{k=1}^M\lambda_{k}(x)(\tilde{V}(x+\zeta_k,s,f,t_{\max})-2V(x,f,s,t_{\max}) V(x+\zeta_k,s,f,t_{\max})+\tilde{V}(x,f,s,t_{\max}))\nonumber\\
&= \sum_{k=1}^M\lambda_{k}(x)(V(x+\zeta_k,s,f,t_{\max})-V(x,f,s,t_{\max}))^2\nonumber\\
&= \sum_{k=1}^M\lambda_{k}(x)(\Delta_{\zeta_k}V(x,s,f,t_{\max}))^2.
\end{align}

From equation~\eqref{eq:kh_var}, we also know that $\Delta_{\zeta_k}W(x,s,t_{\max}) =  \Delta_{\zeta_k}V(x,s,\tilde{f}, t_{\max}) -  \Delta_{\zeta_k}\tilde{V}(x,s,f,t_{\max})$. This means that:
\begin{equation}
\label{eq:var_part2_bis}
\begin{aligned}
 \mathbb{A}V(x,\tilde{f},s,t_{\max}) - \mathbb{A}\tilde{V}(x,f,s,t_{\max}) &= \sum_{k=1}^M \lambda_{k}(x)(\Delta_{\zeta_k}V(x,\tilde{f},s,t_{\max})-  \Delta_{\zeta_k}\tilde{V}(x,s,f,t_{\max}))\\
 &= \mathbb{A}W(x,s,t_{\max}).
\end{aligned}
\end{equation}

\textbf{Step 4.} Using equations~\eqref{eq:var_part1_bis} and~\eqref{eq:var_part2_bis} in equation~\eqref{eq:var_intermediate}, we get that:
\begin{equation}
\label{eq:var_intermediate_bis}
0 = \sum_{k=1}^M\lambda_{k}(x)(\Delta_{\zeta_k}V(x,s,f,t_{\max}))^2 + \mathbb{A}W(x,\tilde{f},s,t_{\max})+ \frac{\partial W}{\partial s}(x,s,t_{\max}).
\end{equation}

Introduce in equation~\eqref{eq:var_intermediate_bis}:
\begin{equation}
\label{eq:forcing_term_for_variance}
f(x, s, t_{\max}) \coloneqq \sum_{k=1}^M\lambda_{k}(x)(\Delta_{\zeta_k}V(x,s,f,t_{\max}))^2.
\end{equation}

Equation~\eqref{eq:var_intermediate_bis} can finally be rewritten as:
\begin{equation}
\label{eq:feynman_kac_for_variance}
0 =  f(x,s,t_{\max})  + \mathbb{A}W(x,s,t_{\max})+ \frac{\partial W}{\partial s}(x,s,t_{\max}),  
\end{equation}

with the terminal condition $W(x, t_{\max}, t_{\max}) = 0$. Using the implication \eqref{eq:fc_starting_point_implication} $\!\implies\!$ \eqref{eq:integral_output} concludes the proof.
\end{proof}

\begin{proof}[Proof of equation~\eqref{eq:variance_int_rep}] In most practical scenarios, $(\tilde{R}_k(t))$ is a martingale with zero mean. Taking the expectation on both sides of equation~\eqref{eq:fc_int_var} leads to equation~\eqref{eq:variance_int_rep}.
\end{proof}

\subsection{Proofs for section~\ref{supp_section:dlmc}}

These proofs draw on the isometry property of stochastic integrals, the Central Limit Theorem~(CLT) for Continuous Markov Chains~(CTMCs), the Poisson equation, and the Chemical Master Equation~(CME).

\subsubsection{Proof for the SSA with Deep Control Variates estimator}

\begin{proof}[Proof of equation~\eqref{eq:variance_cv}] We know from the almost sure relationship in equation~\eqref{eq:fc_int_mean} that:
\begin{equation}
f(X(T)) = U(x, T) + \sum_{k=1}^{M}  \int_{0}^{T}\Delta_{\zeta_k} U(X(s), T-s) d\tilde{R}_k(s) 
\iff U(x, T) - f(X(T)) = - \sum_{k=1}^{M}  \int_{0}^{T}\Delta_{\zeta_k} U(X(s), T-s) d\tilde{R}_k(s). 
\end{equation}

This means that:
\begin{equation}
\label{eq:fc_int_mean_proof_error} 
\begin{split}
U(x, T) - \bigg(f(X(T)) - \sum_{k=1}^{M}  \int_{0}^{T}\Delta_{\zeta_k} \widehat{U}_{\eta}(X(s), T-s) &d\tilde{R}_k(s)\bigg)\\ 
&= - \sum_{k=1}^{M}  \int_{0}^{T}\left(\Delta_{\zeta_k} U(X(s), T-s)-\Delta_{\zeta_k} \widehat{U}_{\eta}(X(s), T-s)\right) d\tilde{R}_k(s)\\
\iff U(x, T) - E^{\text{CV}}_{\eta}(T,T) = - \sum_{k=1}^{M}  \int_{0}^{T}&\Delta_{\zeta_k}\varepsilon(X(s), T-s) d\tilde{R}_k(s).
\end{split}
\end{equation}

Squaring equation~\eqref{eq:fc_int_mean_proof_error} and taking its expectation gives:
\begin{equation}
\text{Var}_{x}\Big(E_{\eta}^{\text{CV}}(T,T)\Big) = \mathbb{E}_{x}\bigg[\Big(U(x, T) - E^{\text{CV}}_{\eta}(f,T))\Big)^2\bigg] = \mathbb{E}_{x}\Bigg[\bigg(\sum_{k=1}^{M} \int_{0}^{T}  \Delta_{\zeta_k}\varepsilon(X(s), T-s)d\tilde{R}_k(s)\bigg)^2\Bigg].
\end{equation}

The isometry property of stochastic integrals allows us to conclude that~(see chapter~3 in ref.~\cite{bremaud1981point}):
\begin{equation}
\text{Var}_{x}\Big(E_{\eta}^{\text{CV}}(T,T)\Big) = \mathbb{E}_x\Bigg[\sum_{k=1}^{M}\int_{0}^{T} \Big(\Delta_{\zeta_k}\varepsilon(X(s), T-s)\Big)^2\lambda_{k}(X(s))ds\Bigg].
\end{equation}

\end{proof}

\subsubsection{Proof for the Deep Time Average Variance Constant estimator}

\begin{proof}[Proof of equation~\eqref{eq:tavc}] Given two functions $f$ and $g$ defined on $\mathbb{N}^{N}$ and taking values in $\mathbb{R}$, we define the inner product $\langle \cdot, \cdot \rangle$ as:
\begin{equation}
\langle f, g \rangle \coloneqq \sum_{x\in\mathbb{N}^{N}} f(x) g(x).   
\end{equation}

We know from the CLT for CTMCs that~\cite{milias2014fast,bhattacharya1982functional}:
\begin{equation}
\label{eq:tavc_clt}
\text{TAVC}(f) = -2 \langle F\mathbb{A}F, \pi\rangle,
\end{equation}

where $F$ is a solution to the Poisson equation~\eqref{eq:poisson_eq}. The product rule for the generator $\mathbb{A}$ states that~\cite{gupta2022frequency}:
\begin{equation}
\label{eq:tavc_product_rule}
\mathbb{A}\left(F^2\right) = 2F\mathbb{A}F + \sum_{k=1}^{M} \lambda_{k}(\cdot)\left(\Delta_{\zeta_{k}}F\right)^2 \iff 2F\mathbb{A}F = \mathbb{A}\left(F^2\right) - \sum_{k=1}^{M} \lambda_{k}(\cdot)\left(\Delta_{\zeta_{k}}F\right)^2.
\end{equation}

Also recall that $\mathbb{A}^{*}$ is the adjoint of $\mathbb{Q}$, \emph{i.e.} $\mathbb{A}^{*} = \mathbb{Q}$, and that, at stationarity, the CME gives $\mathbb{Q}\pi = 0$~(see equation~\eqref{eq:cme}). Combined with the bilinearity of the inner product and equation~\eqref{eq:tavc_product_rule},  equation~\eqref{eq:tavc_clt} implies that:
\begin{align}
\text{TAVC}(f) &= -\langle 2  F\mathbb{A}F, \pi\rangle\nonumber\\
&= -\langle \mathbb{A}\left(F^2\right), \pi\rangle + \left\langle \sum_{k=1}^{M} \lambda_{k}(\cdot)\left(\Delta_{\zeta_{k}}F\right)^2, \pi\right\rangle\nonumber\\
&= -\langle F^2, \mathbb{A}^{*}\pi\rangle + \left\langle \sum_{k=1}^{M} \lambda_{k}(\cdot)\left(\Delta_{\zeta_{k}}F\right)^2, \pi\right\rangle\nonumber\\
&= \left\langle \sum_{k=1}^{M} \lambda_{k}(\cdot)\left(\Delta_{\zeta_{k}}F\right)^2, \pi\right\rangle.
\end{align}

\end{proof}
\printbibliography[heading=bibintoc, title={Supplementary references}]
\end{refsection}

\end{document}